\documentclass[a4paper,11pt]{article}
\pdfoutput=1 

\usepackage{jinstpub} 

\usepackage{latexsym,amsmath,subfigure} 
\usepackage[noabbrev,capitalize]{cleveref}

\title{Neutron Tagging following Atmospheric Neutrino Events in 
  a Water Cherenkov Detector}

\newcommand{\AFFicrr}{\affiliation[1]{Kamioka Observatory, Institute for Cosmic Ray Research, University of Tokyo, Kamioka, Gifu 506-1205, Japan}}
\newcommand{\AFFkashiwa}{\affiliation[2]{Research Center for Cosmic Neutrinos, Institute for Cosmic Ray Research, University of Tokyo, Kashiwa, Chiba 277-8582, Japan}}
\newcommand{\AFFicrronly}{\affiliation[3]{Institute for Cosmic Ray Research, University of Tokyo, Kashiwa, Chiba 277-8582, Japan}}
\newcommand{\AFFmad}{\affiliation[4]{Department of Theoretical Physics, University Autonoma Madrid, 28049 Madrid, Spain}}
\newcommand{\AFFbnl}{\affiliation[5]{Physics Department, Brookhaven National Laboratory, Upton, NY 11973, USA}}
\newcommand{\AFFbu}{\affiliation[6]{Department of Physics, Boston University, Boston, MA 02215, USA}}
\newcommand{\AFFbcit}{\affiliation[7]{Department of Physics, British Columbia Institute of Technology, Burnaby, BC, V5G 3H2, Canada }}
\newcommand{\AFFuci}{\affiliation[8]{Department of Physics and Astronomy, University of California, Irvine, Irvine, CA 92697-4575, USA }}
\newcommand{\AFFcsu}{\affiliation[9]{Department of Physics, California State University, Dominguez Hills, Carson, CA 90747, USA}}
\newcommand{\AFFcnm}{\affiliation[10]{Institute for Universe and Elementary Particles, Chonnam National University, Gwangju 61186, Korea}}
\newcommand{\AFFduke}{\affiliation[11]{Department of Physics, Duke University, Durham NC 27708, USA}}
\newcommand{\AFFllr}{\affiliation[12]{Ecole Polytechnique, IN2P3-CNRS, Laboratoire Leprince-Ringuet, F-91120 Palaiseau, France }}
\newcommand{\AFFfukuoka}{\affiliation[13]{Junior College, Fukuoka Institute of Technology, Fukuoka, Fukuoka 811-0295, Japan}}
\newcommand{\AFFgifu}{\affiliation[14]{Department of Physics, Gifu University, Gifu, Gifu 501-1193, Japan}}
\newcommand{\AFFgist}{\affiliation[15]{GIST College, Gwangju Institute of Science and Technology, Gwangju 500-712, Korea}}
\newcommand{\AFFuh}{\affiliation[16]{Department of Physics and Astronomy, University of Hawaii, Honolulu, HI 96822, USA}}
\newcommand{\AFFicl}{\affiliation[17]{Department of Physics, Imperial College London , London, SW7 2AZ, United Kingdom }}
\newcommand{\AFFbari}{\affiliation[18]{ Dipartimento Interuniversitario di Fisica, INFN Sezione di Bari and Universit\`a e Politecnico di Bari, I-70125, Bari, Italy}}
\newcommand{\AFFnapoli}{\affiliation[19]{Dipartimento di Fisica, INFN Sezione di Napoli and Universit\`a di Napoli, I-80126, Napoli, Italy}}
\newcommand{\AFFpadova}{\affiliation[20]{Dipartimento di Fisica, INFN Sezione di Padova and Universit\`a di Padova, I-35131, Padova, Italy}}
\newcommand{\AFFroma}{\affiliation[21]{INFN Sezione di Roma and Universit\`a di Roma ``La Sapienza'', I-00185, Roma, Italy}}
\newcommand{\AFFilance}{\affiliation[22]{ILANCE, CNRS - University of Tokyo International Research Laboratory, Kashiwa, Chiba 277-8582, Japan}}
\newcommand{\AFFkeio}{\affiliation[23]{Department of Physics, Keio University, Yokohama, Kanagawa, 223-8522, Japan}}
\newcommand{\AFFkek}{\affiliation[24]{High Energy Accelerator Research Organization (KEK), Tsukuba, Ibaraki 305-0801, Japan }}
\newcommand{\AFFkcl}{\affiliation[25]{Department of Physics, King's College London, London, WC2R 2LS, UK }}
\newcommand{\AFFkobe}{\affiliation[26]{Department of Physics, Kobe University, Kobe, Hyogo 657-8501, Japan}}
\newcommand{\AFFkyoto}{\affiliation[27]{Department of Physics, Kyoto University, Kyoto, Kyoto 606-8502, Japan}}
\newcommand{\AFFliv}{\affiliation[28]{Department of Physics, University of Liverpool, Liverpool, L69 7ZE, United Kingdom}}
\newcommand{\AFFmiyagi}{\affiliation[29]{Department of Physics, Miyagi University of Education, Sendai, Miyagi 980-0845, Japan}}
\newcommand{\AFFnagoya}{\affiliation[30]{Institute for Space-Earth Environmental Research, Nagoya University, Nagoya, Aichi 464-8602, Japan}}

\newcommand{\AFFpol}{\affiliation[32]{National Centre For Nuclear Research, 02-093 Warsaw, Poland}}
\newcommand{\AFFsuny}{\affiliation[33]{Department of Physics and Astronomy, State University of New York at Stony Brook, NY 11794-3800, USA}}
\newcommand{\AFFokayama}{\affiliation[34]{Department of Physics, Okayama University, Okayama, Okayama 700-8530, Japan }}
\newcommand{\AFFosaka}{\affiliation[35]{Department of Physics, Osaka University, Toyonaka, Osaka 560-0043, Japan}}
\newcommand{\AFFox}{\affiliation[36]{Department of Physics, Oxford University, Oxford, OX1 3PU, United Kingdom}}
\newcommand{\AFFral}{\affiliation[37]{Rutherford Appleton Laboratory, Harwell, Oxford, OX11 0QX, UK }}
\newcommand{\AFFregina}{\affiliation[38]{Department of Physics, University of Regina, 3737 Wascana Parkway, Regina, SK, S4SOA2, Canada}}
\newcommand{\AFFseoul}{\affiliation[39]{Department of Physics, Seoul National University, Seoul 151-742, Korea}}
\newcommand{\AFFsheff}{\affiliation[40]{Department of Physics and Astronomy, University of Sheffield, S3 7RH, Sheffield, United Kingdom}}
\newcommand{\AFFshizuokasc}{\affiliation[41]{Department of Informatics in Social Welfare, Shizuoka University of Welfare, Yaizu, Shizuoka, 425-8611, Japan}}
\newcommand{\AFFstfc}{\affiliation[42]{STFC, Rutherford Appleton Laboratory, Harwell Oxford, and Daresbury Laboratory, Warrington, OX11 0QX, United Kingdom}}
\newcommand{\AFFskk}{\affiliation[43]{Department of Physics, Sungkyunkwan University, Suwon 440-746, Korea}}
\newcommand{\AFFtohoku}{\affiliation[44]{Department of Physics, Faculty of Science, Tohoku University, Sendai, Miyagi 980-8578, Japan}}
\newcommand{\AFFtokai}{\affiliation[45]{Department of Physics, Tokai University, Hiratsuka, Kanagawa 259-1292, Japan}}
\newcommand{\AFFipmu}{\affiliation[46]{Kavli Institute for the Physics and Mathematics of the Universe (WPI), The University of Tokyo Institutes for Advanced Study, University of Tokyo, Kashiwa, Chiba 277-8583, Japan }}
\newcommand{\AFFtokyo}{\affiliation[47]{The University of Tokyo, Bunkyo, Tokyo 113-0033, Japan }}
\newcommand{\AFFtodai}{\affiliation[48]{Department of Physics, University of Tokyo, Bunkyo, Tokyo 113-0033, Japan }}
\newcommand{\AFFtit}{\affiliation[49]{Department of Physics,Tokyo Institute of Technology, Meguro, Tokyo 152-8551, Japan }}
\newcommand{\AFFtus}{\affiliation[50]{Department of Physics, Faculty of Science and Technology, Tokyo University of Science, Noda, Chiba 278-8510, Japan }}
\newcommand{\AFFtoronto}{\affiliation[51]{Department of Physics, University of Toronto, ON, M5S 1A7, Canada }}
\newcommand{\AFFtriumf}{\affiliation[52]{TRIUMF, 4004 Wesbrook Mall, Vancouver, BC, V6T2A3, Canada }}
\newcommand{\AFFtsinghua}{\affiliation[53]{Department of Engineering Physics, Tsinghua University, Beijing, 100084, China}}
\newcommand{\AFFubc}{\affiliation[54]{Department of Physics and Astronomy, University of British Columbia, Vancouver, BC, V6T1Z4, Canada}}
\newcommand{\AFFuw}{\affiliation[55]{Department of Physics, University of Washington, Seattle, WA 98195-1560, USA}}
\newcommand{\AFFwu}{\affiliation[56]{Faculty of Physics, University of Warsaw, Warsaw, 02-093, Poland }}
\newcommand{\AFFwarwick}{\affiliation[57]{Department of Physics, University of Warwick, Coventry, CV4 7AL, UK }}
\newcommand{\AFFwinnipeg}{\affiliation[58]{Department of Physics, University of Winnipeg, MB R3J 3L8, Canada }}
\newcommand{\AFFynu}{\affiliation[59]{Faculty of Engineering, Yokohama National University, Yokohama, Kanagawa, 240-8501, Japan}}
\newcommand{\AFFdeceased}{\affiliation[*]{deceased}}
\author[1,46]{K.~Abe}

\author[1]{Y.~Haga}
\author[1,46]{Y.~Hayato}
\author[1,46]{K.~Hiraide}
\author[1]{K.~Ieki}
\author[1]{M.~Ikeda}
\author[1]{S.~Imaizumi}
\author[1]{K.~Iyogi}
\author[1,46]{J.~Kameda}
\author[1]{Y.~Kanemura}
% ...
\collaboration[60]{Super-Kamiokande Collaboration}

\emailAdd{hayato@icrr.u-tokyo.ac.jp}

\abstract{
We present the development of neutron-tagging techniques in
Super-Kamiokande IV using a neural network analysis. 
The detection efficiency of neutron capture on hydrogen
is estimated to be 
\color{black}26\%, \color{black}
with a mis-tag
rate of 0.016 per neutrino event. 
The uncertainty of the tagging efficiency is estimated to be 9.0\%.
Measurement of the tagging efficiency with data from 
an Americium-Beryllium calibration agrees with this value within 10\%. 
The tagging procedure was performed on 3,244.4 days of SK-IV atmospheric 
neutrino data,
identifying 18,091 neutrons in 26,473 neutrino events. 
The fitted neutron capture lifetime was measured as 218~$\pm$~9$~\mu$s.
}
\begin{document}
\maketitle
\flushbottom
\section{Introduction}

The Super-Kamiokande (SK) water Cherenkov detector is utilized to
study a wide range of physics; it has measured neutrinos from various
sources (solar~\cite{sk4solar}, atmospheric~\cite{skatm2018}, and
accelerator~\cite{t2knue}), while searching for nucleon
decay~\cite{skndecay} and supernova neutrinos~\cite{sksnr}.  
While SK efficiently detects  relativistic 
charged particles with small masses, like electrons, muons, and pions,
heavy particles with low momentum or no charge, such as
protons and neutrons, produce little or no Cherenkov light and 
cannot be easily detected.
However, the ability to detect neutrons, though challenging, 
 is expected to improve the sensitivity of various
analyses~\cite{SKGDProposal}.
As an example, the detection of neutrons can improve the
statistical separation of neutrinos and anti-neutrinos  
since neutrino events are expected to produce fewer neutrons than anti-neutrino
events.
The clearest example is the anti-neutrino charged current
quasi-elastic (CCQE) interaction, which produces neutron 
in the final state but the neutrino CCQE produces proton 
instead.
Improving this separation can enhance sensitivity to the neutrino mass ordering 
via analysis of atmospheric neutrino oscillations. 
Further, the observed number of neutrons is correlated with 
the incident neutrino energy, making it possible to improve estimations 
of the parent energy in atmospheric neutrino interactions.
Detection of neutrons can also help to reduce backgrounds to
nucleon decay searches, since their main backgrounds, 
atmospheric neutrino events, 
are frequently associated with neutrons, while neutron ejection 
from nucleon decay in oxygen is expected to be rare. 
Neutron tagging has been demonstrated as a powerful tool for background reduction in recent nucleon decay searches~\cite{Takenaka:2020vqy}.

\begin{figure}
\begin{center}
\includegraphics[width=0.7\textwidth]{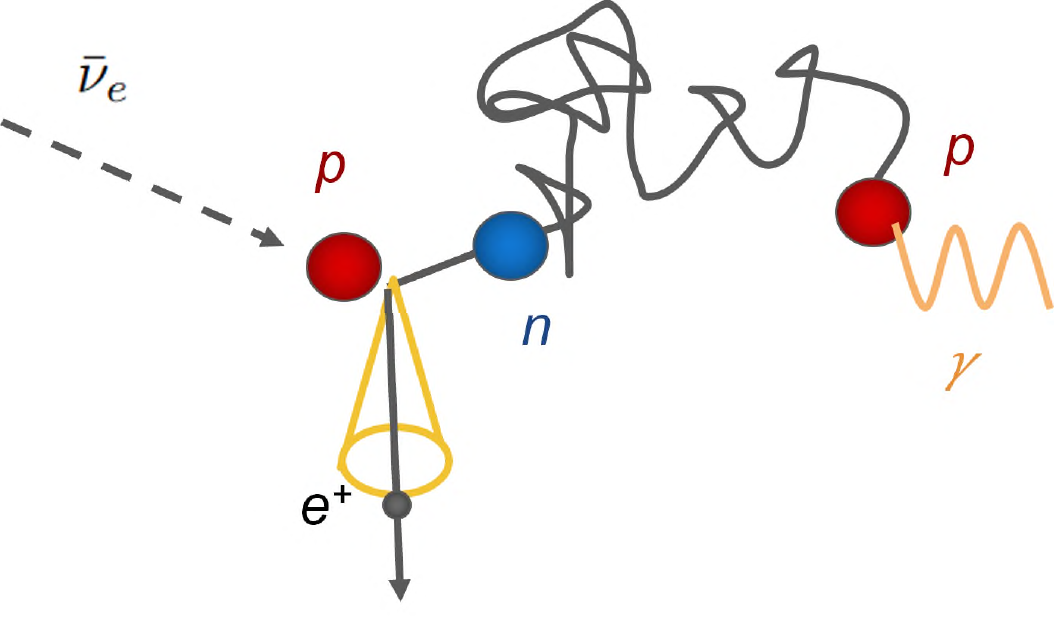}
\caption{A sketch of neutron production and capture by hydrogen.
Anti-neutrino charged current quasi-elastic scattering produces
a neutron. The neutron travels in the SK detector (water), thermalized, 
and eventually capture by hydrogen. When the neutron is captured by
hydrogen, 2.2 MeV gamma-ray is emitted.}
\label{fig:neutron_tag_image} 
\end{center}
\end{figure}

The neutron detection method presented here relies on observing 
the gamma ray produced in neutron capture on hydrogen.
Neutrino or anti-neutrino interaction produce neutrons and
the produced neutrons travel in the SK water and thermalized.
The thermalized neutron will eventually be captured by
an oxygen or hydrogen nucleus, with capture cross sections 
of 0.19 mb and 0.33 b, respectively. 
Therefore, almost all the neutrons are captured by hydrogen, with a characteristic capture time of
$204.8\pm0.4~\mu$s~\cite{nlife}.  
This results in the emission of a 2.2~MeV gamma ray,
\begin{equation}
n + p \rightarrow d + \gamma \text{ (2.2~MeV)}
\end{equation}
as shown in \cref{fig:neutron_tag_image}.

The gamma ray may then scatter electrons (Compton scattering) in the water,
accelerating some of them above the Cherenkov threshold.  
Identifying the light from those electrons can be used to infer the 
presence of the gamma ray and hence its parent neutron.
However, it is not simple to identify 2.2~MeV gamma rays
from neutron captures on hydrogen; the conventional event
reconstruction threshold at SK is $\sim$4MeV, which corresponds
to 34 hit PMTs, while the typical number of hits from
a 2.2~MeV gamma ray is 7. As a result of our inability to
fully reconstruct neutron captures on hydrogen, we instead
developed dedicated neutron tagging techniques.
This neutron tagging technique was first demonstrated in SK in
2009~\cite{ntagpaper} as a means to reduce background events 
in the search for supernova relic neutrinos~\cite{sksnr}. 
\color{black}
Neutron tagging was used in previous atmospheric neutrino and proton 
decay analyses by 
Super-Kamiokande~\cite{Miura:2016krn,TheSuper-Kamiokande:2017tit,Kachulis:2017nci}.
However, a detailed description of the neural network has not yet been 
published. In this paper we provide that description. We also document 
improvements and extensions being used for current analyses. 
Differences between the present method and that used in previous 
analysis’s are described in ~\cref{section:previous_method}. 
In this paper we also demonstrate the performance of the algorithm 
by applying our tagging algorithm to atmospheric neutrino MC and data.
\color{black}

\section{Super-Kamiokande}
\subsection{Detector}
Super-Kamiokande is located 1,000 meters (2,700 meters-water-equivalent)
below Mount Ikenoyama in Gifu,
Japan~\cite{skcalib}.  
It consists of a 50~kiloton (kton) cylindrical tank 
filled with ultra-pure water, which is divided into a 32~kton inner detector (ID) surrounded
by an 18~kton outer detector (OD).  
The ID and the OD are optically separated by Tyvek sheeting. 
The ID is observed by 11,129 inward-facing 20-inch
photo multiplier tubes (PMTs), while the OD is observed by 1,885
outward-facing 8-inch PMTs.  The ID provides most of the information
used in event reconstruction, while the OD is used as an active cosmic
ray veto and to provide information regarding particles escaping from the ID.

\subsection{Event categories}
Generally speaking, events in SK
with more than 100 MeV of deposited energy in the detector are
separated into three categories during data reduction.
Fully-contained (FC) events are events with PMT activity only in the ID.
Partially-contained (PC) events have activity in both the ID and the
OD but are reconstructed to have originated 
from inside the ID.
Upward-going muon (UPMU) events are events with both ID and OD
activity but which are reconstructed to have originated 
from outside the ID and enter from below the horizon.
Accordingly, nearly all UPMU events are muons from neutrino interactions in the rock around
the detector. In this paper, the neutron tagging method is only
applied to the FC event sample.

\subsection{Triggers and events}
The SK data acquisition system was upgraded for the start of the SK-IV 
run period from October 6th, 2008~\cite{sk4elec}. 
The hardware trigger used during the previous periods,  
SK I-III, was replaced with a software trigger in SK-IV. 
This new software trigger allows different timing gate widths to be set 
depending on the nature of a particular event.  
There are five standard
triggers used in SK-IV, which are summarized in ~\cref{triggertable4}.  
For example, the SHE (Super High Energy) trigger threshold of
70 hits (reduced to 58 hits later on in SK-IV) in a 200 ns window corresponds
to the number of hits an electron of about 10~MeV (8~MeV) produces in
the detector. 
 While the 40 $\mu$s gate width of the SHE trigger is
long enough to record the hits from relativistic particles and their
decay products, the longer neutron capture lifetime means that only
around 15\% of thermal neutrons are captured before the end of the SHE
gate.  
In order to detect hits from later neutron captures, all
SHE triggers which do not have a corresponding OD trigger are followed
by an additional trigger called AFT (AFTer trigger), which records 
from 35~$\mu$s to 535~$\mu$s after the SHE trigger was issued. 
At the beginning of SK-IV, the event gate width was up to 385~$\mu$s, instead of 
535~$\mu$s. During this period, the efficiency is $\sim14$\% lower
than the later data taking period with the longer event gate width.
However, this configuration was used 
for just 30 days, representing less than 1\% of the current data set. 
Therefore, the effect of this period is neglected in the
discussions that follow. In order to reduce the
total amount of data, the AFT trigger is not issued following an OD
trigger, because an OD trigger indicates that the event is not 
categorized as FC but as cosmic ray, PC, or UPMU.
As a result, the neutron tagging method can only be applied 
to FC atmospheric neutrino or nucleon decay candidate events.
We estimate that the combined SHE and AFT triggers cover 93\% of 
neutron captures from FC events.

\begin{table}[!ht]
\centering
\caption{Trigger information for SK-IV. The abbreviations are as
  follows: OD (Outer Detector), SLE (Super Low Energy), HE (High
  Energy), SHE 
(Super High Energy)
  and AFT (After). 
  There are
  $\sim$9 hits of dark noise in 200~ns, and 6 hits 
  correspond to  $\sim$1~MeV electron-equivalent energy.
\color{black}
  There are two trigger threshold values for SLE and SHE
  in the table. In the beginning of SK4, we set the threshold
  of SLE to 34 hits but later the threshold value was lowered 
  by 3 hits. Similarly, the SHE threshold was changed from
  70 to 58 during SK4.
\color{black}
}
\label{triggertable4}
\begin{tabular}{|c|c|c|}
\hline 
\hline 
SK-IV Triggers & Hits/200 ns Threshold & Event Width ($\mu$s) \\ 
\hline 
OD & 22 (in OD) & $-5\rightarrow$ 35\\ 
SLE & 34 -- 31 &
$-0.5\rightarrow$ 1.0 \\ 
HE & 50 & $-5\rightarrow$ 35 \\ 
SHE & 70 -- 58 & $-5\rightarrow$ 35\\ 
AFT & SHE, no OD & 35 $\rightarrow$ 535\\ 
\hline
\end{tabular}
\end{table}

\section{Simulation}
\label{neutronsimulation}
The atmospheric neutrino Monte-Carlo (MC), corresponding to 
a 500 years exposure of the detector, is 
used to optimize and estimate the performance of the
neutron-tagging software.  
The atmospheric neutrino flux is provided
by Honda {\it et al.}~\cite{honda11}. The interactions of neutrinos in
water are simulated using the NEUT simulation software 
(v5.3.6)~\cite{neut1,skatm2018}.
NEUT also simulates nuclear interactions to propagate
particles through the nuclei in which they were created.  
Particles are then propagated through the detector; 
its response 
is simulated by a GEANT3~\cite{Brun:1987ma} based simulation of SK called
SKDETSIM~\cite{Ashie:2005ik}. 
Hadrons except for low momentum pions are simulated by the GEANT3
interface with the CALOR~\cite{Zeitnitz:1994bs} package, which uses
HETC~\cite{calorhetc} for hadrons below 10~GeV, FLUKA
(GFLUKA)~\cite{calorfluka} for hadrons above 10~GeV, and
MICAP~\cite{calormicap} for neutrons below 20~MeV.
The propagation of pions below 500 MeV/c in the water are simulated
by NEUT. The low energy cutoff for neutral hadrons is set to $10^{-5}$~eV so that neutrons
continue to be simulated until they are captured.

\begin{figure}
\includegraphics[width=\textwidth]{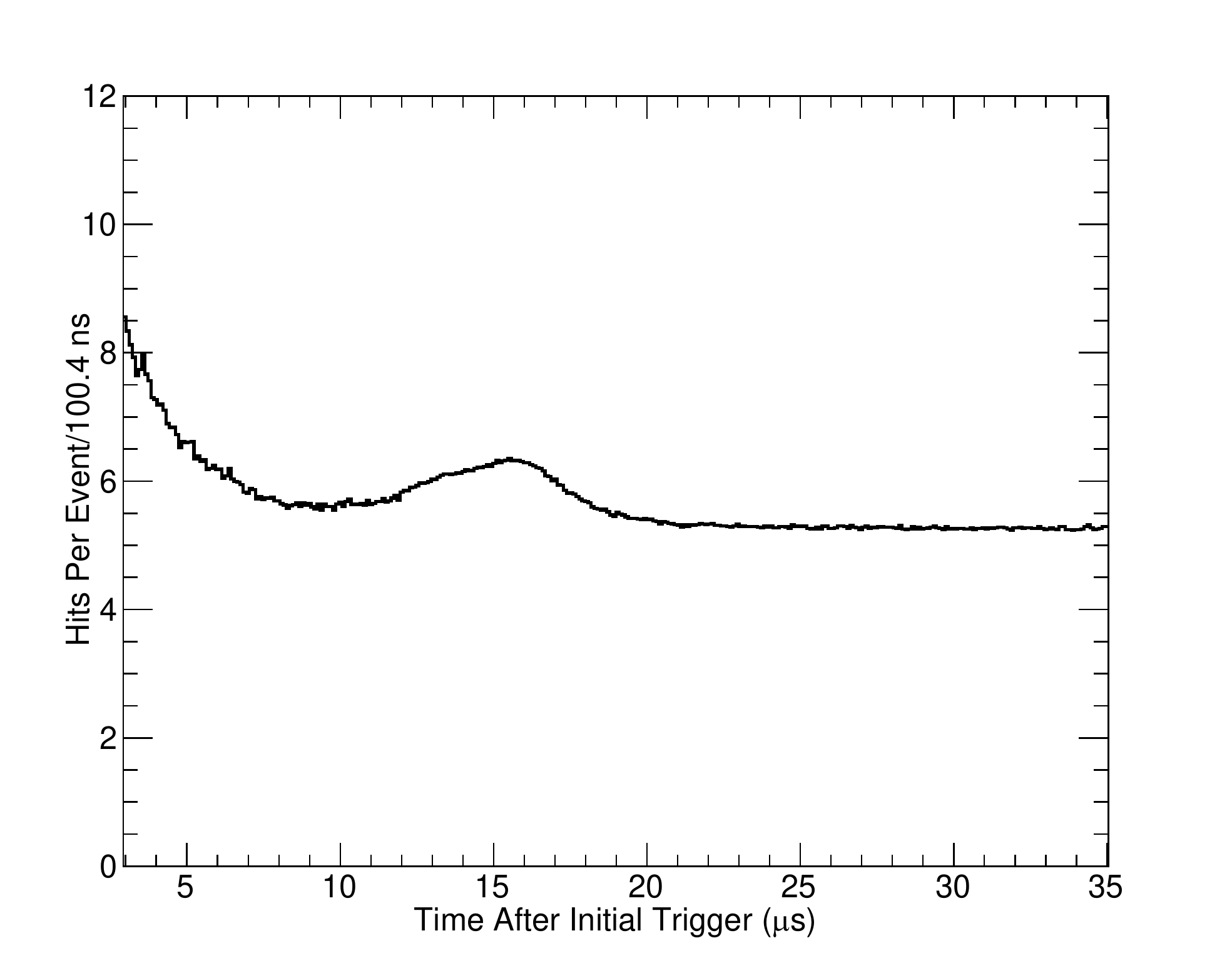}
\caption{Hit rates over the course of an event, taken from the average
  of SK-IV events.  Note the suppressed zero on the x-axis.  The
  falling exponential on the left of the plot is due to decay
  electrons.  The increase due to PMT after-pulsing in the 12 to 18
  $\mu$s region can be clearly seen.  After the PMT after-pulsing, the
  hit rate is flat for the rest of the event
  window.}
\label{fig:event_hit_timing} 
\end{figure}

Uncorrelated PMT dark noise is the only source of background hits
simulated by SKDETSIM. 
PMT after-pulsing, which occurs between 12 and
18 $\mu$s, 
creates a slight increase
in the hit rate as seen in ~\cref{fig:event_hit_timing}, 
but it is not modeled in SKDETSIM.
Therefore, the search for neutron capture begins 18 $\mu$s after the
primary trigger in order to avoid possible biases due to this 
after-pulsing. Based on the MC result, assuming 
the previously
measured capture lifetime of neutrons in water, $204.8~\mu$s~\cite{nlife},
this reduces the maximum efficiency for
neutron capture from 93\% to 84\%.  
Other low energy sources in the detector, such as radioactive decays from
the surrounding rock, radon contamination in the water, and
radioactive contaminants in the tank structure~\cite{NAKANO2017108,NAKANO2020164297} also
generate random background hits.
These backgrounds do not affect the reconstruction of higher energy particles, so
they have been neglected in the simulation program for the atmospheric
neutrino samples. In addition, hits from low energy sources are correlated in space and time, yet the
low energy sources are difficult to model.
Background hits from low energy backgrounds could mimic
a 2.2~MeV gamma ray signal from neutron capture, which only produces
around 7 hits in the detector. Therefore, it is 
necessary to accurately account for them.
Our method is to overlay randomly triggered data events to account for both 
noise from the PMTs and these unmodeled processes.

\begin{figure}[!ht]
	\includegraphics[width=1.\textwidth]{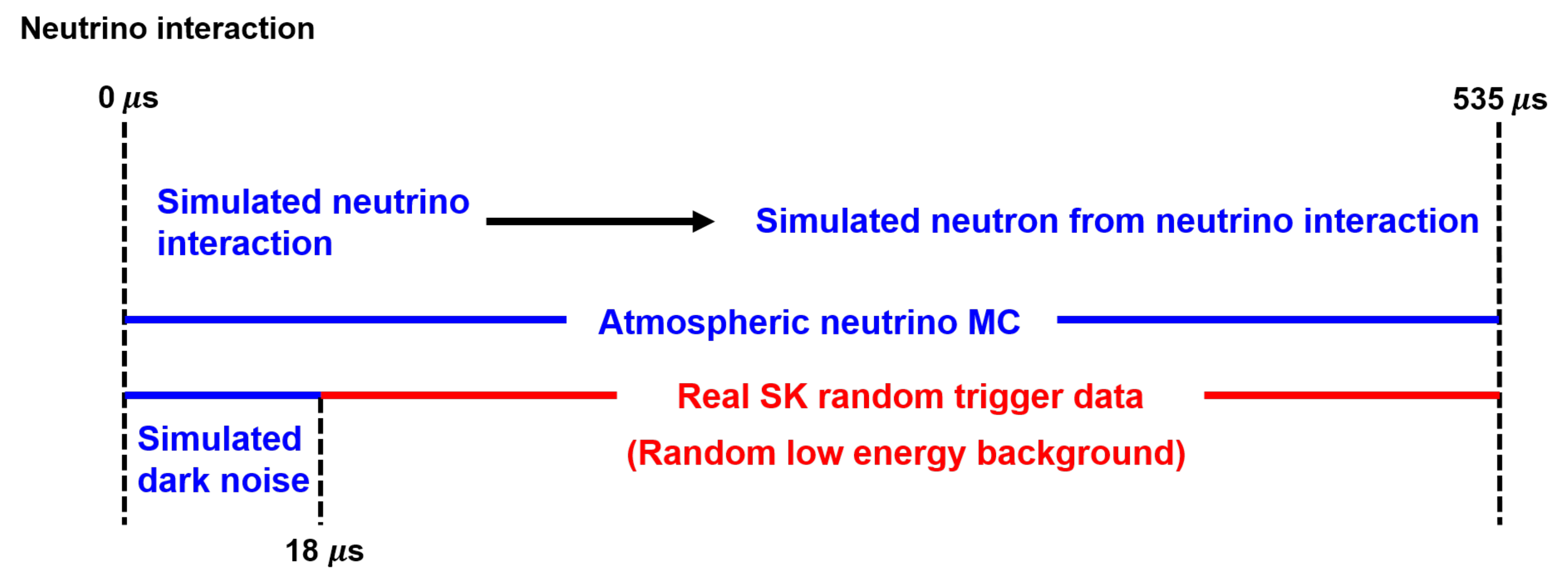}
    \centering
	\caption{Diagram of MC construction procedure. Random trigger data 
      are superimposed on simulated PMT hits from neutron capture events 
      18~$\mu$s after the primary interaction.} 
	\label{fig:MC_gen_pic}
\end{figure}

In total, about 1.9 million random trigger events with a gate width of 
1\ ms were collected in 2009.
When a MC event is produced, PMT dark noise 
is simulated by SKDETSIM up to 18 $\mu$s and thereafter 
the dark noise is provided by real hits 
from the random trigger events.
The 900 ns
channel dead-time of the SK-IV digitizer is modeled by removing any
hit occurring less than 900 ns after a previous hit. 
This hybrid MC technique is illustrated in ~\cref{fig:MC_gen_pic}.  
Our 500 year MC samples contain  about 2.5 million FC events,
which means that some random trigger events are shared between 
two MC events due to the shortage of the random trigger events.
In these cases the first 517 $\mu$s of a random
trigger event is used for one event and the hits from 483 $\mu$s
to 1 ms are used for the other event, a slight overlap with negligible effect on the analysis. 

\section{Neutron Tagging Algorithm}
\begin{figure}[!ht]
	\includegraphics[width=1.\textwidth]{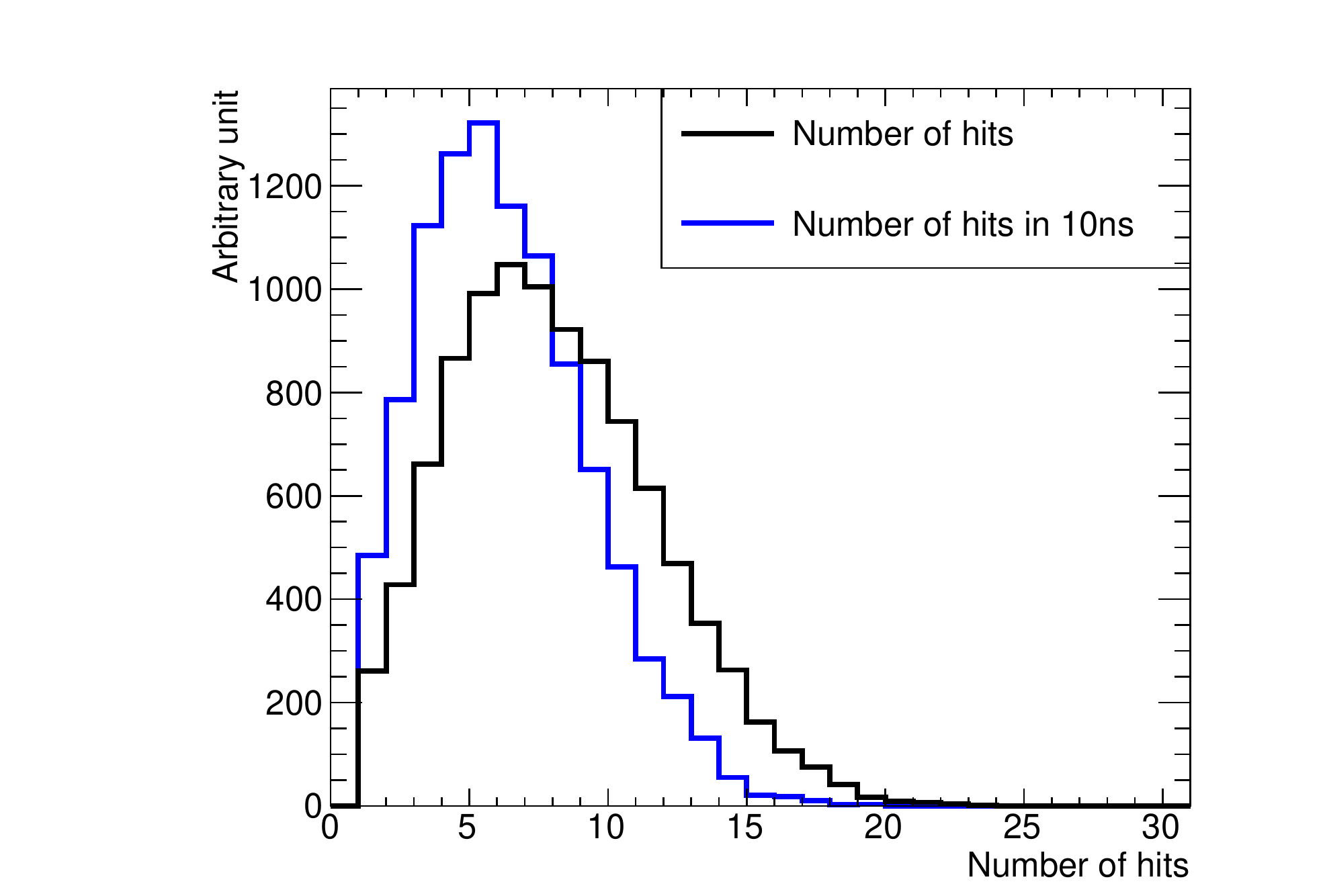}
    \centering
	\caption{Number of hit PMTs in SK for 2.2 MeV $\gamma$ using the
      SK detector simulation program. No noise are simulated and only
      the Cherenkov photons from the 2.2 MeV $\gamma$ are recorded.
      The black histogram shows the total number of PMT hits. The
      blue histogram shows the largest number of hit PMTs in 10~ns
      obtained by using the 10~ns sliding window. In searching for
      the largest number of hit PMTs, the time-of-flight (ToF) from 
      the generated point of 2.2 MeV $\gamma$ is subtracted from each 
      timing of PMT hit.} 
	\label{fig:nhits_gamma_pic}
\end{figure}

Before the neutron tagging is performed, all FC events are analyzed using
the standard SK atmospheric neutrino event reconstruction software
(APFit)~\cite{Ashie:2005ik}.  
This software reconstructs 
information associated with the primary event, including finding
its vertex position in the detector, counting the number of the 
Cherenkov rings, identifying each Cherenkov ring as showering-like or non-showering-like,
reconstructing its momentum, and finding $\pi^0$ candidates and 
decay electrons. 
Neutron tagging is then performed as a two-step process.
\color{black}
Approximately seven Cherenkov photons are  expected to be detected from a 2.2~MeV
$\gamma$, as shown in ~\cref{fig:nhits_gamma_pic}.
The number of hits in 10 ns after time-of-flight (ToF) subtraction 
is slightly smaller than the total number of PMT hits even though
the duration of the Cherenkov photon emission is nearly instantaneous. 
This is because some of the Cherenkov photons are scattered in the detector and 
as a result travel a longer distance before detection.
\color{black}
Therefore, the first step in the neutron tagging
process is to search in time for clusters of hit PMTs.  
These clusters are chosen as candidate neutron captures. 
During the second step, a neural network is used to differentiate 
real neutron capture candidates from backgrounds. 
Details of the neural network are described in ~\cref{sec:neural_net}.

\subsection{Step one: Initial Candidate Selection}
\color{black}
The MC sample shows that 70\%
\color{black}
of neutrons are captured 
within 200 cm of the initial interaction vertex.
In the search for clusters of hits in time, each PMT's hit timing is corrected for each photon's
ToF from the primary event vertex (neutrino vertex), which is reconstructed 
by APFit, to give a residual time.


A 10~ns sliding window is then used to search for clusters of hits 
in the residual time. 
\color{black}
This width was selected to take into
account the distance from the actual neutron capture to the neutrino 
vertex, the 2.2~ns timing resolution of the PMT, and 
contamination from accidental background hits, which is expected to be $\sim$0.5 hits per event in a 10~ns window. 
\color{black}
If there are five 
or more hits in the 10~ns window, the
cluster is selected as a possible neutron candidate. 
The number of hits 
in the 10 ns sliding window is defined as $N^\text{RAW}_{10}$, and the 
earliest timing of the hit in the 10~ns sliding window is defined as 
$t_0$. If multiple candidates are found with
their $t_0$ values within 20~ns of each other, only the candidate with 
the largest $N^\text{RAW}_{10}$ is considered. 
If there are two or more candidates
which have the same 
$N^\text{RAW}_{10}$, the last one is taken. 
This procedure avoids
double counting the same neutron capture as multiple candidates.
The candidate is rejected if $N^\text{RAW}_{10}$ is larger than 
50,
or the number of hits in a 200~ns window around the candidate ($N_{200}$) 
is larger than 200. 
\color{black}
The maximum distance of a point in the detector volume 
to a PMT is 50~meters. Therefore, 200~ns is sufficient to collect 
most of the photons reaching a PMT without scattering. 
Accordingly,  $N_{200}$ 
is a good variable to roughly estimate the total visible energy
in the detector. 

\begin{figure}[!ht]
	\includegraphics[width=0.49\textwidth]{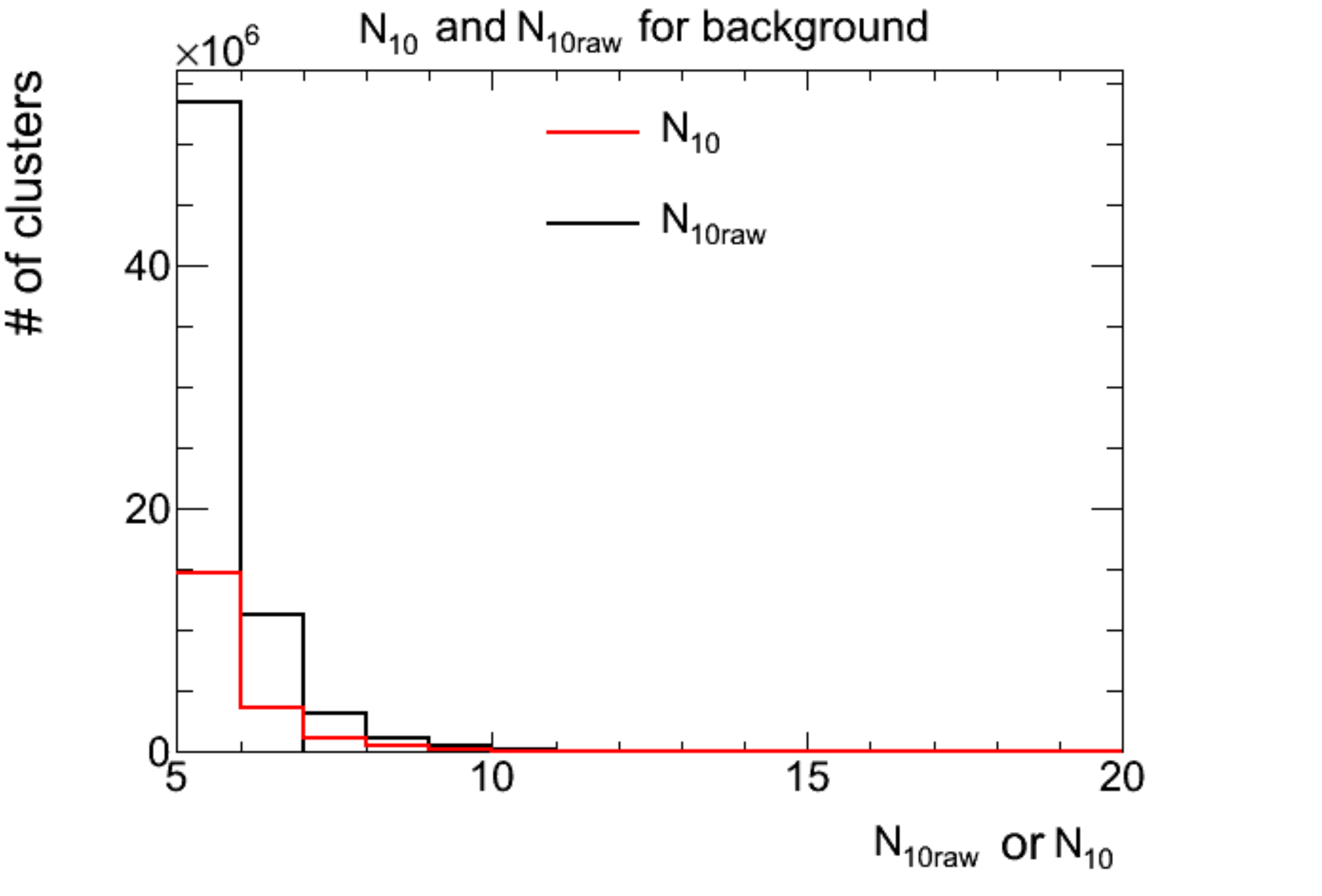}
	\includegraphics[width=0.49\textwidth]{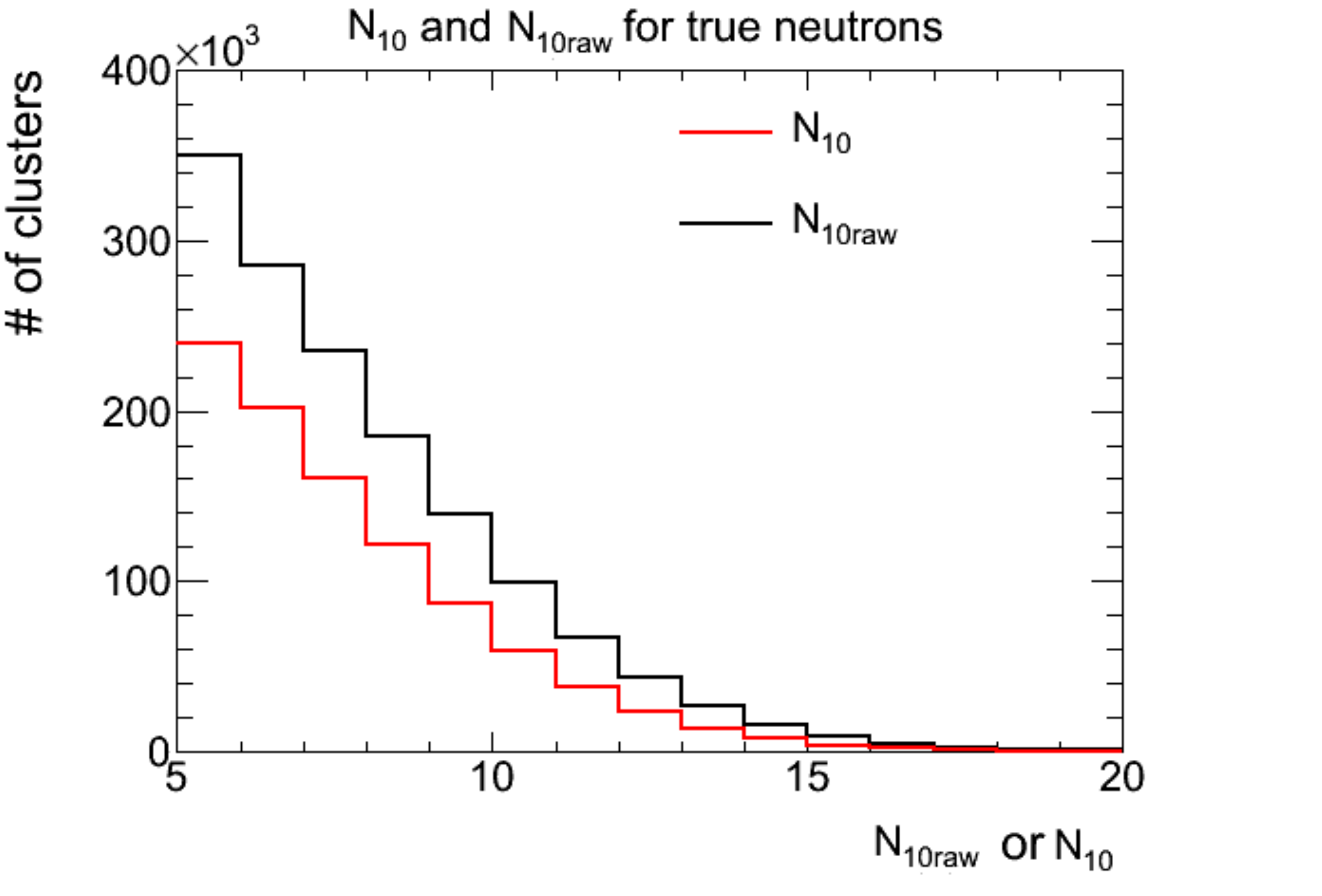} \\
    \hspace*{0.22\textwidth}(a)\hspace{0.47\textwidth}(b)
	\caption{Distributions of $N^\text{RAW}_{10}$ and $N_{10}$ (see text for definitions) for 
      background clusters (left) and true neutron clusters (right).
      Both histograms were made using
      the same number of primary atmospheric neutrino MC events.} 
\label{fig:n10_cut}
\end{figure}

\color{black}
Scintillation from radioactivity in the PMT glass creates a time-clustered noise signal, which can potentially  increase the number of observed hits as described 
in ~\cref{section:PMTNoise}.
\color{black}
Such spurious hits are removed using the variable, $N_{10}$,
which is similar to $N^\text{RAW}_{10}$.
Here, $N_{10}$ is the maximum number of hits in a 10~ns sliding window 
around the 2.2~MeV~$\gamma$ candidate after removing the
time-clustered noise hits.
In order to eliminate the time-clustered noise, hits are removed 
when there are multiple hits from the same PMT within 12 $\mu$s when $N^\text{RAW}_{10}$ is smaller than 7,
and when there are multiple hits from the same PMT within 6 $\mu$s otherwise. 
A hit cluster, with $N_{10}$ 5 or greater, is regarded as a neutron candidate.
~\cref{fig:n10_cut} shows the distributions of $N^\text{RAW}_{10}$ and
$N_{10}$ for the neutron signal and the background from the simulation.
Removing noise hits is shown to be effective in reducing background clusters 
without a significant loss of signal clusters.

\begin{figure}
  \begin{center}
    \includegraphics[width=0.8\textwidth]{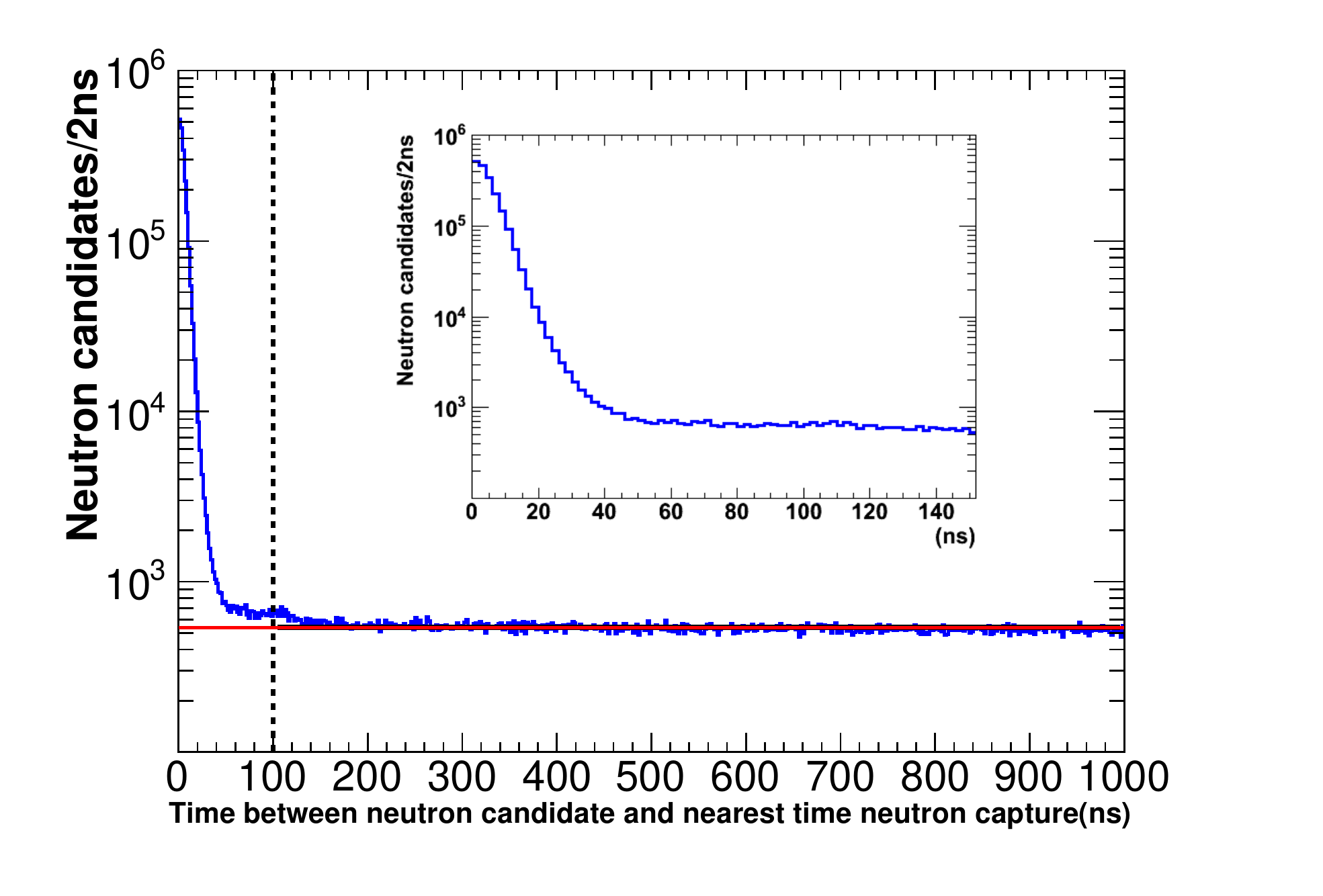}
  \end{center}
  \caption{Absolute time difference between neutron candidates 
   and the nearest true neutron capture in MC.  The flat tail is assumed 
   to be the background candidate event rate,
    the spike above the background rate near zero is from true neutron
    candidates.  The dotted black line represents the cut dividing true
    neutron candidates from fake neutron candidates.}
  \label{fig:candidate_timediff_criteria}
\end{figure}

 In order to evaluate the efficiency and background rate of 
the neutron tagging method
using the MC sample, the time difference between the $t_0$
obtained from the neutron tagging algorithm
and the true capture time of the neutron
is used. 
If the time difference is less than 100 ns, the candidate is
labeled as correctly-tagged, otherwise, it is labeled as
fake. 
~\cref{fig:candidate_timediff_criteria} shows the absolute time difference 
between each neutron candidate and the nearest true neutron capture in the MC.  
The flat tail is used to estimate the background
rate of fake candidates, while the excess on top of this flat rate
corresponds to true neutron candidates. 
 By extrapolating the stable background rate into the region around zero
where the true neutron candidates appear, it is estimated that about
1.3\% of neutrons labeled in MC truth as true candidates are in fact
fake, while 0.07\% of true neutron candidates are labeled in MC truth
as fake.  Based on MC truth, the simulation indicates an efficiency of
\color{black}
49\% 
\color{black}
for the candidate selection with a background rate of 
\color{black}
22\% 
\color{black}
fake candidates per event.

\subsection{Step two: Final Candidate Selection with Neural Network}
\label{sec:neural_net}

Following the initial candidate selection, a neural network is used to
separate the neutron capture signal 
from various backgrounds.
In the field
of particle physics, neural
networks are commonly used as a tool for signal-background
classification, and in this analysis, 
TMulitLayerPerception(TMLP) library 
in ROOT software framework~\cite{Brun:1997pa}
is used to
implement a feed-forward Multi-Layer Perceptron (MLP).
In total, twenty-three variables are used as the inputs to the neural network. 
The neural network was trained on a 250-year-equivalent atmospheric neutrino MC
using the MLP method.
A separate statistically independent 250-year-equivalent MC data set is then used for testing
the trained neural network.
After the initial candidate selection process, there are about 1.7 million
true neutron candidates, and about 15 million background candidates in the
full atmospheric neutrino MC.

%
%

\begin{figure}[!ht]
	\includegraphics[width=0.4\textwidth]{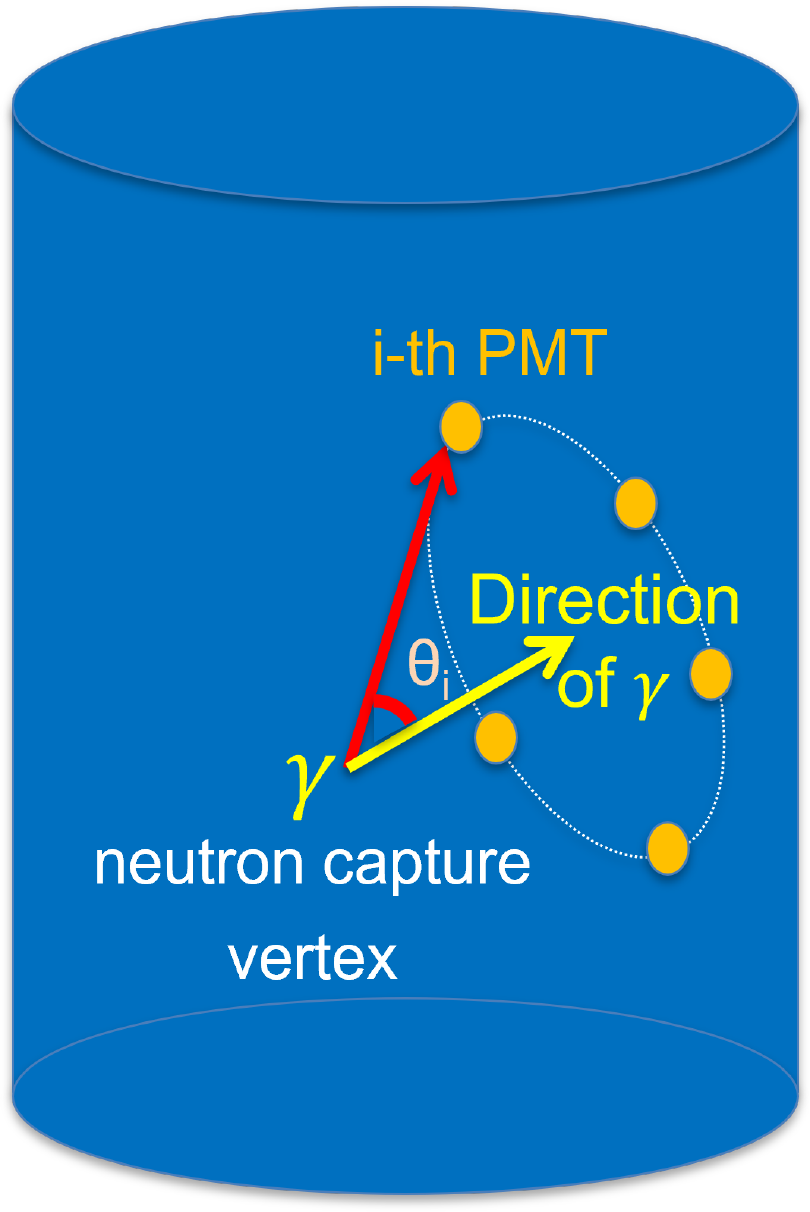}
    \centering
	\caption{Image of Cherenkov photon emission from a 2.2~MeV $\gamma$. The yellow line shows the direction of $\gamma$, whose starting point is the neutron capture vertex, the orange filled circles show the hit PMTs, the red line shows the direction to the i-th PMT viewed from the neutron capture vertex, and $\theta_i$ is the angle between the $\gamma$ direction and the direction to the i-th PMT (opening angle), respectively.}
	\label{fig:gamma_SK_geometry}
\end{figure}

\color{black}
Neutrons are typically captured within a few meters of the primary 
neutrino vertex and a 2.2~MeV $\gamma$ ray is emitted if proton
captures a neutron.
The 2.2~MeV $\gamma$ signal produces $\sim$7 PMT hits with a timing distribution whose width is typically less than 10 ns after applying the ToF correction.
In SK, Cherenkov photons from highly relativistic particles are 
emitted on a cone with a 42 
degree opening angle with respect to the direction of the incident 
particle, while the azimuthal distribution is uniform
as shown in \cref{fig:gamma_SK_geometry}

Based on these general characteristics 
we have selected 23 variables to distinquish neutron-induced signals from backg
rounds as summarized in~\cref{ListOfNNVariables}.

These variables are classified into three 
categories, those related to the timing distribution of the PMT hits, those related to the spatial
distribution of the hits, and those related to the reconstructed event's vertex and energy.

The variables related to the timing distribution of the PMT hits are:
the number of hits in 10 ns after ToF correction ($N_{10}$), 
the number of hits in 300 ns without ToF correction ($N_{300})$,
the root-mean-square of hit timing after ToF correction ($t_{\rm rms}$),
the minimum root-mean-square of the hit timing distribution after ToF correction 
(min($t_{\rm rms}$)),
the difference of $N_{10}$ using the reconstructed primary neutrino 
interaction vertex and the neutron vertex ($\Delta N_{10}$), and
the difference of $t_{\rm rms}$ using the reconstructed primary neutrino 
interaction vertex and the neutron vertex ($\Delta t_{\rm rms}$).

Variables that describe the event topology, such as the spatial charge distribution are:
the mean opening angle ($\theta_{\rm mean}$) of PMT hits,
the root-mean-square of the azimuthal angle ($\phi_{\rm rms}$) of the PMT hits,
the number of clustered hits ($N_{\rm c}$),
the acceptance parameter ($P_{\rm{Acceptance}}$),
the Cherenkov angle likelihood parameter ($L_\text{Cherenkov}$),
the isotropy parameter ($\beta_l$), and 
the number of hits on low-probability PMTs ($N_{\rm low}$).

Finally, variables related to the event reconstruction are:
the reconstructed energy using the BONSAI fitter~\cite{bonsai} ($BS_{\rm energy}$),
the reconstructed neutron capture vertex position using BONSAI
($BS_{\rm wall}$),
the reconstructed neutron capture vertex position using the Neut-Fit fitter
$NF_{\rm wall}$,
the distance from the reconstructed primary neutrino interaction vertex 
and the reconstructed neutron capture vertex ($(NF-AP)_{\rm dis}$),
the agreement of the reconstructed neutron capture positions of
the two different reconstruction algorithms  $(NF-BS)_{\rm dis}$), and
the vertex distance to the ID wall ($L_\text{towall}$).
 Details of the two vertex fitters, BONSAI and Neut-Fit, are described in 
~\cref{section:LE_reconstruction_algorithms}.

Among the input parameters, $N_{10}$ and $N_c$ have the first 
and the second most significant correlation to the Neural Net output. 

\begin{table}[hb]
\caption{List of neural net input variables}
\label{ListOfNNVariables}
\begin{center}
\begin{tabular}{|l|l|}
\hline
\multicolumn{2}{|l|}{Timing distribution of the PMT hits related parameters}\\
\hline
$N_{10}$      & the number of hits in 10 ns after ToF correction\\
$N_{300}$     & the number of hits in 300 ns without ToF correction \\
$t_{\rm rms}$ & the root-mean-square of hit timing after ToF correction \\
min($t_{\rm rms}$) & the minimum root-mean-square of the hit timing \\
                   & distribution after ToF correction \\
$\Delta N_{10}$ & the difference of $N_{10}$ using the reconstructed primary \\
                & neutrino interaction vertex and the neutron vertex \\
$\Delta t_{\rm rms}$ 
              & the difference of $t_{\rm rms}$ using the reconstructed primary \\
              & neutrino interaction vertex and the neutron vertex \\
\hline
\multicolumn{2}{|l|}{Event topology related parameters} \\
\hline
$\theta_{\rm mean}$ & the mean opening angle of PMT hits \\
$\phi_{\rm rms}$    & the root-mean-square of the azimuthal angle of PMT hits \\
$N_{\rm c}$         & the number of clustered hits \\
$P_{\rm{Acceptance}}$ & the acceptance parameter \\
$L_\text{Cherenkov}$ & the Cherenkov angle likelihood parameter \\
$\beta_l$           & the isotropy parameter \\
$N_{\rm low}$       & the number of hits on low-probability PMTs \\
\hline
\multicolumn{2}{|l|}{Event reconstruction related parameters} \\
\hline
$BS_{\rm energy}$   & the reconstructed energy using the BONSAI fitter \\
$BS_{\rm wall}$     & the reconstructed neutron capture vertex position \\
                    & using BONSAI fitter \\
$NF_{\rm wall}$     & the reconstructed neutron capture vertex position \\
                    & using the Neut-Fit fitter \\
$(NF-AP)_{\rm dis}$ & the distance from the reconstructed primary neutrino \\
                    & interaction vertex and the reconstructed neutron capture \\
					& vertex \\
$(NF-BS)_{\rm dis}$ & the agreement of the reconstructed neutron capture \\
                    & positions of the two different reconstruction algorithms \\
$L_\text{towall}$   & the vertex distance to the ID wall \\
\hline
\end{tabular}
\end{center}
\end{table}

\clearpage

In the following subsections all neural network input variables are reviewed. Comparisons of data and the Monte-Carlo simulation for each variable are summarized in ~\cref{section:NN_Input_Distributions}.
\color{black}

\subsubsection{Number of Hits in 10 ns: $N_{10}$}
 This is the same variable which is used to search for the initial
 candidates of the 2.2~MeV $\gamma$ cluster.  As above, $N_{10}$ is the maximum
 number of hits in a 10~ns sliding window around the
 2.2~MeV~$\gamma$ candidate after the time-clustered noise cuts.
 The signal events tend to give larger $N_{10}$ compared
 to the background as shown in the left plot of 
 ~\cref{fig:neutron4x.n10_n300}.

\subsubsection{Number of Hits in 300 ns: $N_{300}$}
\begin{figure}[!ht]
	\includegraphics[width=0.6\textwidth]{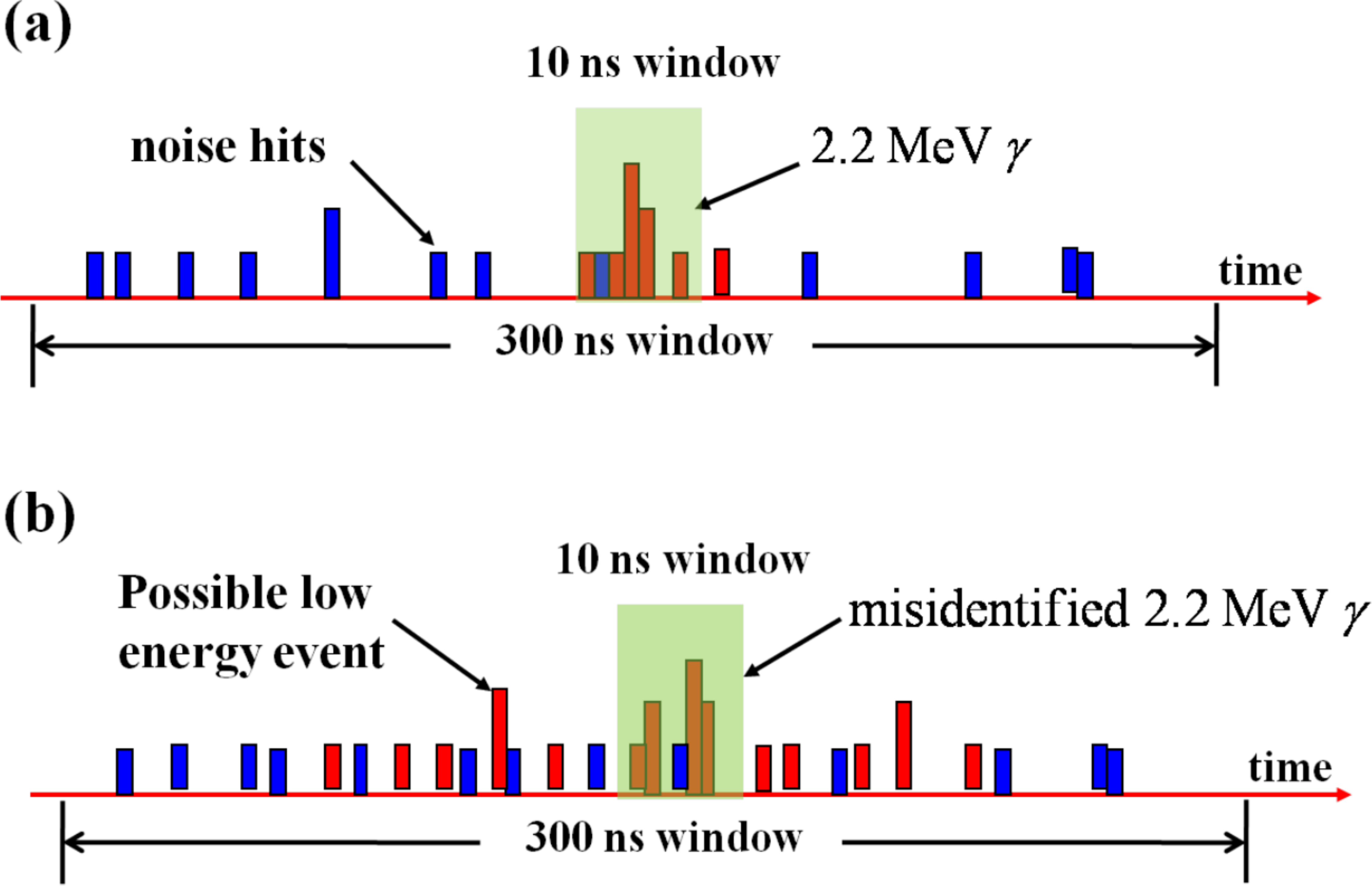}
    \centering
	\caption{The usefulness of the $N_{300}$ cut: (a) shows a typical
      neutron capture event and (b) shows a hypothetical background
      signal to be rejected. The blue bars show dark noise hits
      and red bars show the hits from Cherenkov photons.}
	\label{fig:n300pic}
\end{figure}

Cherenkov photons from a true neutron capture are almost
completely contained in a 10 ns window in residual time.  However, a
fake peak could be detected if there are a sufficient number of 
Cherenkov photons emitted from a higher energy particle at a
different vertex (as explained in ~\cref{fig:n300pic}).
\color{black}
\noindent 
Therefore, if there is a coincident higher energy event the reconstructed vertex may not be
correct. Considering the size of the detector and the effect 
of the `incorrect' ToF subtraction, we have decided to use
300~ns as the timing window for this variable.
\color{black}
In order to reject these fake candidates, $N_{300}$ is defined as the
total number of hits in a $\pm$150~ns window around the candidate peak.
Then, the variable $N_{300} - N_{10}$ is used as an input for the neural 
network because $N_{10}$ is expected to have a larger fraction of $N_{300}$ 
for the signal compared to the background, as shown in the right plot 
of ~\cref{fig:neutron4x.n10_n300}.

\subsubsection{Root-Mean-Square of Hit Timing: $t_{\rm rms}$}
The residual timing distribution is expected to have a narrower peak
for signal events than for background events. Therefore, the
root-mean-square of the candidate hit timing, which is defined in
~\cref{eq:trms}, is selected as one of the neural network input
variables.
In order to obtain the ToF-corrected timing, we used the reconstructed
vertex obtained by the Neut-Fit.
The signal events tend to give smaller min$(t_{\rm rms})$
compared to the background as shown in the top plot
of ~\cref{fig:neutron4x.trmsold_mintrms}.

\subsubsection{Minimum Root-Mean-Square of Hit Timing: min($t_{\rm rms}$)}
\begin{figure}[!ht]
	\includegraphics[width=1.\textwidth]{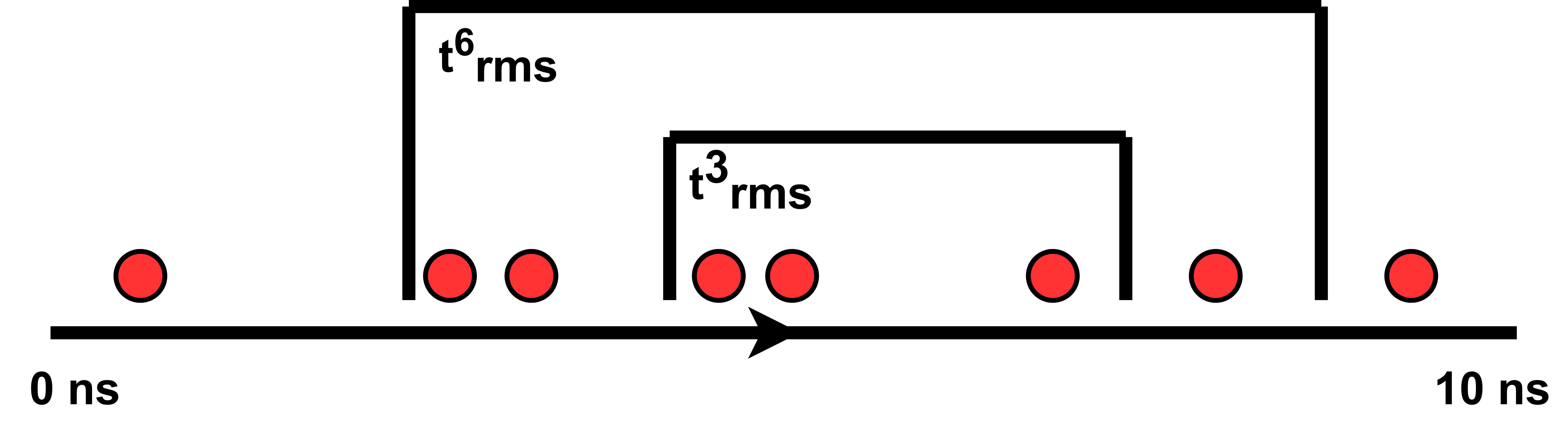}
    \centering
	\caption{The selection of $min(t_{\rm rms})$ hit clusters. Optimal
      selections for clusters of 3-6 hits are shown for an example
      candidate where $N_{10} = 8$. Possible background hits on the
      fringes of the candidate time are ignored.}
	\label{fig:mintrms}
\end{figure}

Background PMT hits can
sometimes occur in the 10~ns residual time window along with the hits
from a true neutron capture. 
For further reduction of these background hits, the RMS of the hit
timings is calculated for every set of three consecutive hits in the
10~ns window, as shown in ~\cref{fig:mintrms}).
Here, the Neut-Fit vertex is used to correct for the ToF.

The smallest value of all the sets is passed on as $\min(t^3_\text{rms})$. 
A similar quantity is also calculated for sets of six hits and defined as
$\min(t^6_\text{rms})$ and these two variables are used as the neural
network inputs. The hits from the signal are expected to be concentrated 
in time, and thus these variables are smaller for the signal compared to 
the background, as shown in the bottom plots of 
~\cref{fig:neutron4x.trmsold_mintrms}.

\subsubsection{Neut-Fit Root-Mean-Square of Hit Timing Variable and Number of Hits in 10 ns : $\Delta t_{\rm rms}$, $\Delta N_{10}$}
The variable $t_{\rm rms}$ is calculated twice, once with the hits
which are ToF-corrected to the primary event vertex, and again 
with the hits which are ToF-corrected to the Neut-Fit vertex.
The difference between the two
$t_{\rm rms}$ values is defined as $\Delta t_{\rm rms}$, and is used
in the neural network.  
Note that when
$t_{\rm rms}$ is recalculated using the Neut-Fit vertex, additional
hits can be moved into the 10 ns window and can have the effect of
increasing $t_{\rm rms}$.  This means that sometimes 
$\Delta t_{\rm rms}$ can be negative.
Instead of using the reconstructed vertex from APFit, $N_{10n}$ is
defined as the value of $N_{10}$ recalculated using the vertex from
Neut-Fit.  The difference, $\Delta N_{10}$, is defined as $N_{10n}-N_{10}$.  
These two variables are expected to be close to 0 for the 
signal because the vertexes between the two reconstructions 
are expected to be close for signal but not necessarily same
for the background, as shown in ~\cref{fig:neutron4x.trmsdiff_n10d}.

\subsubsection{Mean Opening Angle: $\theta_{\rm mean}$}
In water, Cherenkov photons from highly relativistic particles are
emitted on a cone with a 42 degrees opening angle with respect to the
direction of the incident particle.  
Therefore, Cherenkov photons from
electrons that are Compton-scattered by a 2.2~MeV $\gamma$ are expected to
have a peak opening angle around 42 degrees. In contrast, background
candidates are not expected to form such a peak.  The direction of the
Compton-scattered electron is reconstructed as the vector sum of the
directions from the Neut-Fit vertex to each PMT hit in the 10 ns window.
Here, the timing of each hit is ToF-corrected using the
Neut-Fit vertex, and opening angle to each hit PMT is then calculated 
from this direction. 
The mean value of these angles is used as an input variable.
The top left plot of ~\cref{fig:neutron4x.theta_phi_nc} shows 
distributions of the mean opening angles for the signal and 
the background; a clear peak around 42 degrees is observed for the 
signal.

\subsubsection{Hit Vector Root-Mean-Square of the azimuthal angle: $\phi_{\rm rms}$}
Cherenkov hits from a true neutron capture are expected to be
distributed uniformly in the azimuth of the reconstructed direction
of the Compton-scattered electron.
Background candidates, on the other hand, are often geometrically 
concentrated and form compact clusters of hits.
The variable $\phi_{\rm rms}$ is
computed by calculating the azimuthal angle of each hit with respect
to the reconstructed direction of the Compton-scattered electron,
where the definition of the direction is same as the one used to 
calculate $\theta_{\rm mean}$.
Angles
between consecutive hits in the azimuth are then
calculated such that the variable $\phi_{\rm rms}$ is 
the root-mean-square of these angular differences.
This variable is expected to be small for a true neutron capture,
since the steps between consecutive hits are fairly uniform.
For background events with spatial clusters, this variable is larger,
since the steps are small when stepping through hits in a
spatial cluster and then become larger when moving away from the cluster 
to other hits in the event.  
The top right plot of ~\cref{fig:neutron4x.theta_phi_nc} shows the signal
and the background distributions for $\phi_{\rm rms}$.

\subsubsection{Number of Clustered Hits: $N_{\rm c}$}
Background candidates are often found to have geometrically-clustered
PMT hits.  
Radioactive contaminants in the PMT glass could be the source
of these backgrounds, and since radioactive products emit weak
Cherenkov light, they may be detected by  nearby PMTs.
Conversely, because the primary event vertex is required to be at
least 200~cm from the wall, hits from a true neutron capture are
not expected to be clustered tightly together.  This clustering
tendency can thus be used to separate the signal from the background.

Clusters are defined based on the opening angles between hits, viewed
from the Neut-Fit reconstructed vertex.  Clusters are built
starting with a single hit and hits are then added to the cluster
iteratively according to the following rule: if a hit is within
14.1 degrees of any hit in a cluster, it is added to the cluster.  The
number of clustered hits, $N_c$, is defined as the total number of hits
in clusters of 3 or more hits.

The neural network uses $N_{10} - N_{\rm c}$ as an input variable.
Since the spatial distribution of PMTs for signal events is expected to be
broad, $N_{10}$ is much larger than $N_c$. On the other hand,
$N_{10} - N_{\rm c}$ is expected to be small for background
events, as shown in ~\cref{fig:neutron4x.theta_phi_nc}.

\subsubsection{Acceptance parameter: $P_{\rm{Acceptance}}$}
The probability to detect photons from a signal $\gamma$ depends
on the position of the PMT and its relative orientation to the incoming 
photon.
Most noise hits on the other hand do not have this dependency. 
Therefore, it is possible to 
discriminate the signal from noise using this difference.
First, we define the probability $P_{i}$ for each PMT to detect 
Cherenkov photons using the distance from the neutron capture 
position to the PMT and the direction. Then, the
acceptance parameter, $P_{\rm{Acceptance}}$, is obtained by multiplying 
the probability $P_i$ for all the PMTs used 
to calculate $N_{10}$. 
Here $P_{i}$ and $P_{\rm{Acceptance}}$ are defined
as follows:

\begin{eqnarray}
A_{i} &=& \frac{F(\theta_{i})}{R_{i}^{2}}e^{-R_{i}/L}, \\ 
A_{\rm Total}&=&\sum_i A_{i}, \\ 
P_{i} &=&\frac{A_{i}}{A_{\rm Total}} \label{eqn:probability} \\
P_{\rm{Acceptance}} &=& \frac{\log(\prod_i^{N_{10n}}P_i)}{N_{10n}},
\label{eqn:acceptance}
\end{eqnarray}
where $F(\theta_{i})$ encodes the angular dependence of the PMT
detection efficiency, $R_{i}$ is the distance from the captured
neutron position to the $i$-th PMT, and $L$ is the light attenuation length
in water. The Neut-Fit vertex is used in the calculation
of distance ($R_{i}$) and angle $\theta_{i}$.

Acceptance values for signal neutron events are expected to
be larger compared to those for the backgrounds, as is 
shown in the left plot of ~\cref{fig:neutron4x.accep_likelihood}.

\subsubsection{Cherenkov angle likelihood parameter: $L_\text{Cherenkov}$}
 Signal photons are expected to be distributed near 42 degrees from 
 the direction of the relativistic electron that produced them. 
 In order to quantify this characteristic, a Cherenkov angle likelihood parameter
 ($L_\text{Cherenkov}$) is defined as follows.
 First, cones are defined using combinations of three PMT positions from $N_{10n}$
 to specify a base, and the Neut-Fit vertex is used for the apex.
 The opening angle of a given cone is defined as $\theta_{i}$ for the $i$-th
 PMT combination. The likelihood function $L_\text{bkg}$ and
 $L_\text{sig}$ are constructed as functions of $\theta_{i}$, $N_{10n}$ 
and the products of
 probabilities defined in ~\cref{eqn:probability} for those three hits 
 ( $\prod_{j=1}^{3}P_j$ ).
 Then the Cherenkov angle likelihood parameter ($L_\text{Cherenkov}$)
 is defined as:
 \begin{equation}
   L_\text{Cherenkov} = \sum_{i=0}^{_{N10n}\text{C}_3}(\log(L_\text{bkg}(N_{10n},\theta_i,\prod_{j=1}^3P_j))
   - \log(L_\text{sig}(N_{10n},\theta_i,\prod_{j=1}^3P_j))).
   \label{eqn:CherenkovLikelihood}
 \end{equation}
   
The distribution of this likelihood parameter for the signal and 
the background is shown in ~\cref{fig:neutron4x.accep_likelihood},
where the values for the signal events are clearly smaller compared 
to the ones for the background events.

\subsubsection{Isotropy parameter: $\beta_l$}
Isotropy parameters are introduced to characterize the spatial 
distribution of the detected photons. First, $\beta_l$ is defined 
as follows:
\begin{equation}
  \beta_l = 
  \frac{2}{N_{10n}(N_{10n}-1)}{\sum_{i=1}^{N_{10n}-1}\sum_{j=i+1}^{N_{10n}}P_l(\cos\theta_{ij})},
\label{eqn:betal}
\end{equation}
where $\theta_{ij}$ is the opening angle between two PMTs 
with hits as viewed from the reconstructed neutron 
vertex with Neut-Fit. Here $l$ is a natural number which 
has been chosen to be smaller than 6 in this analysis and $P_{l}$ is 
formed from spherical harmonics. 
These $\beta$ parameters are constructed as follows.
At first, we define a function $f(\theta,\phi)$, which gives 
1 when there is a PMT which has detected a photon and gives 0 otherwise.
Here, $\theta$ and $\phi$ are zenith and 
azimuthal angles of the PMT viewed from the Neut-Fit vertex.
The $f(\theta,\phi)$ is expressed using the spherical harmonic
function $Y^*_{lm}(\theta,\phi)$,
\begin{equation}
  f(\theta,\phi) 
  = \sum_{l=0}^{\infty}\sum_{m=-l}^{l}\alpha_{lm}Y^*_{lm}(\theta,\phi).
\end{equation}
\noindent Here we define the position of $i$-th hit PMT as $(\theta_i,\phi_i)$,
then $\alpha_{lm}$ is 
\begin{eqnarray}
  \alpha_{lm} 
  &=& \int\int f(\theta,\phi)Y_{lm}(\theta,\phi)d\theta d\phi \\
  &=& \sum_{i=1}^{N_{10n}}Y_{lm}(\theta_i,\phi_i).
\end{eqnarray}  

Next we define the rotationally invariant variable $\beta'_l$,
\begin{eqnarray}
  \beta'_l &=& \sum_m|\alpha_{lm}|^2 \\
  &=& \sum_{i,j}\sum_{m}Y_{lm}(\theta_i,\phi_i)Y^*_{lm}(\theta_j,\phi_j) \\
  &=& \frac{2l+1}{4\pi}\sum_{i,j}P_l(cos\theta_{ij}).
\end{eqnarray}
$\beta'_l$ depends on the number of PMT hits and also, same combinations
of PMTs are used multiple times. Therefore, we use $\beta_l$, which is
defined as shown in ~\cref{eqn:betal}, instead of $\beta'_l$.

The distributions of $\beta_l$ used in the neural network 
are shown in ~\cref{fig:Isotropy}. As shown in these figures, the 
signal events concentrate around 0.35 for $\beta_1$ and around 
0 for other values of $l$, while backgrounds have broader distributions
and have peaks at around 1.

\subsubsection{Alternative Fit Variables: $BS_{\rm wall}$, $NF_{\rm wall}$ and $BS_{\rm energy}$}
It is known that Radon emanates from the SK detector components, 
in particular the PMTs, their signal and HV cables, and the 
anti-implosion PMT housings. 
The subsequent decay of Radon produces background events
concentrated near these materials and hence close to the inner
detector wall.
\color{black}
Therefore, the distance between the neutron
vertexes reconstructed by alternative event reconstruction tools
and the nearest inner detector wall is selected to separate the 
signal from the background. We have used two event reconstruction
tools, BONSAI and Neut-Fit.
The variables $BS_{\rm wall}$ and $NF_{\rm wall}$ are calculated 
using BONSAI and Neut-Fit, respectively and the distributions
are shown in the top plots of ~\cref{fig:neutron4x.bswall_fwall_bse}.
\color{black}
BONSAI also gives the reconstructed energy and this variable 
($BS_{\rm energy}$) is expected to be around 2.2~MeV for true 
neutron captures. Therefore, this variable is also used as one of 
the inputs. 
The bottom plot of ~\cref{fig:neutron4x.bswall_fwall_bse} shows the distribution of $BS_{\rm energy}$ there is a clear peak at 2.2 MeV 
for signal but much broader distribution for background.

\subsubsection{Fit Agreement Variables: $(NF-BS)_{\rm dis}$, $(NF-AP)_{\rm dis}$}
\label{subsec:fit_agreement}
 Neut-Fit and BONSAI use different fit criteria and procedures to search
 for low energy event vertexes.  When they agree well on the location
 of an event, it is likely to be a real low energy event, as opposed
 to simply PMT noise.  Therefore, the distance between these two
 vertexes $(NF-BS)_{\rm dis}$ is also selected as an input to the
 neural network and the distribution is shown as the left plot of
 ~\cref{fig:neutron4x.bfdis_fpdis}.
Furthermore, neutrons are expected to travel no further than a few 
meters from their production points before thermalizing and being 
captured, as shown in the right plot of ~\cref{fig:neutron4x.bfdis_fpdis}.
Therefore, the distance between the Neut-Fit
neutron vertex and APFit primary vertex, $(NF-AP)_{\rm dis}$ is also
used as an input to the neural network.  

\subsubsection{Distance to the ID wall parameter: $L_\text{towall}$}
We define the distance from Neut-Fit's reconstructed neutron vertex
to the ID wall in the direction of the particle 
as ``to wall'' 
($L_\text{towall}$). 
The direction of the particle is defined by
the sum of the vectors from the reconstructed neutron vertex
to each hit PMT.
The distribution of $L_\text{towall}$ is
shown in ~\cref{fig:neutron4x.towall}. As shown in the figure, the noise
distribution has a larger mean value compared to the signal.

\subsection{Neural Network Results}

\begin{figure}[!ht]
	\includegraphics[width=0.49\textwidth]{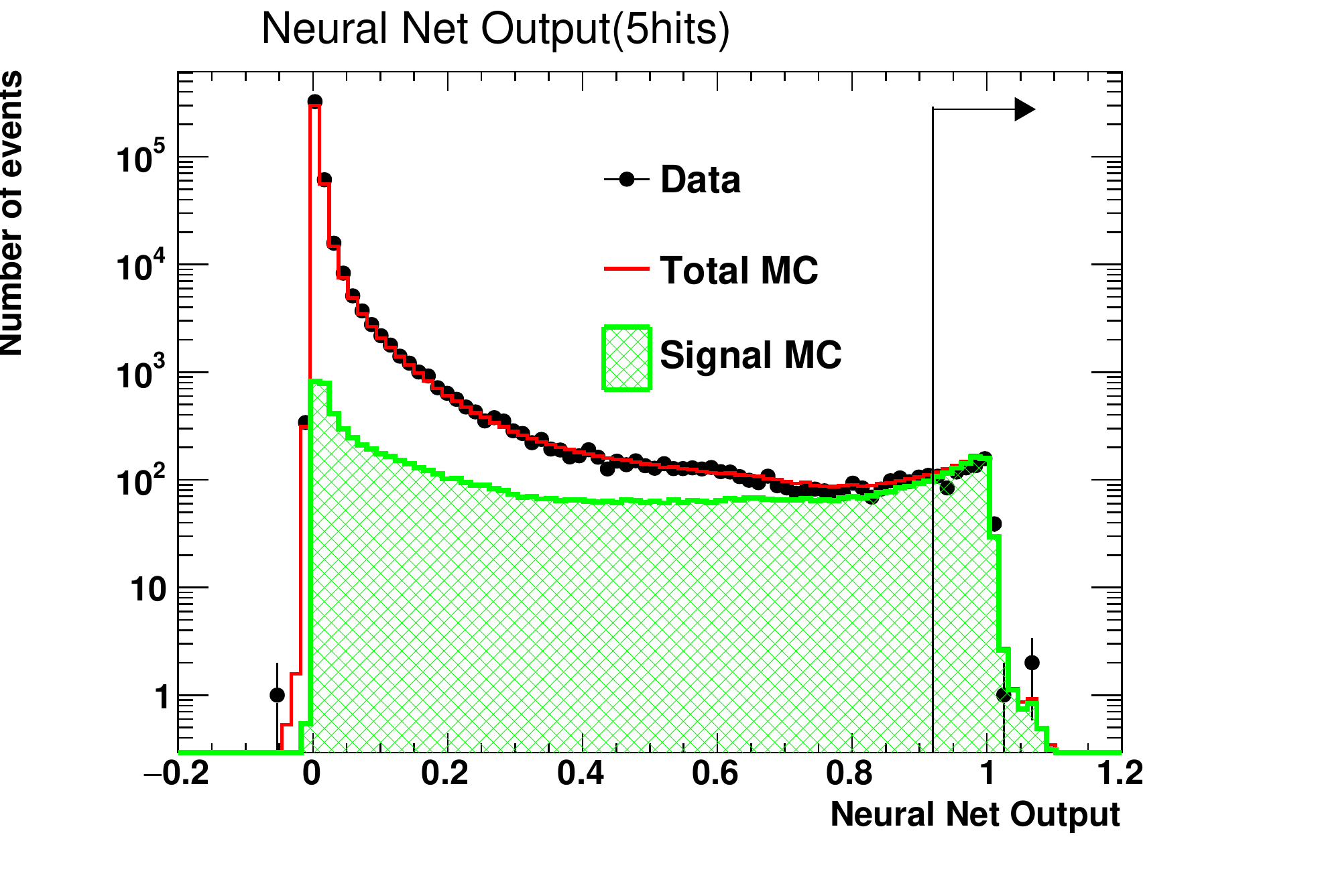}
	\includegraphics[width=0.49\textwidth]{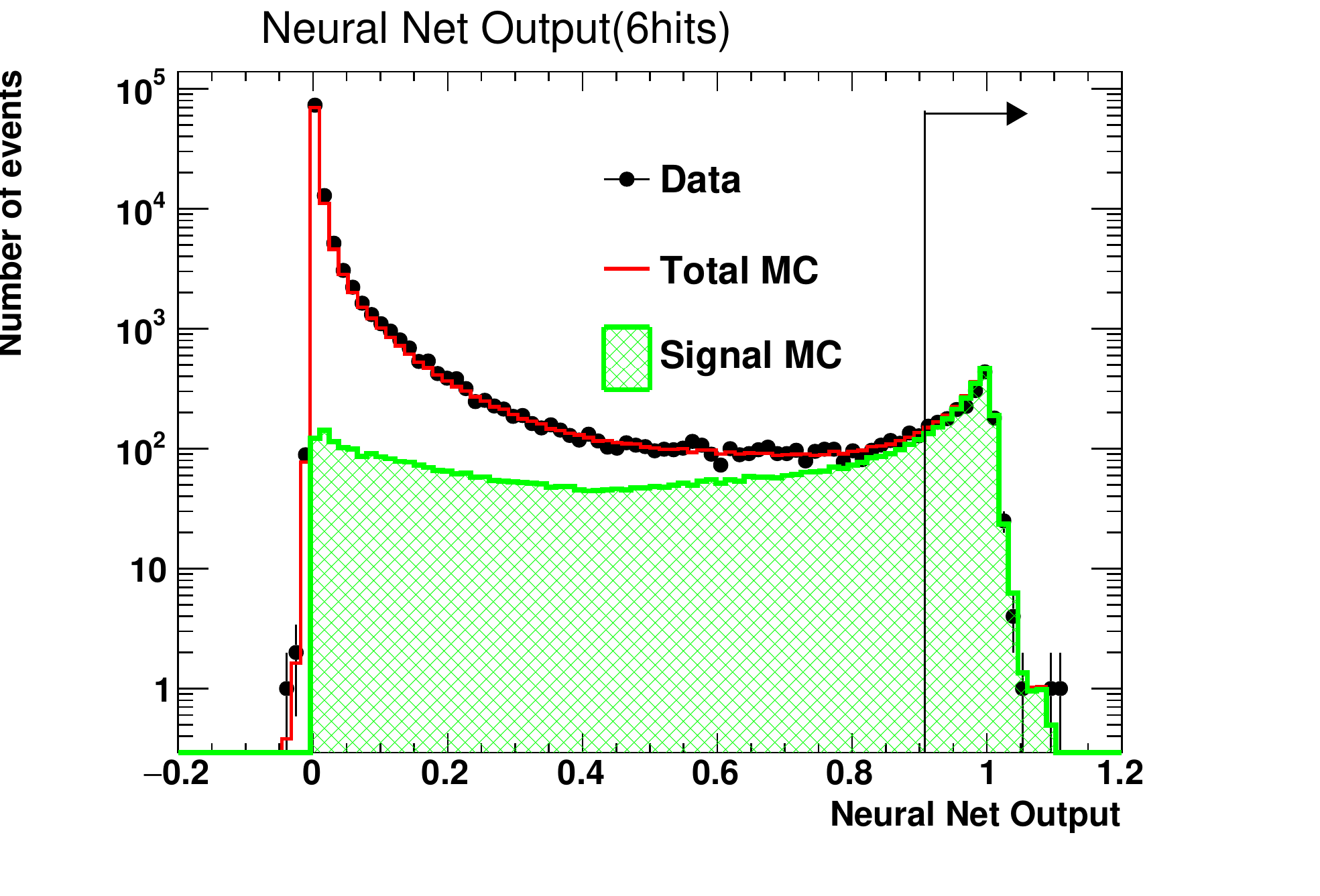}
	\includegraphics[width=0.49\textwidth]{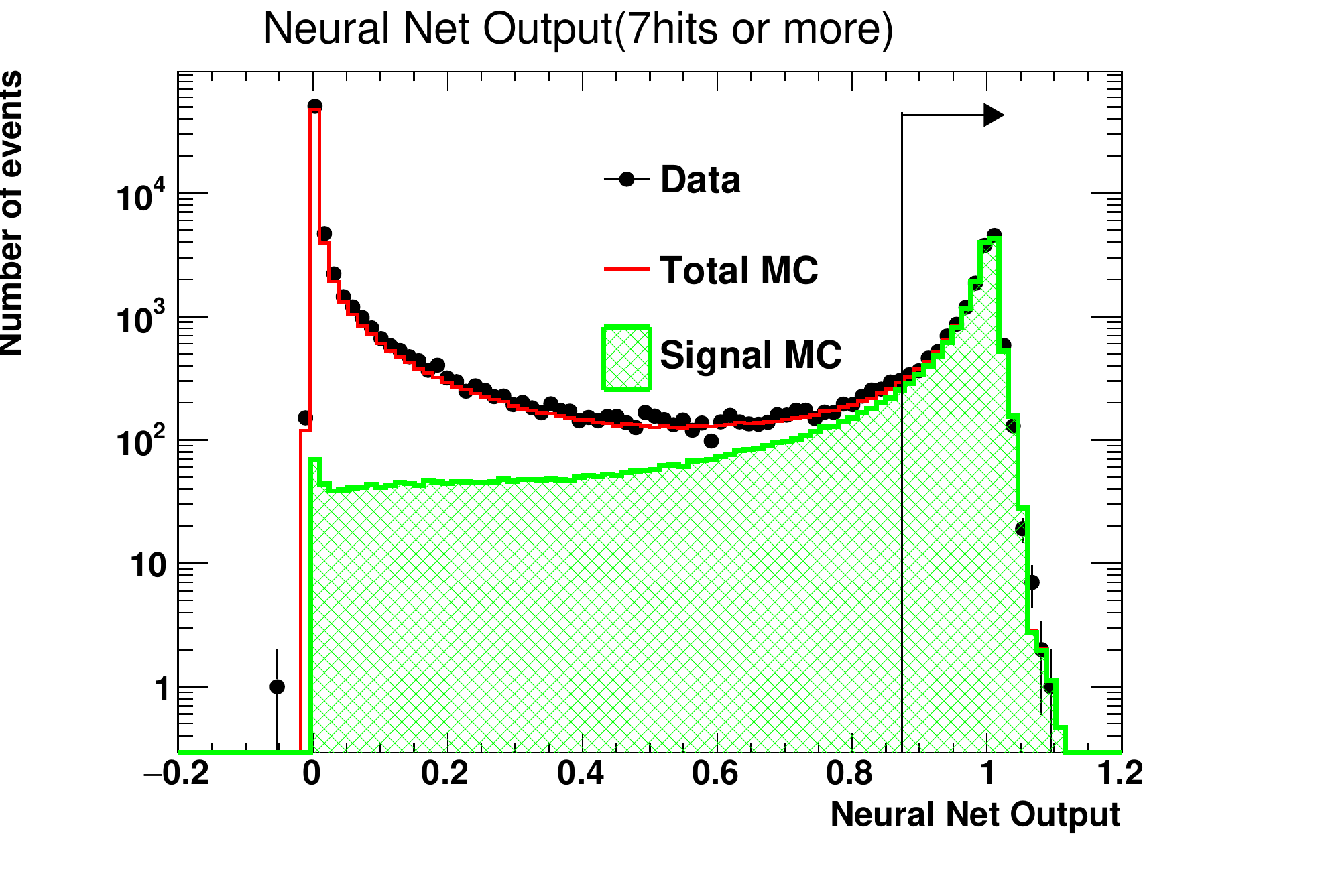}
    \centering
	\caption{Neural network output for selecting final neutron
      candidates.
      The plots show the output for different values of $N_{10}$: 
      5 in the top left panel, 6 in the top right, and 7 or more in the bottom panel.
      The green histogram corresponds to the neutron 
      capture signal,
      the red histogram shows the total MC, and the overlaid black points 
      show the data, respectively.
      The red histogram is normalized to the number of neutron 
      capture events in the atmospheric neutrino event samples from 
      the full SK-IV data set, which spanned 3,244.4 days between 
      October 2008 and May 2018.
      Black lines and arrows in each plot show the threshold values 
      for selecting candidates.
      Since noise contamination is larger for small $N_{10}$ candidates, 
      different threshold values were chosen to ensure the  
      mis-identification rate is below 0.018 per neutrino interaction.
      The threshold values are 0.92 when $N_{10}$ is 5, 
      0.908 when it is 6, and is 0.874 for 7 or more.}
	\label{fig:neutron4x.nntest}
\end{figure}

The output of the neural network is shown in
~\cref{fig:neutron4x.nntest} and is defined such 
that it is approximately equal to the likelihood of that candidate being a true neutron capture.  
The neural network successfully separates the
true neutron candidates from the fake neutron candidates.  
As described above, a neutron candidate is identified as a ``true candidate'' 
if the time difference between it and 
the true capture time of a MC neutron is less than 100 ns.
Threshold values for selecting neutron
candidates are chosen such that the mis-identification of noise events 
is less than 0.018 per neutrino interaction.
Since the fraction of noise events 
increases dramatically when $N_{10}$ decreases,
the chosen thresholds vary based on this parameter.
The thresholds are 0.92 for $N_{10}=5$, 0.908 for $N_{10}=6$
and 0.874 when $N_{10}>6$.
When applied to the 500 year atmospheric neutrino MC data, these cut values
give a final neutron tagging efficiency of 
\color{black}
26\% 
\color{black} 
with 0.016
background neutron tags per primary event as shown in ~\cref{ntageff}.
\color{black}
Here, there are sufficient MC statistics to make  statistical error negligibly small.
\color{black}
\begin{table}[!ht]
\centering
\caption{Final efficiency and background rate of the neutron tag
  algorithm after each stage of the selection.}
\label{ntageff}
\begin{tabular}{|c||c|c|}
\hline
Selection stage      & Efficiency & Background / Event \\ \hline 
\color{black} Initial Selection    & 49$\%$   & 22   \color{black}\\ 
\color{black} After Neural Network & 26$\%$   & 0.016 \color{black}\\
\hline
\end{tabular}
\end{table}

\begin{figure}[!ht]
	\begin{center}
 		\begin{tabular}{cc}
  			\begin{minipage}{.5\linewidth}
  				\includegraphics[width=1.\textwidth]{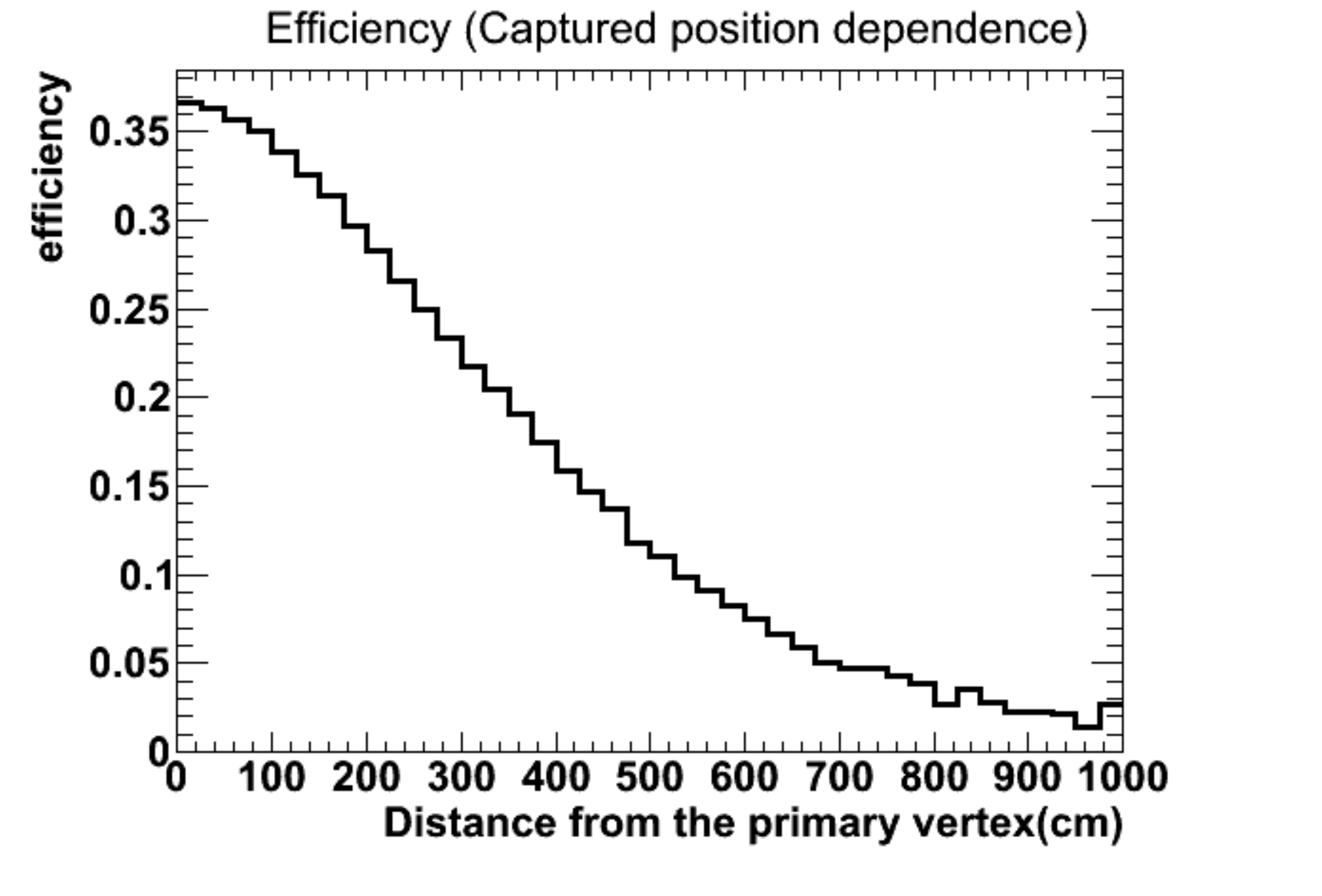}
			\end{minipage}
  		&
  			\begin{minipage}{.5\linewidth}
  				\includegraphics[width=1.\textwidth]{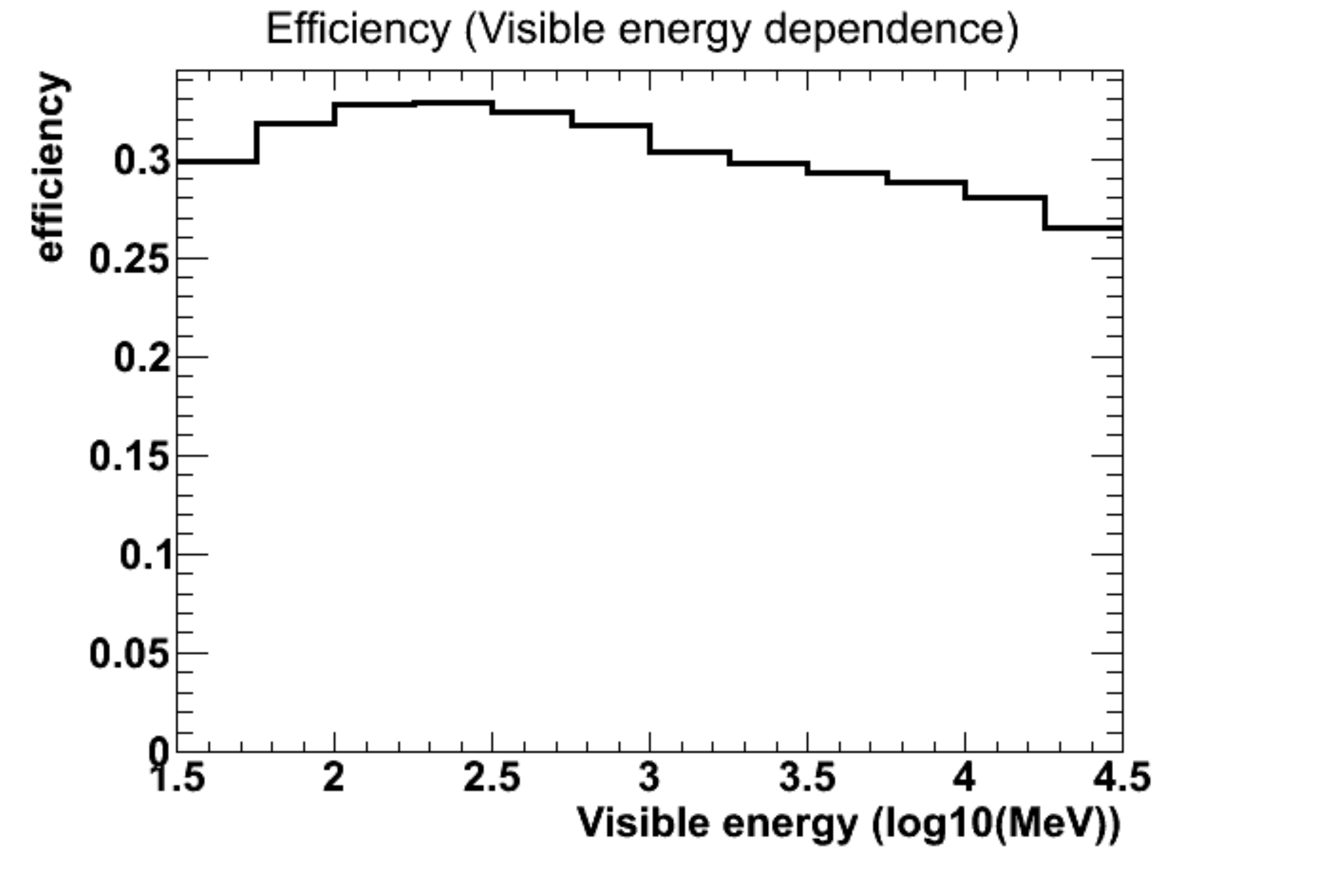}
			\end{minipage}
 		\end{tabular}
	\includegraphics[width=0.5\textwidth]{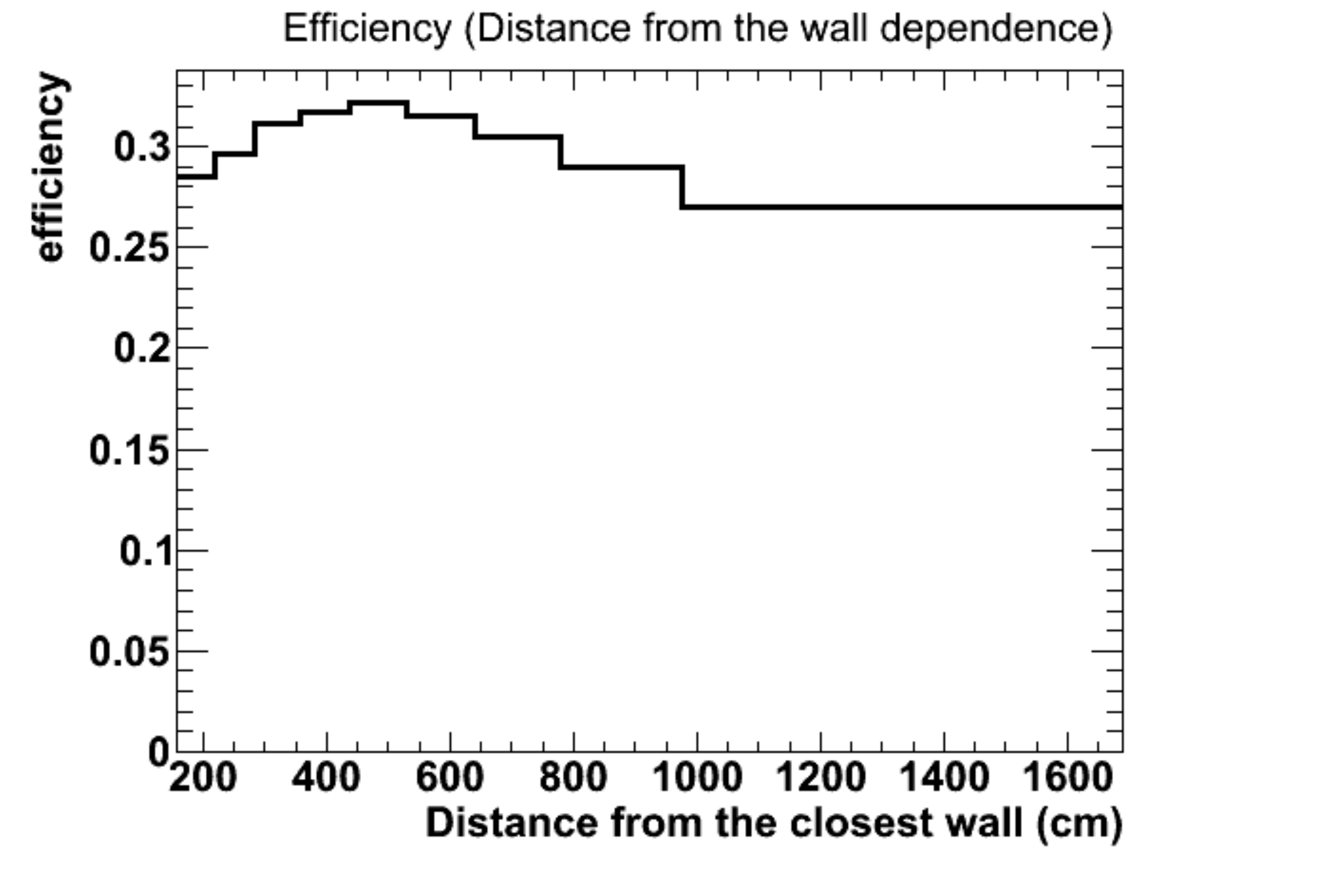}
 		\caption{Dependencies of the 2.2~MeV detection efficiency on
          the distance traveled by the neutron (top left), the visible 
          energy of the primary neutrino interaction (top right) and
          the efficiency of detecting neutrons as a function of
        distance from the closest ID wall (bottom).
          The efficiency is defined
          as the number of selected candidates divided by the number
          of true neutron capture events in each bin.}
 		\label{fig:effevis_dwall}
 	\end{center}
\end{figure}

The efficiency is heavily dependent on the distance between the
neutrino interaction and the neutron vertex and is mildly dependent on the
energy deposited in the detector, as shown in the top two plots in 
~\cref{fig:effevis_dwall}.
Here, the total energy deposit in the detector is defined as the
electron-equivalent energy ($E_{\rm vis}$).
It also depends on the vertex position of the primary event in the detector. 
 If the primary event is close to the center of the tank, 
the attenuation in the water reduces the amount of light reaching 
the PMTs and the detection efficiency decreases.
However, if the event is too close to the tank wall, 
the PMT acceptance is reduced and the efficiency also decreases. 
As shown in the bottom plot in ~\cref{fig:effevis_dwall}, the efficiency
is maximal for events in the region between the center and edge
of the fiducial volume.

\section{Comparison with SK-IV data}

 The neutron tagging algorithm was applied to 3,244.4 days
 of SK-IV FC data and compared to a 250 years sample 
 of atmospheric neutrino MC data, which included oscillations. 
 This SK-IV data set corresponds 
 to 26,473 FC events whose reconstructed vertexes are at least 200~cm from
 the ID wall.  The details of the atmospheric neutrino
 event selection are discussed in a previous article~\cite{skatm2018}.
 The MC sample is livetime-normalized and oscillated using a
 two-flavor oscillation approximation with $\Delta m^{2} = 2.5 \times
 10^{-3}~\text{~eV}^{2}$ and $\sin^{2}2\theta = 1.0$.  A summary of
 the comparison is shown in ~\cref{atmpdneutronsummary}.
 The agreement between the observed number of atmospheric neutrino
 events and the MC prediction without any corrections is typically 
 within $\sim$5\%.
 In other analyses at SK, nuisance parameters modeling flux 
 or neutrino interaction uncertainties are applied 
 to modify the MC expectation to agree with the data.  
 However, the present study does not include these corrections.
Note that only events whose electron-equivalent energy ($E_{\rm vis}$) 
is smaller than 30~ GeV are used here. 
At higher energies, the flux is highly suppressed and 
neutrino interactions have a large neutron multiplicity, which is not 
modeled well in our Monte-Carlo simulation programs, NEUT and SKDETSIM.

\begin{table}[!ht]
\centering
\caption{A comparison of the expected and measured number of neutron capture
  events in the SK-IV atmospheric neutrino SK-IV data.
The MC data sample is normalized to the data livetime and oscillated under a 
two-flavor approximation: $\Delta m^{2} = 2.5 \times 10^{-3}$~eV$^{2}$ and
$\sin^{2} 2\theta = 1.0$.
}
\label{atmpdneutronsummary}
\begin{tabular}{|c||c|c|}
\hline
Sample                & SK-IV Data (3,244.4 days) & MC        \\ 
\hline
Fully contained events                  & 26,473 & 25,845    \\
Total neutrons tagged                   & 18,091 & 18,288    \\ 
Events with at least one tagged neutron &  9,327 &  8,912.7  \\ 
Events with exactly one tagged neutron  &  5,676 &  5,138.3  \\
\hline
\end{tabular}
\end{table}

\begin{figure}[!ht]
	\includegraphics[width=0.6\textwidth]{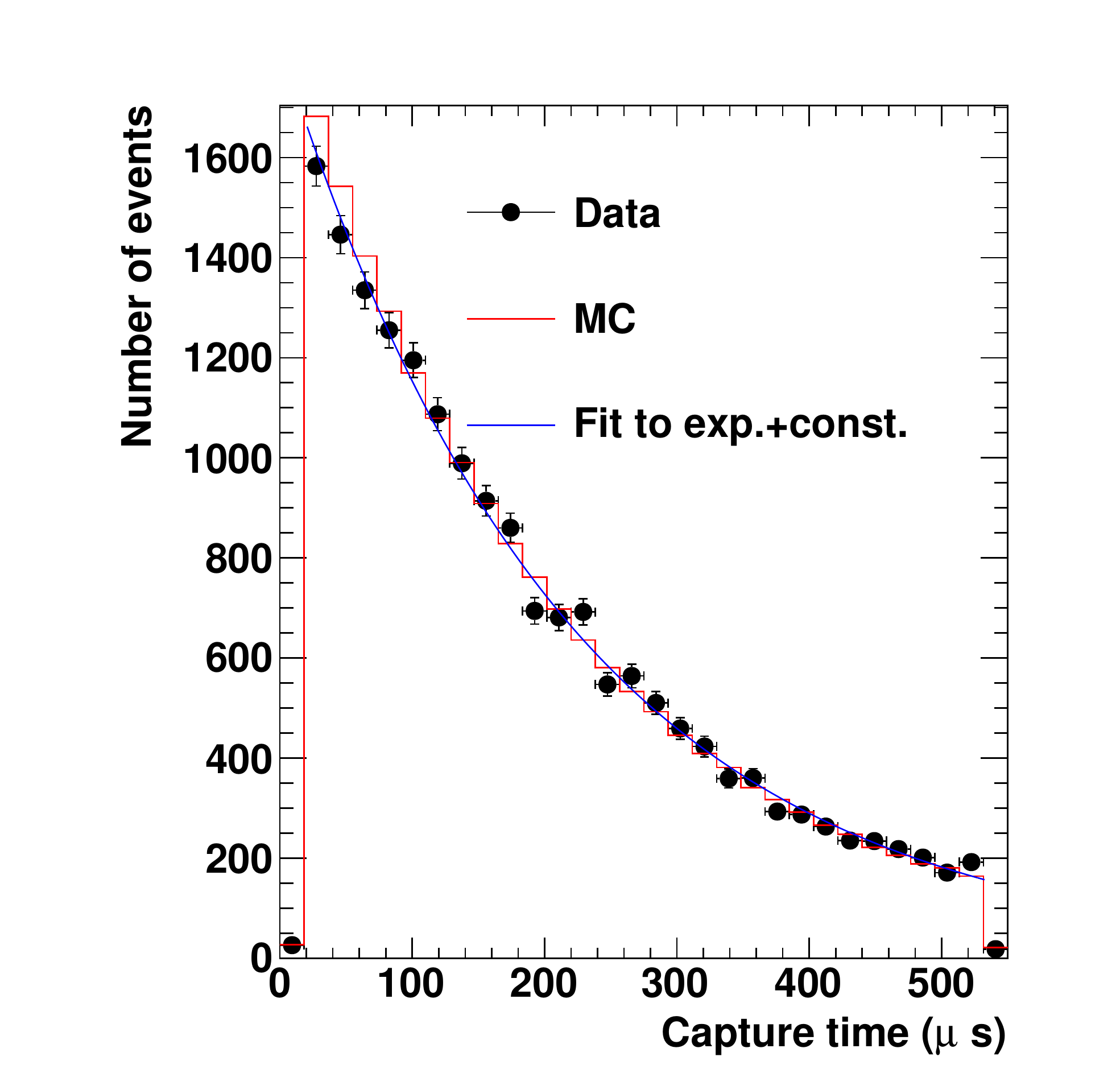}
    \centering
	\caption{Timing distribution of selected 2.2~MeV $\gamma$ candidates.
      The primary trigger is at $t=0$. The blue curve is
      the best fit to an exponential function with a constant offset.  
      Data are taken from 3,244.4 days of SK-IV and the MC is normalized as in ~\cref{atmpdneutronsummary}.
}
	\label{fig:dt}
\end{figure}

\begin{figure}[!ht]
	\begin{center}
 		\begin{tabular}{cc}
  			\begin{minipage}{.45\linewidth}
  				\includegraphics[width=1.\textwidth]{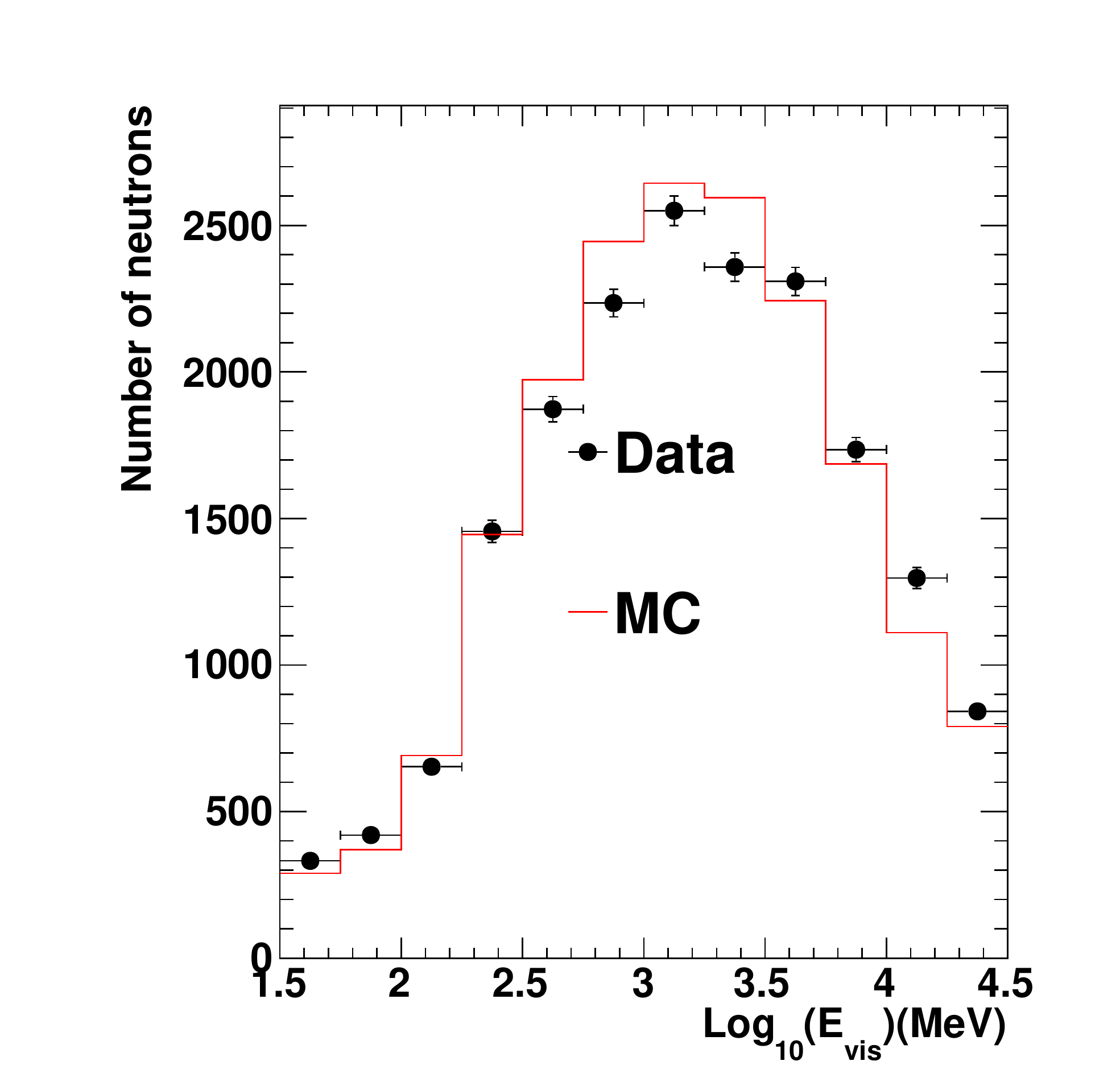}
			\end{minipage}
  		&
  			\begin{minipage}{.45\linewidth}
  				\includegraphics[width=1.\textwidth]{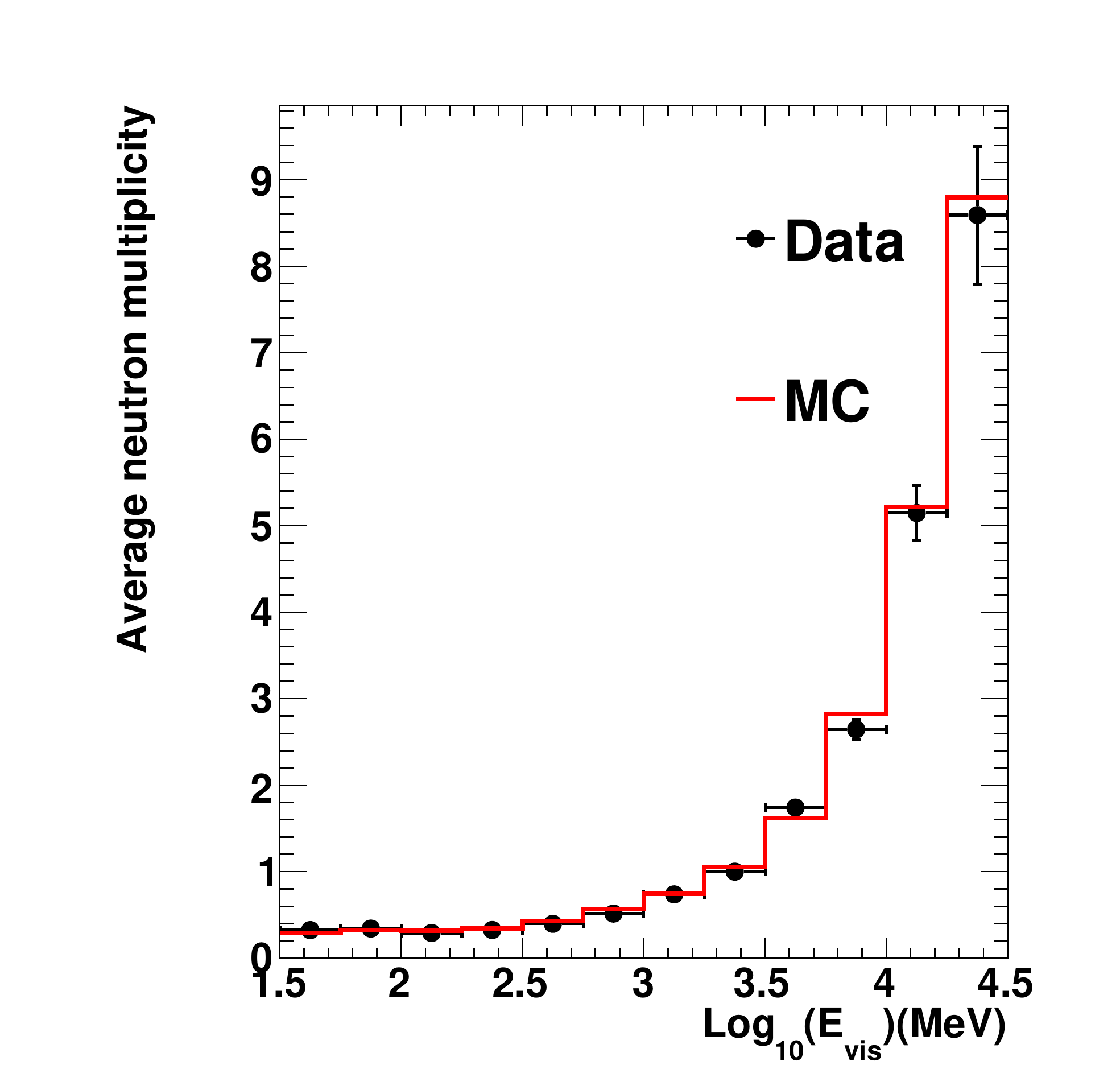}
			\end{minipage}
 		
\\
		\begin{minipage}{.45\linewidth}
  				\includegraphics[width=1.\textwidth]{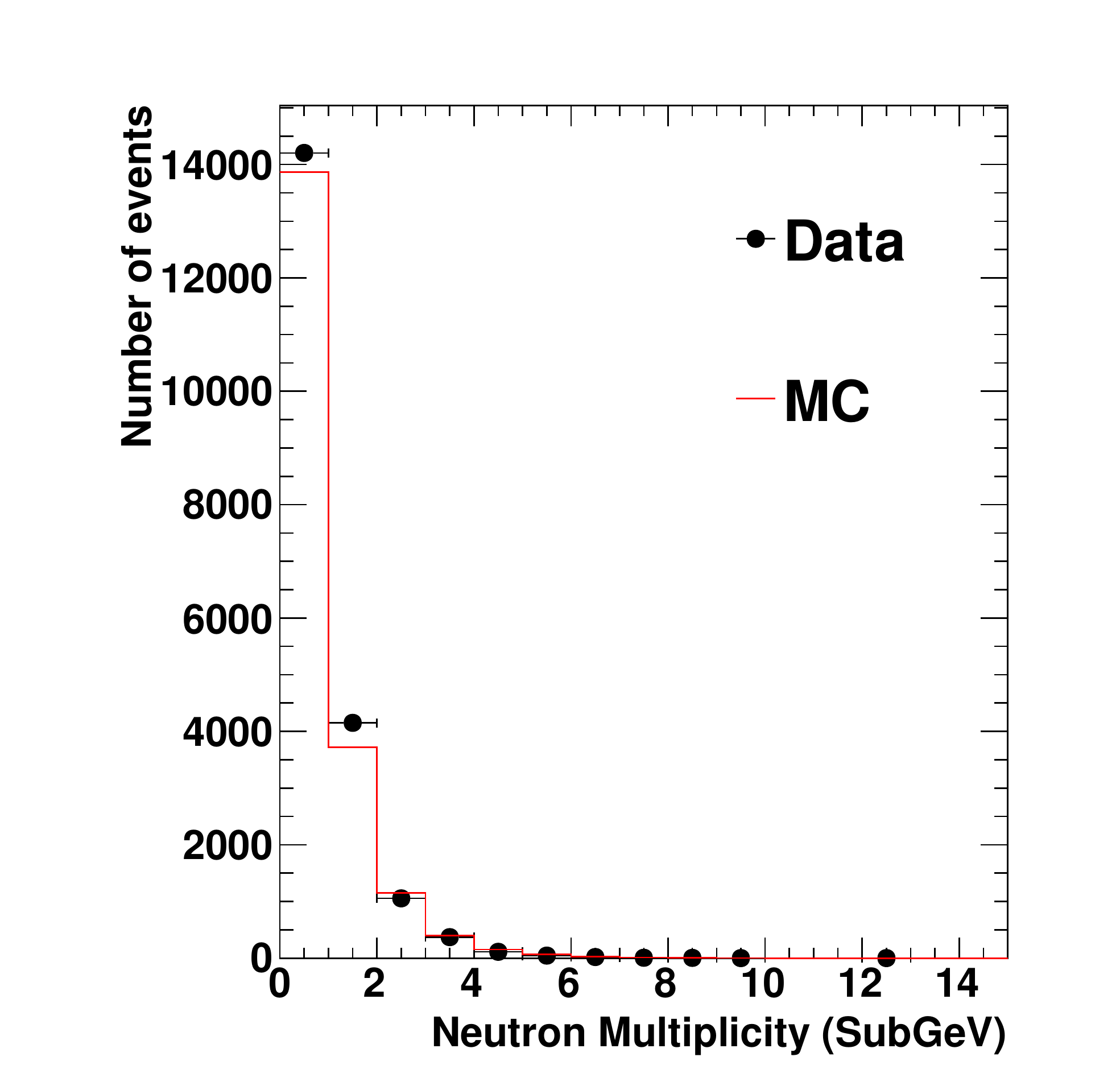}
			\end{minipage}
  		&
  			\begin{minipage}{.45\linewidth}
  				\includegraphics[width=1.\textwidth]{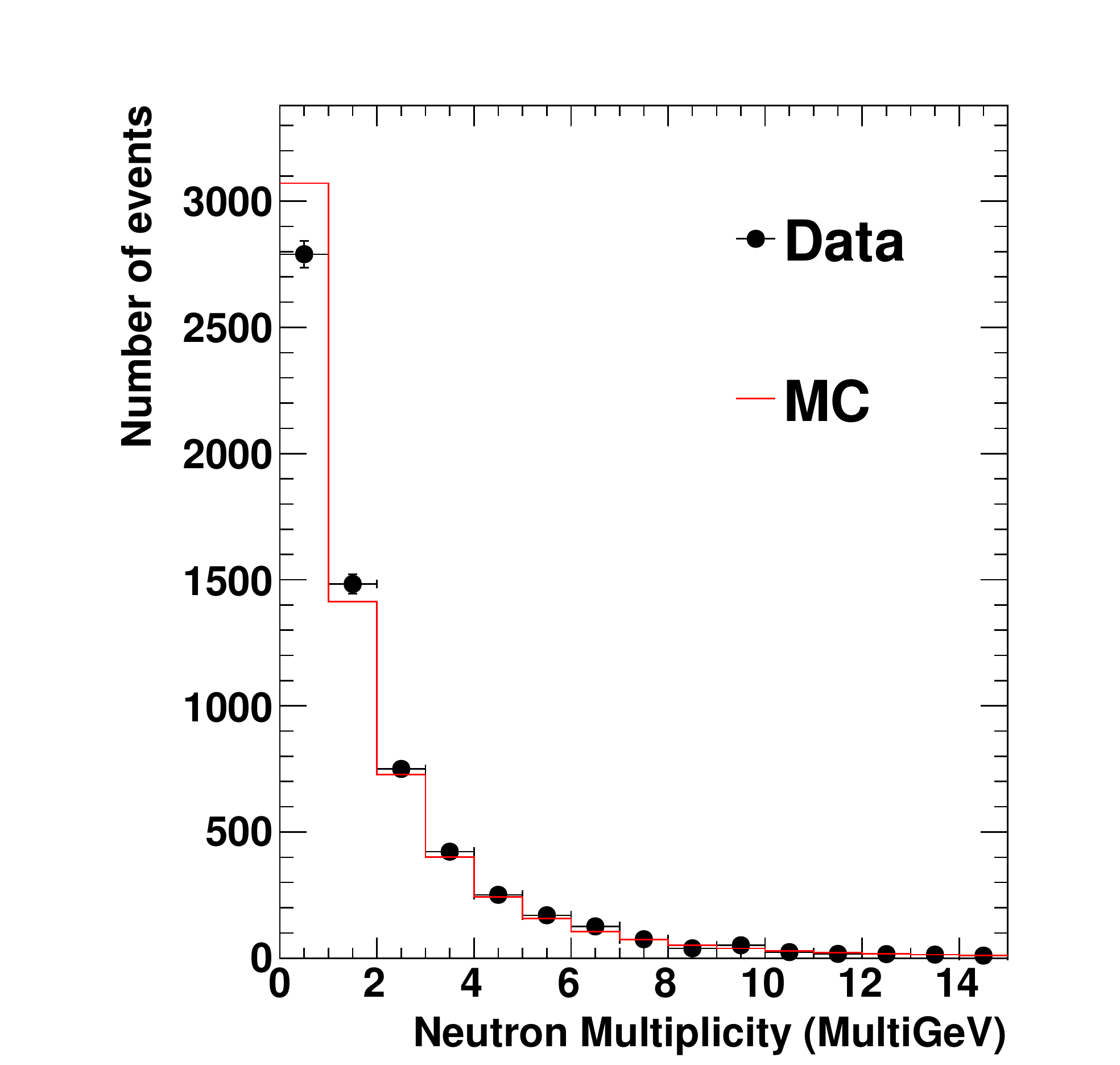}
			\end{minipage}
		\end{tabular}
 		\caption{Comparison of data and MC for tagged neutrons in the 
          SK-IV atmospheric neutrino data.
          The top left (right) plot shows the total number of neutrons 
          (average neutron multiplicity) 
          as a function of visible energy ($E_{\rm vis}$).
          The bottom left plot shows the neutron multiplicity for
          sub-GeV events ($E_{\rm vis} < 1.33$~GeV) and the bottom
          right plot shows that for multi-GeV events
          ($E_{\rm vis} \geq 1.33$~GeV). These plots are normalized
          to the number of neutrino events observed in data.  Only
          statistical errors are shown.}
 		\label{fig:atmpdntag}
 	\end{center}
\end{figure}

The difference between each tagged neutron's timing and the primary event timing 
is shown in
~\cref{fig:dt}.  By fitting the data to a falling exponential with a
constant offset, a capture lifetime of $218\pm9~\mu$s was
extracted
\color{black}
with a $\chi^2$ of 33.1 for 25 degrees of freedom.
\color{black}
This is consistent with the previously measured lifetime of
$204.8\pm0.4~\mu$s~\cite{nlife}.  
By assuming that the falling
exponential part of the fit is from true neutrons while the flat
constant is from backgrounds, it is estimated that 100\%$\pm$3.23\% of
tagged neutrons correspond to true neutron captures. 
Performing the same analysis on MC resulted in 97.9\%$\pm$3.94\%,
which is consistent with the results from the data. 
The total number of events with neutrons, the average multiplicity, and the neutron multiplicity
subdivided into
sub-GeV ($E_{\rm vis} < 1,330$ MeV) and multi-GeV ($E_{\rm vis} > 1,330$ MeV) 
samples are shown in ~\cref{fig:atmpdntag}.
These plots are normalized to the data livetime and the MC includes oscillations as described above.

\section{Systematic error study using the americium-beryllium source}

\begin{figure}[!htb]
	\includegraphics[width=0.7\textwidth]{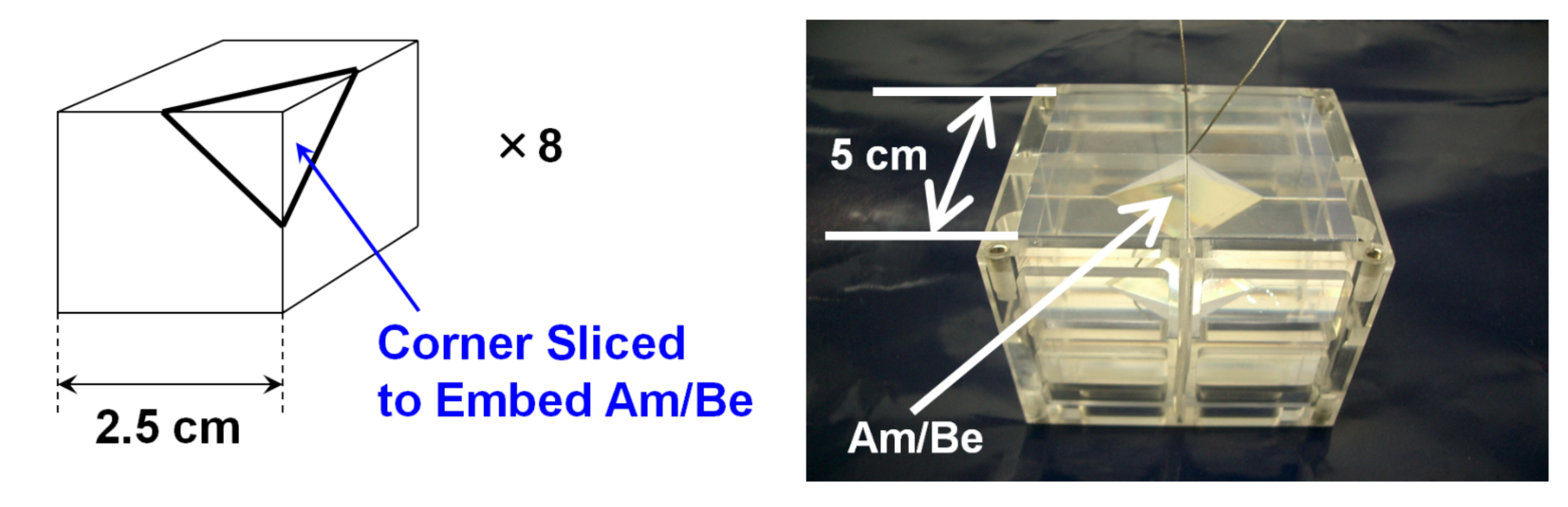}
    \centering
	\caption{Am-Be crystal embedded in a 5~cm cube of BGO
      scintillator. This is held in an acrylic case.}
	\label{fig:ambesetup}
\end{figure}

In order to study the efficiency of neutron tagging with a well-defined control sample,
calibration data sets were collected in 2016 by deploying an
americium-beryllium (Am-Be) source in the SK detector.  The
${}^{241}$Am emits an $\alpha$-particle which interacts with ${}^{9}$Be 
and emits a neutron as follows:
\begin{equation}
  \alpha + {}^{9}\text{Be} \rightarrow  {}^{12}\text{C}^{*} + n 
  \label{eqn:alpha_capture1}
\end{equation}
\begin{equation}
{}^{12}\text{C}^{*} \rightarrow  {}^{12}\text{C} + \gamma~ \text{
  (4.43~MeV)}
  \label{eqn:alpha_capture2}
\end{equation}
\begin{equation}
\text{or} \nonumber
\end{equation}
\begin{equation}
\alpha + {}^{9}\text{Be} \rightarrow  {}^{12}\text{C} + n~ \text{(ground state)}
  \label{eqn:alpha_capture3}
\end{equation}
The intensity of the ${}^{241}$Am source was 97~$\mu$Ci and
the emission rate of 4.43~MeV $\gamma$ was measured to be 87~Hz.
\color{black}
From this measurement the ground state transition rate was estimated to
be 76~Hz~\cite{ntagpaper}.
\color{black}
As shown in ~\cref{fig:ambesetup}, the Am-Be
source is embedded in a 5~cm cube of bismuth germanium oxide (BGO)
scintillator to amplify the light released by the 4.43~MeV
$\gamma$.  
This scintillation light is used to trigger the SK
detector and initiate a search for a subsequent neutron capture
signal.  The ground-state transition produces an irreducible constant
background of neutrons.  Upon triggering, an extended time window of
$-5\rightarrow835~\mu$s is stored in order to study the detection
efficiency.

The Am-Be source was placed at three different locations in the SK
tank: the position called Center is $(35.3, -70.7, 0)$~cm, the 
position called Y12 is $(35.3, -1201.9, 0)$~cm, which is 
close to the barrel wall, and the position called Z15 is 
$(35.3, -70.7, 1500.0)$~cm, which is close to the top of 
the tank. In order to estimate the accidental
background, we have also collected randomly triggered data with this source.

\subsection{Am-Be Data Selection}

\begin{figure}[!ht]
	\includegraphics[width=0.9\textwidth]{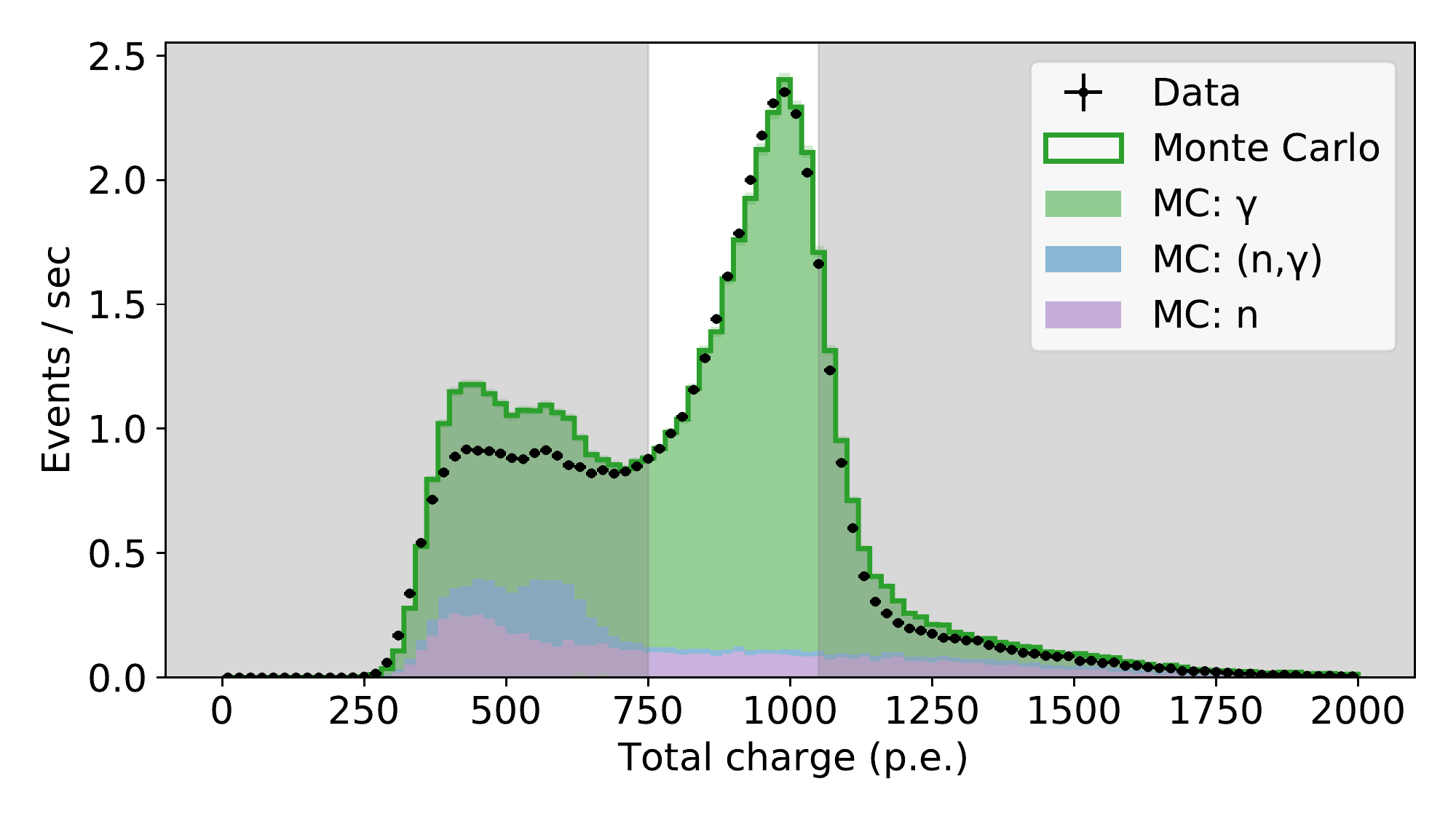}
    \centering
	\caption{Total photo-electrons from 4.43~MeV $\gamma$
      scintillation events with the Am-Be source positioned at the
      center of the tank. The black points show the data. The histogram shows the distribution obtained by the Monte-Carlo simulation. The green shaded area corresponds to the events originated by $\gamma$, the blue shaded area corresponds to the events originated by neutron and $\gamma$, and the purple shaded area corresponds to the events originated by neutron. Events with $750 < \text{p.e.} < 1,050$ were selected to trigger 4.43 MeV $\gamma$ events. The Monte-Carlo histogram is normalized to the number of events of the data, whose total charge is $750 < \text{p.e.} < 1,050$. The events in the gray shaded area are not used for the analysis.}
	\label{fig:totpe_c}
\end{figure}

The primary selection criteria for 4.43~MeV~$\gamma$ events is
based on the number of p.e. detected in the ID.
The total observed p.e. from the BGO
scintillator placed at the center of SK is shown in ~\cref{fig:totpe_c}. 
\color{black}
We have studied the shape of the p.e. distribution from 
the BGO scintillator with a dedicated Monte-Carlo
simulation program based on GEANT4. 
A qualitative study of 
the simulation results suggested that the interactions of
neutrons in the BGO scintillator produce a large number of 
p.e. and the shape of the tail agrees quite well with data.
The discrepancy between the data and the simulation in the 
tail region is coming from the uncertainty of the light
emission of BGO for neutron.
In order to select the prompt events triggered
by a 4.43~MeV~$\gamma$, the selection criteria are defined as
follows:
\begin{align*}
\text{Center: } & 750 < \text{p.e.} < 1,050 \\
\text{Y12: }    & 850 < \text{p.e.} < 1,150 \\ 
\text{Z15: }    & 900 < \text{p.e.} < 1,150. \\
\end{align*} 
Since the light attenuation from the source position to the
PMTs is different for different source positions,
 different cut values are used.
\color{black}

In addition, the primary 4.43~MeV $\gamma$ event is required to be
at least $1.5$~ms later than the previous primary event in order to
avoid contamination from prior neutron emissions. 
A search for delayed activity is then performed using the number of hits in a
200~ns sliding time window ($N_{200}$). 
If there is a timing cluster whose
$N_{200}$ is larger than 49 hits, this event is rejected to eliminate
possible contamination from cosmic ray muons or other background
sources.  
In this search, the evaluation of $N_{200}$ starts 
200~ns after the primary event, as the BGO crystal scintillator has a long decay component.

\subsection{Am-Be Data Analysis}

Neutron tagging is applied to the selected Am-Be events and 
a corresponding MC sample. 
Here we use the same neural network
trained for the atmospheric neutrino data analysis.  Since the kinetic
energy is much lower for the neutrons coming from the Am-Be source
(from 2 to 10~MeV) compared to the neutrons produced by the
atmospheric neutrino interactions, the capture positions of the
neutrons are expected to be much closer. Therefore, the location of
the source is used as the capture position for the
ToF corrections for both real data and MC in the analysis.
 The other
difference is that the total number of photons from the primary
4.43~MeV $\gamma$ is small so there is no need to 
take into account the after-pulsing of the PMTs.
Also, the gate width for this data is enlarged to 835~$\mu$s
and the neutron search timing window is accordingly enlarged from 
$18-535$~$\mu$s to $10-835$~$\mu$s.
Simulated events are produced by injecting neutrons
in the detector with an energy spectrum based on 
an $\alpha$-Be capture 
(~\cref{eqn:alpha_capture1,eqn:alpha_capture2,eqn:alpha_capture3}) as 
given in reference~\cite{Renner:2014dhp}.

\subsection{Systematic uncertainty evaluation}

\begin{figure}
  \begin{center}
  \includegraphics[width=0.8\textwidth]{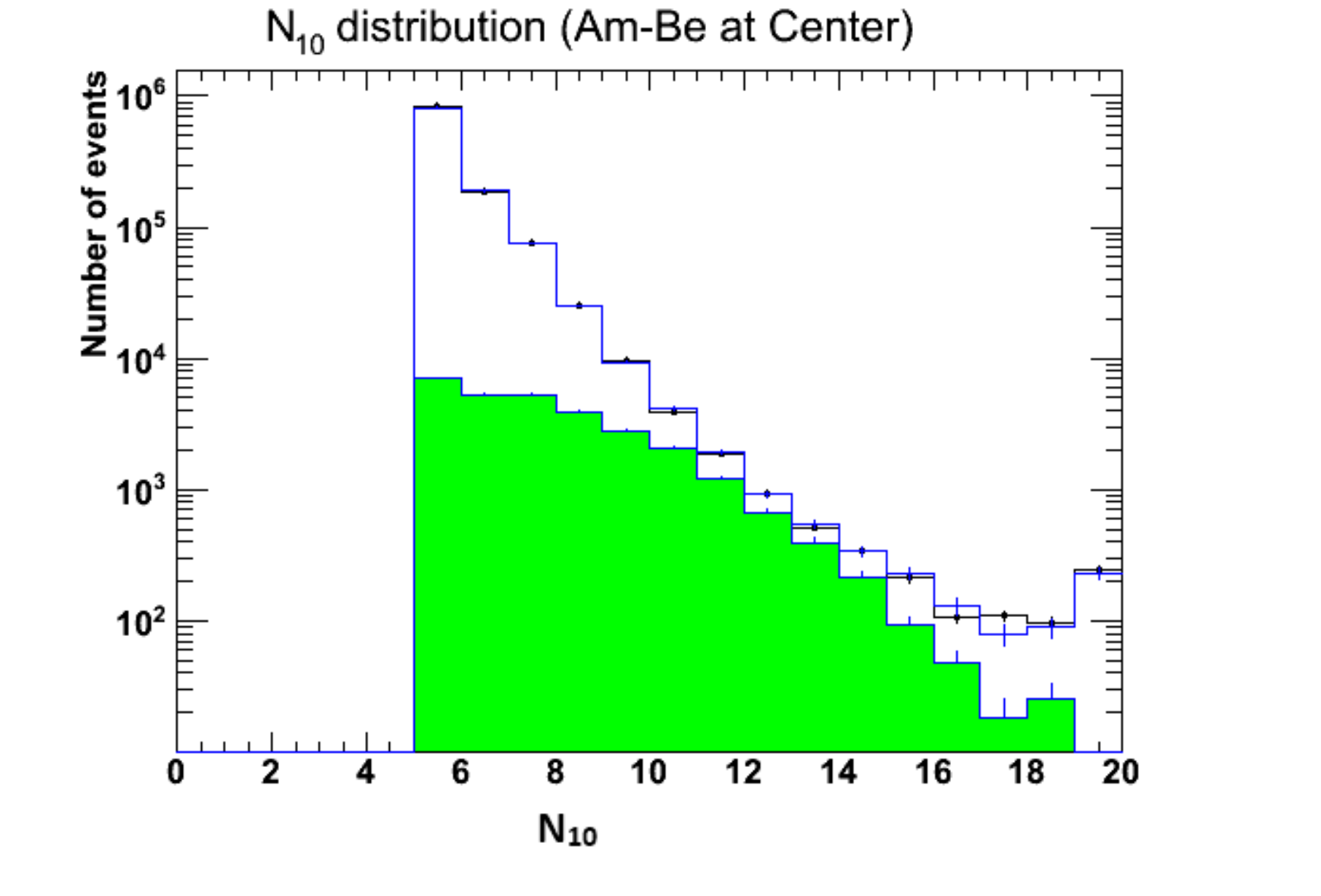}
  \caption{$N_{10}$ after the initial selection of the Am-Be calibration data. The green histogram
    shows true neutron in the MC, the blue histogram shows
    the sum of the true neutrons and background in the MC, 
    and 
    the black filled circles with error bars show the data.
    The blue histogram has been normalized to the number of real data events.
    The last bin contains the overflow events.}  
  \label{fig:AmBeN10}
  \end{center}
\end{figure}

Systematic uncertainties on the neutrino tagging method
are evaluated for the initial and final candidate selection steps 
independently. 
Since the first selection uses $N_{10}$ to select candidates,
a comparison of the MC and data in this variable is used to estimate the
systematic uncertainty. 
This variable is known to have a strong dependence
on the photon detection efficiency of the PMTs, which is the dominant 
source of the uncertainty. 
In the error analysis, simulations with different photon detector efficiencies 
are generated and compared against the background-subtracted $N_{10}$ distribution 
in data. 
~\cref{fig:AmBeN10} shows the $N_{10}$ distributions of 
the data,
together with the true neutron and the background from the MC.
Here, random trigger data taken at the same time as the calibration run 
is used as the background in the MC.
The MC has been normalized to the number of events collected
with the Am-Be source deployed. 
These distributions show good agreement indicating the background can be 
safely subtracted.

\begin{figure}
  \includegraphics[width=0.5\textwidth]{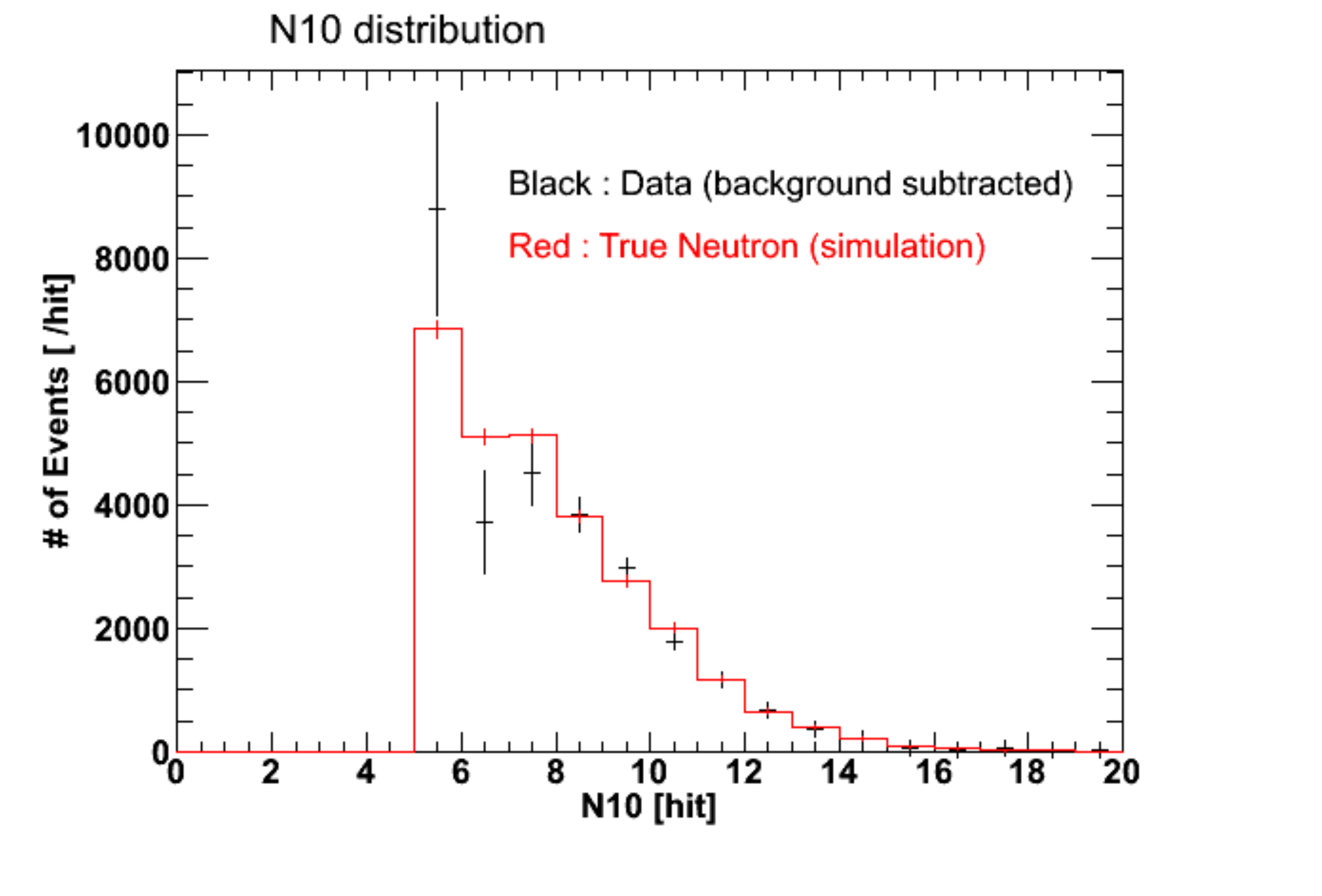}
  \includegraphics[width=0.5\textwidth]{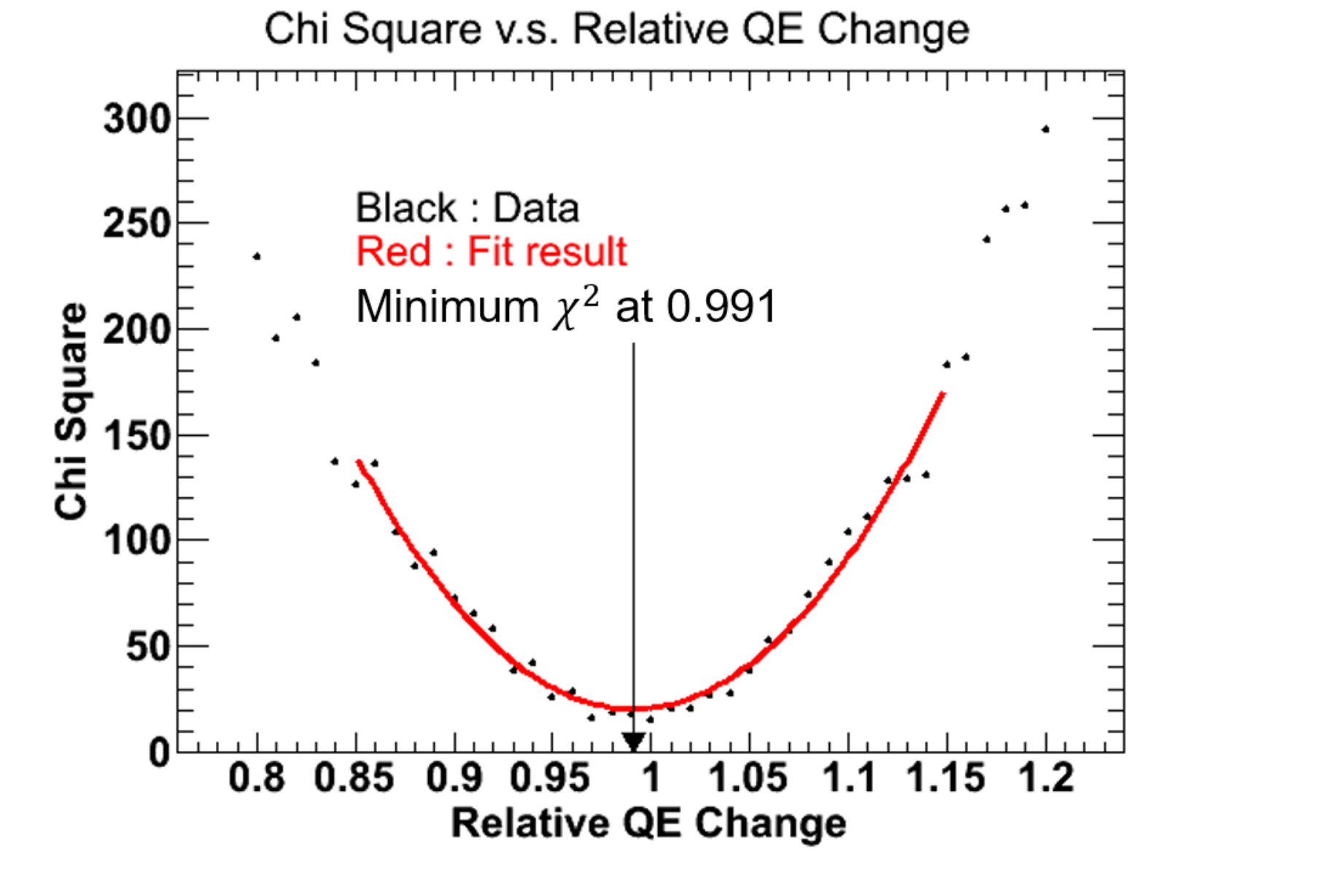}
  \caption{The $N_{10}$ distribution after the initial selection (left).
           Black points show the data after background subtraction 
           and the red histogram shows the true neutron distribution from the Monte-Carlo simulation.
           The photon collection efficiency is set to its nominal value in this figure. 
           The $\chi^{2}$ distribution from a fit to the data as 
           a function of the simulated photon detection
           efficiency (right). Black points show the data and the red
           line shows the result of a fit with a parabolic function.}
  \label{fig:AmBeInitialUncertainty}
\end{figure}

The left plot of ~\cref{fig:AmBeInitialUncertainty} shows 
the distributions of 
$N_{10}$ for the background-subtracted data and simulation,
with the nominal value used for the photon detection efficiency
in the simulation.
Fitting this data assuming different detection efficiencies 
yields the $\chi^{2}$ distribution shown in 
the right plot of ~\cref{fig:AmBeInitialUncertainty}.
The minimum value of the $\chi^2$ was obtained when the photon
detection efficiency is changed by $-0.9$\%.
A 2.2\% change is allowed at $1\sigma$ which 
corresponds to a 1.7\% change in the number of candidates 
passing the initial selection.

\begin{figure}[tb]
  \includegraphics[width=0.5\textwidth]{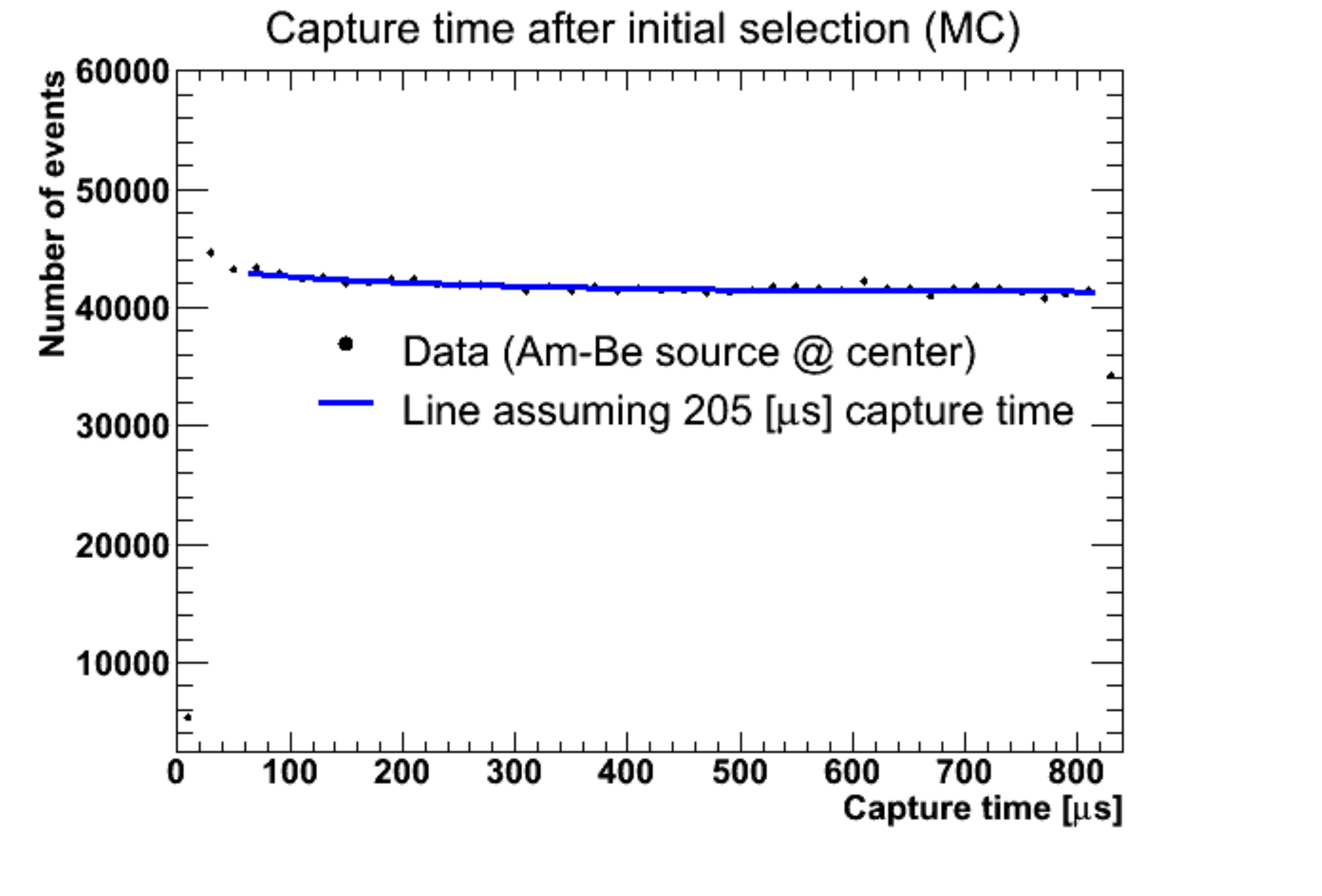}
  \includegraphics[width=0.5\textwidth]{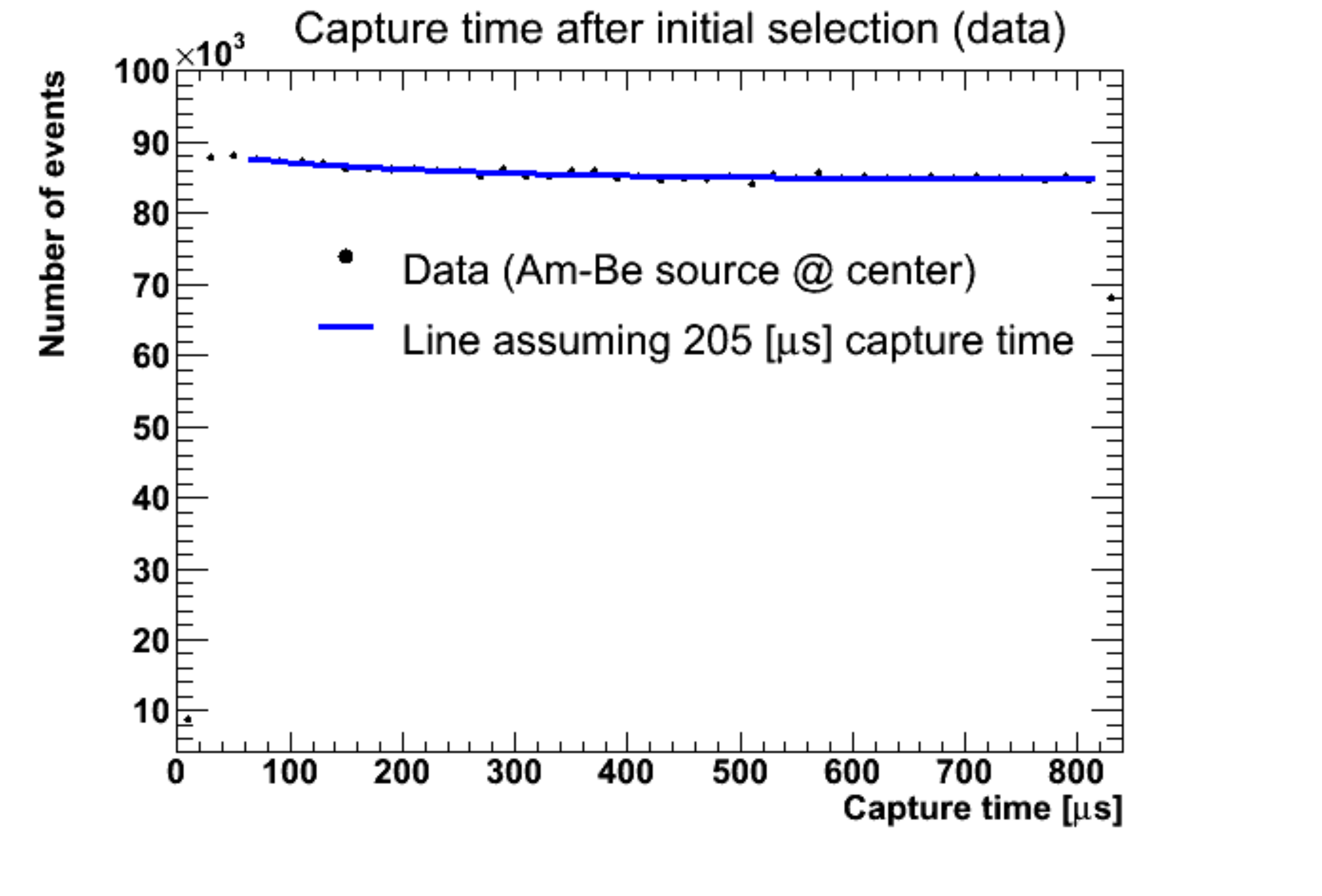}
  \includegraphics[width=0.5\textwidth]{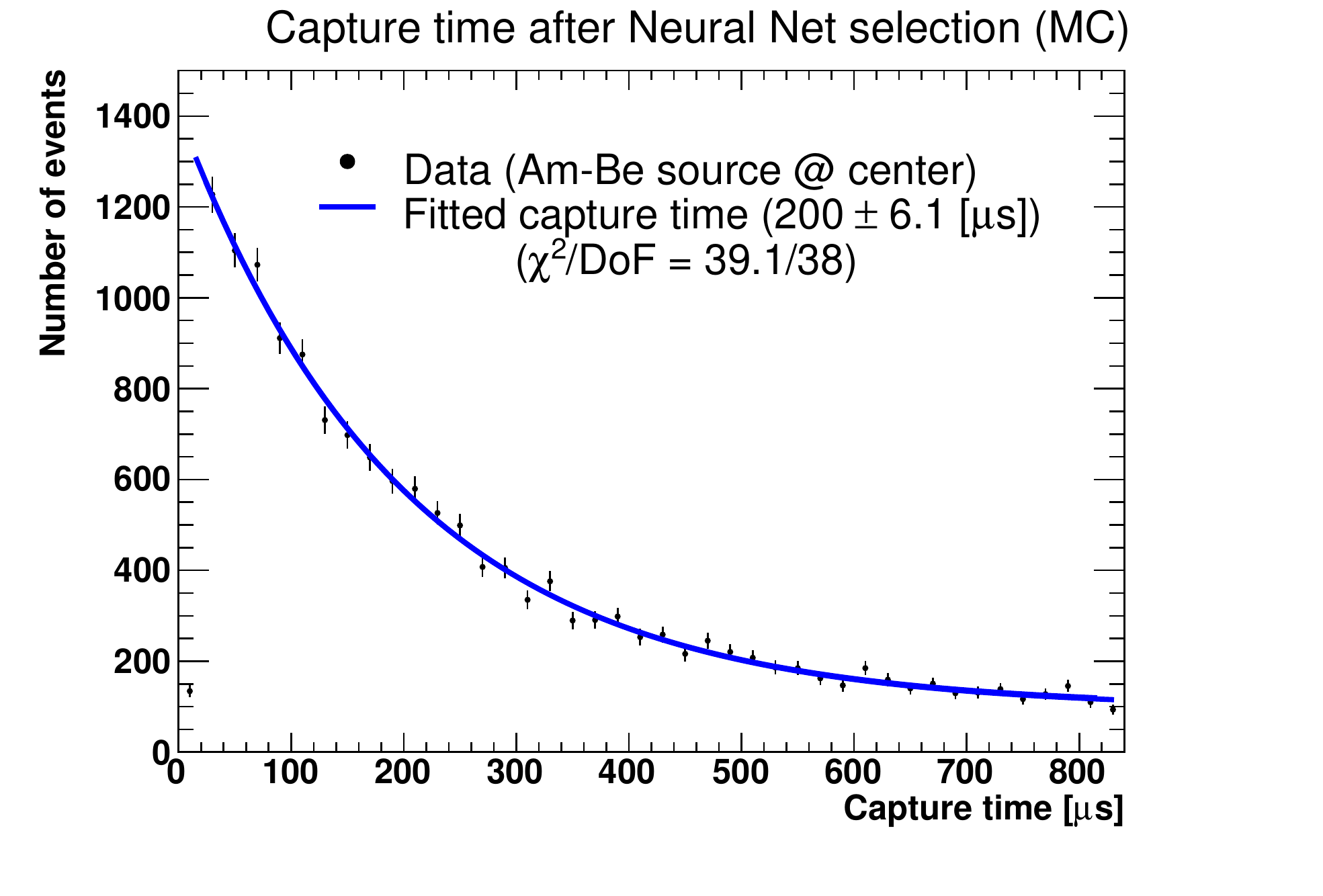}
  \includegraphics[width=0.5\textwidth]{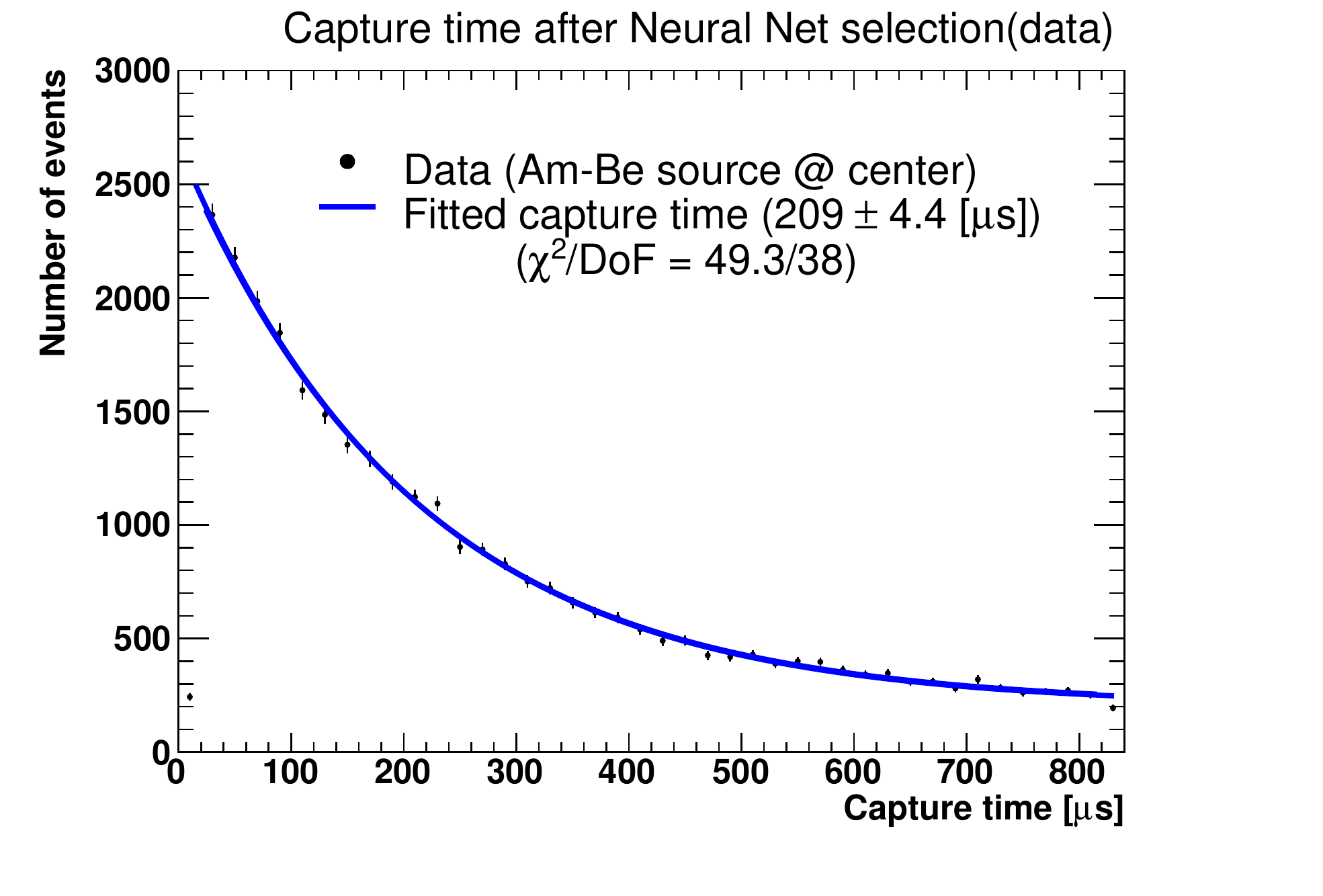}
  \caption{Neutron capture times for data and MC with the Am-Be source.
    The upper two figures show the capture time for events passing 
    the initial selection. The bottom two figures show the capture time after
    the neural net selection. Black dots show the data while the blue
    lines in the upper two plots are reference lines assuming a capture
    time of 205~$\mu$s. Blue lines in the lower two plots show the
    results of fits for the neutron capture time.}
  \label{fig:AmBeCapturetime}
\end{figure}

The final event selection is done using the neural network. 
The systematic uncertainty for this step is evaluated by comparing the relative neutron tagging efficiencies 
between data and MC for candidates that pass the initial selection.
This relative
efficiency ($\epsilon_r$) is defined as 
$\epsilon_r = \epsilon_\text{NN}/\epsilon_\text{IS}$, 
where $\epsilon_\text{IS}$
is the efficiency for identifying the neutron using the initial candidate
selection and $\epsilon_\text{NN}$ is the efficiency for identifying
the neutron using the neural network selection.
Before estimating the efficiencies, we checked whether the neutron 
tagging algorithm is working properly with the Am-Be data.
For this purpose, the neutron capture time was fitted using the  
initial candidates and those passing the neural network. 
~\cref{fig:AmBeCapturetime} shows the distribution of 
the neutron capture times.
All the obtained capture times are close to 200~$\mu$s, which are 
consistent with the past measurements.

Using three data samples taken at the three different 
Am-Be positions, Center, Z15, and Y12, the efficiencies
of the initial selection 
$\epsilon_\text{IS}$ and 
$\epsilon_\text{NN}$ are calculated for both data and MC sample.
The obtained $\epsilon_r$ values for the data and the MC sample
are summarized in ~\cref{table:AmBeEfficiency}
and the maximum efficiency difference is 8.8\%.

\begin{table}[tb]
\centering
\caption{Relative neutron tagging efficiencies 
($\epsilon_\text{NN}/\epsilon_\text{IS}$) in percent obtained for the Am-Be sample.
}
\label{table:AmBeEfficiency}
\begin{tabular}{|c||c|c|c|}
  \hline
         & Data & MC data & (MC - Data)/Data \\
  \hline  
  Center & 62.7\% & 60.0\%        & $-4.5$\%             \\
  Z15    & 62.5\% & 67.0\%        & 8.8\%             \\
  Y12    & 68.1\% & 65.9\%        & $-3.2$\%             \\
  \hline
\end{tabular}
\end{table}

In summary, there is a 1.7\% uncertainty in the initial selection and
an 8.8\% uncertainty in the neural network selection. 
Therefore, we assign $\pm$9.0\% as the neutron detection uncertainty.

\section{Conclusion}
A new neutron tagging technique has been developed for identifying neutrons 
produced in the atmospheric neutrino data sample of Super-Kamiokande-IV.
A tagging efficiency of 26$\%$, with accidental background rate of
0.016 per neutrino event, has been achieved. The error of the tagging
efficiency is estimated to be 9.0\%.
This method was verified with an americium-beryllium neutron source.
Discrepancies of up to $\sim10$\% in the detection efficiencies of data
and MC samples were observed.
These discrepancies seem to
arise from assumptions made in the production of the Monte-Carlo simulation
and from the uncertainties in the 
modeling of the americium-beryllium neutron source itself.
This neutron tagging method was applied to 3,244.4 days of SK-IV data
and 18,091 neutron-capture candidates were identified.
 This agrees well with MC predictions. 
Using the detected neutron candidates, the neutron capture livetime in water was measured to be
218~$\pm$~9$~\mu$s.

This technique provides the ability to utilize additional information in
neutrino interactions.
Measuring neutron multiplicity in an event is
expected to improve the separation of anti-neutrino events from
neutrino events and improve the background rejection in nucleon decay
analyses.
Although neutron information plays a crucial role in 
various physics studies, it is difficult to improve the
efficiency much further due to the limited number of photons emitted
from the scattering of the 2.2 MeV $\gamma$ from neutron capture in water. 
Therefore, the SK collaboration has started to dissolve gadolinium
into the water since it captures neutrons efficiently and emits several 
MeV in multiple $\gamma$ rays.
This will enable more efficient neutron tagging and 
improve the capability of the SK experiment~\cite{SKGDProposal}.

\appendix

\section{Noise characteristics of the 20-inch PMT}\label{section:PMTNoise}
\begin{figure}
\includegraphics[width=\textwidth]{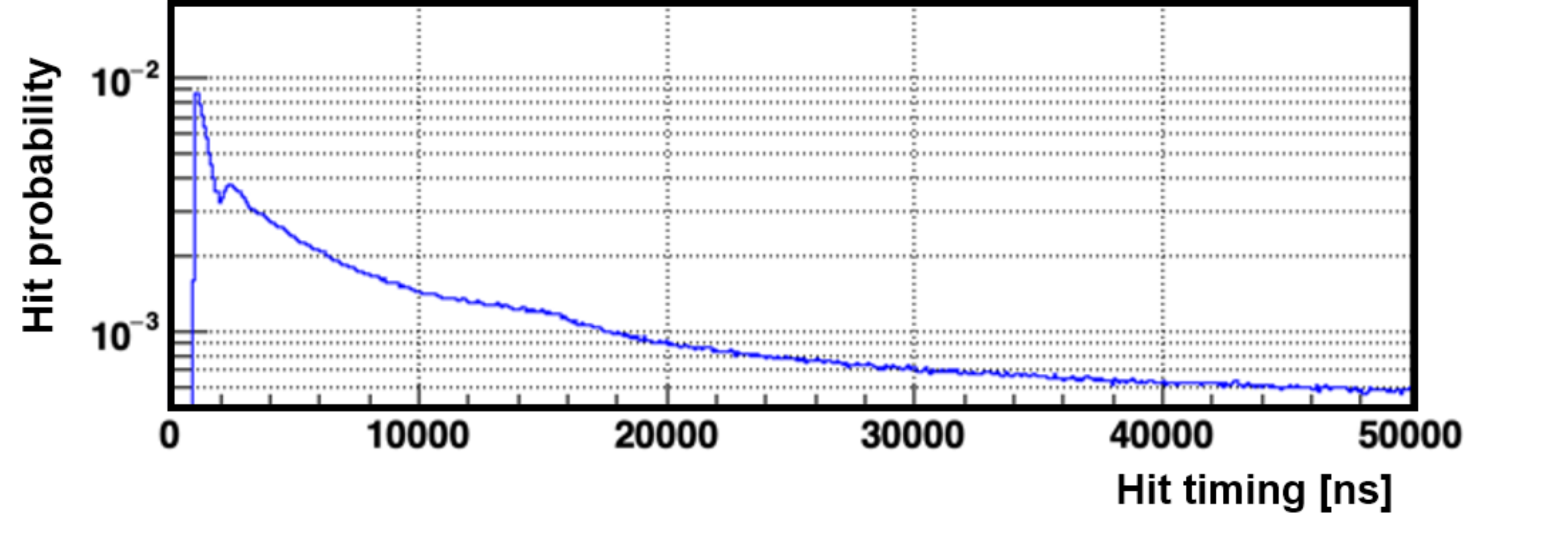}
\caption{Timing distribution of repeated hits (noise rate)
following a signal pulse above the discriminator threshold
($\sim$0.25~p.e.) of a 20-inch PMT. The X axis shows the timing
difference of a repeated hit from the initial hit and
the Y axis shows the probability of the secondary hit.
The gap between 0 to 1~$\mu$s is due 
to the channel dead time of the readout electronics and the peak around 1.5~$\mu$s is due to the reflection of the PMT signal.
The repeat noise rate decreases exponentially in time
and the feature around 12 to 18~$\mu$s 
is due to PMT after-pulsing.
}
\label{fig:TimingCluster2} 
\end{figure}

The typical noise rate of a 20-inch PMT in SK is around 5~kHz at a 0.25
photo-electron (p.e.) threshold.
A careful study of the noise revealed that it consists of two components. 
The first component is distributed 
uniformly in time while the other forms clusters in time.
The timing distribution of repeated hits
following a signal pulse above the discriminator threshold
($\sim$0.25~p.e.) of a 20-inch PMT is shown in ~\cref{fig:TimingCluster2}. 
Further study identified these time-clustered noise hits as caused 
by scintillation light from 
radioactivity in the PMT glass. 
The bump around 15~$\mu$s is caused by ionized residual gas
molecules, which are produced by the collisions with 
the accelerated photo-electrons.
Such pulses are hereafter referred to as after-pulses.

 As the relative ratio of time-clustered noise to uniformly distributed random noise 
is different from PMT to PMT,
this noise is not easy to simulate. 
Therefore, we use randomly triggered data to account for this time-clustered noise in the neutron 
tagging analysis.

\section{Low energy event reconstruction algorithms}\label{section:LE_reconstruction_algorithms}

We used two reconstruction algorithms to estimate the location of 
the neutron capture.  The
first is a standard low energy reconstruction tool, BONSAI, which has been
used for the solar neutrino analyses in SK~\cite{bonsai}.
The BONSAI reconstruction uses timing information from PMT hits
in a 1.3~$\mu$s time window.
It performs an iterative search from a
starting position, with multiple search branches fanning out from that
starting position.  Branches are stopped and pruned when the goodness
of fit drops below a certain level.  In this application, the
reconstructed primary event vertex is used as the starting point for
BONSAI.

The second reconstruction tool is called Neut-Fit, which was developed
specifically for this analysis. Neut-Fit is a simple vertex fitter and
uses the timing information from the hits within a 10 ns time window.
A shrinking grid search method is used to minimize $t_{\rm rms}$,
defined as
\begin{equation}
  t_{\rm rms}(\vec{x}) = \sqrt{\frac{\sum_{i}^{N_{10}}(t_{i} - t_{\rm
        mean})^{2}}{N_{10}}},
  \label{eq:trms}
\end{equation}
where $t_{\rm mean} = \sum_{i}^{N_{10}}t_{i}/N_{10}$, and $t_{i}$ is
the hit timing after ToF subtraction to the vertex $\vec{x}$,
respectively.  
As the search progresses, the search grid shrinks
until the space between points on the grid becomes 0.5 cm.  
The algorithm is applied twice, first with a constraint that 
the reconstructed vertex must be within 200 cm of the primary 
event vertex.
Hits are then ToF-corrected to this constrained neutron vertex,
and these residual times are used for the calculation of neural 
network variables described below.
For example, the number of hits within 10~ns after the Neut-Fit ToF 
correction ($N_{10n}$) is used instead of $N_{10}$.
For the second time, Neut-Fit is applied with no constraint 
other than requiring the vertex being in the SK tank.
This second unconstrained vertex is used for variables 
in the neural network which compare this vertex to the BONSAI vertex 
and the neutrino interaction vertex from APFit
(~\cref{subsec:fit_agreement}).

\section{Distributions of the neural network input variables for neutron tagging}\label{section:NN_Input_Distributions}

\begin{figure}[htb]
  \begin{center}
    \includegraphics[width=0.49\textwidth]{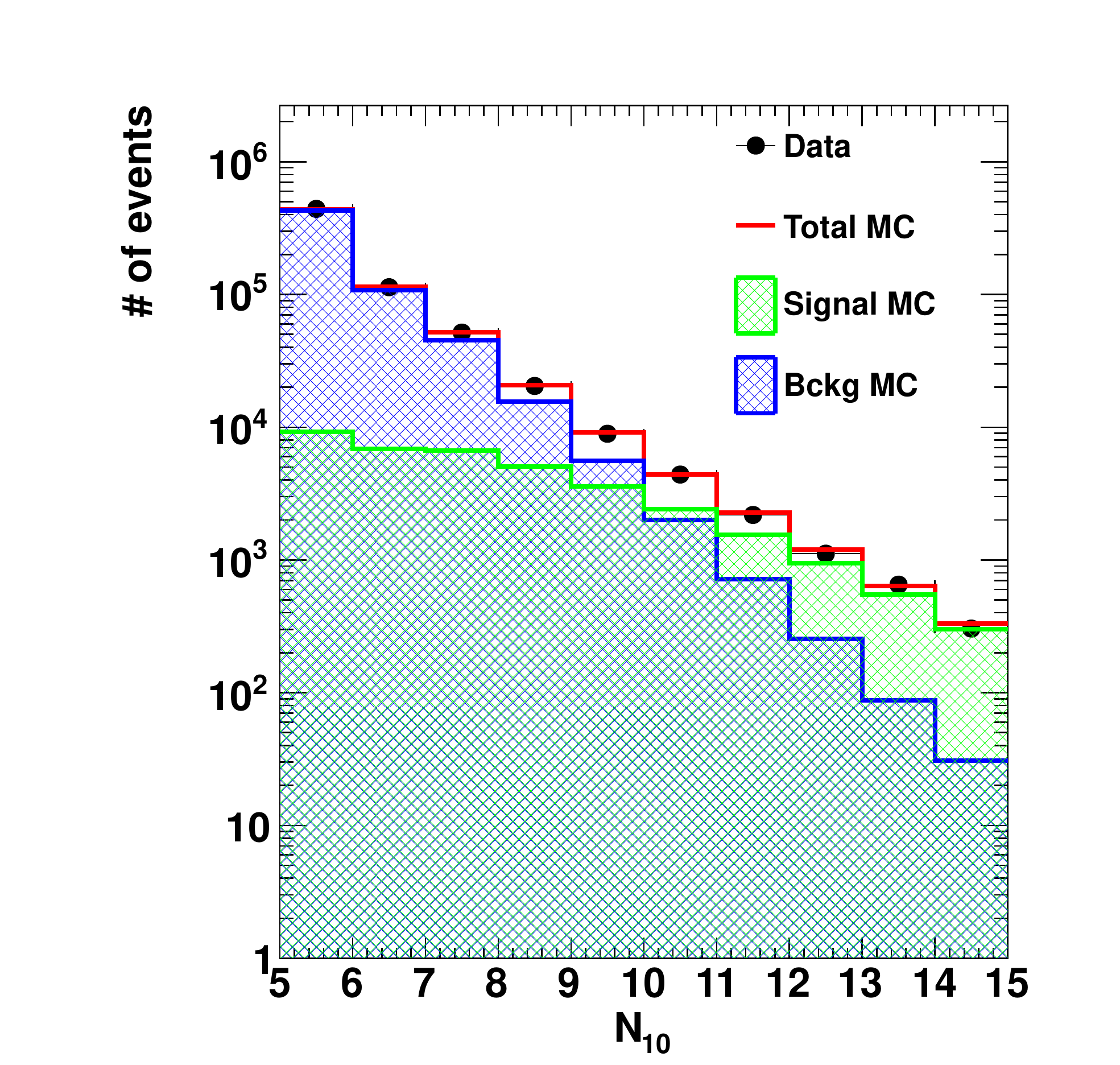}
    \includegraphics[width=0.49\textwidth]{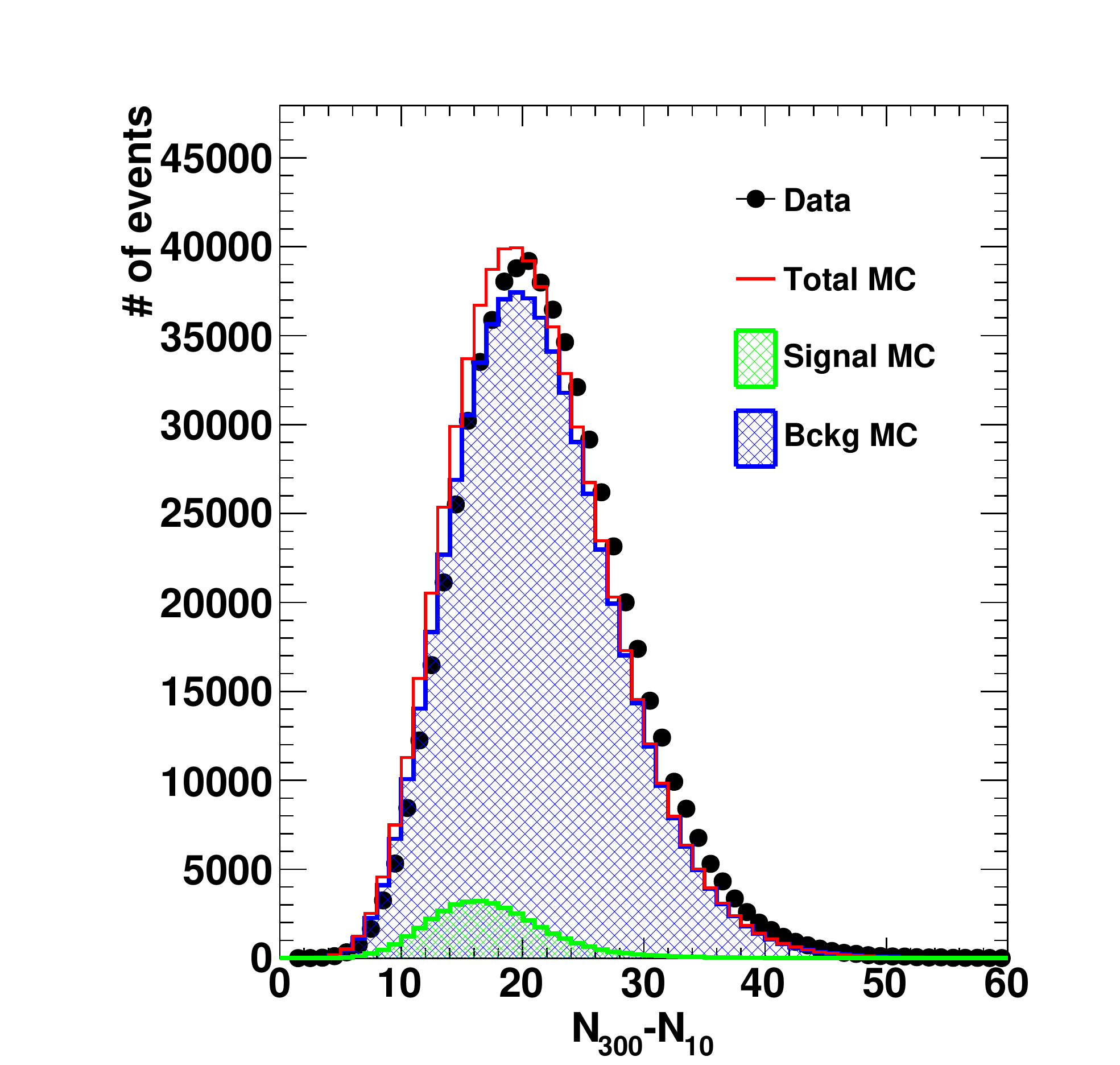}
  \end{center}
  \caption{Distributions of the number of hits in 10~ns variable,
    $N_{10}$ for $N_{10} \geq 5$ (left) and the number of hits 
    in 300~ns variable, $N_{300} - N_{10}$ (right).
    The signal events tend to give larger $N_{10}$ compared
    to the background.
    $N_{10}$ is expected to have a larger fraction of
    $N_{300}$ for the signal compared to the background. 
  }
  \label{fig:neutron4x.n10_n300}
\end{figure}

\begin{figure}[!ht]
  \begin{center}
    \includegraphics[width=0.49\textwidth]{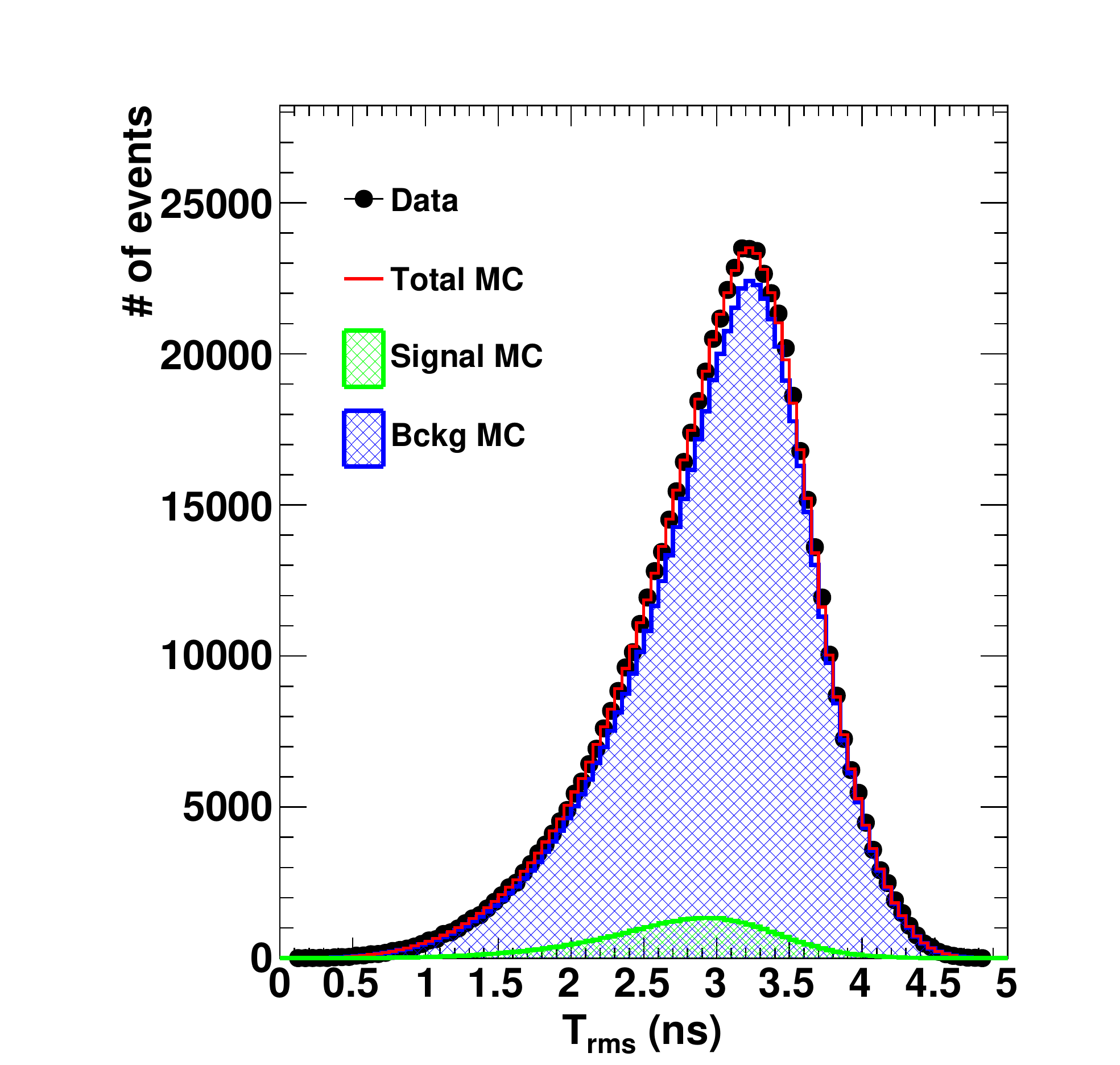}\\
    \includegraphics[width=0.49\textwidth]{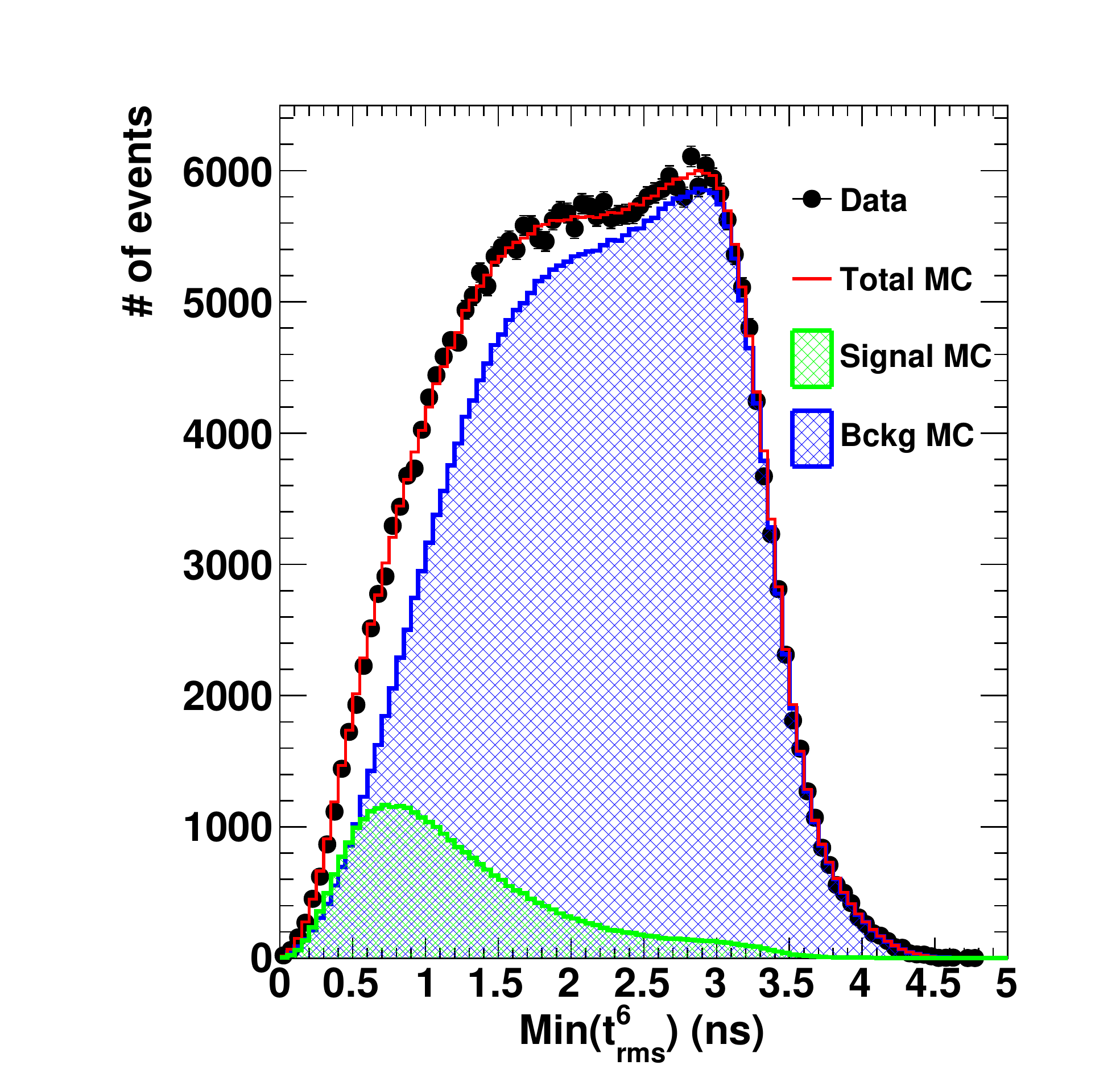}
    \includegraphics[width=0.49\textwidth]{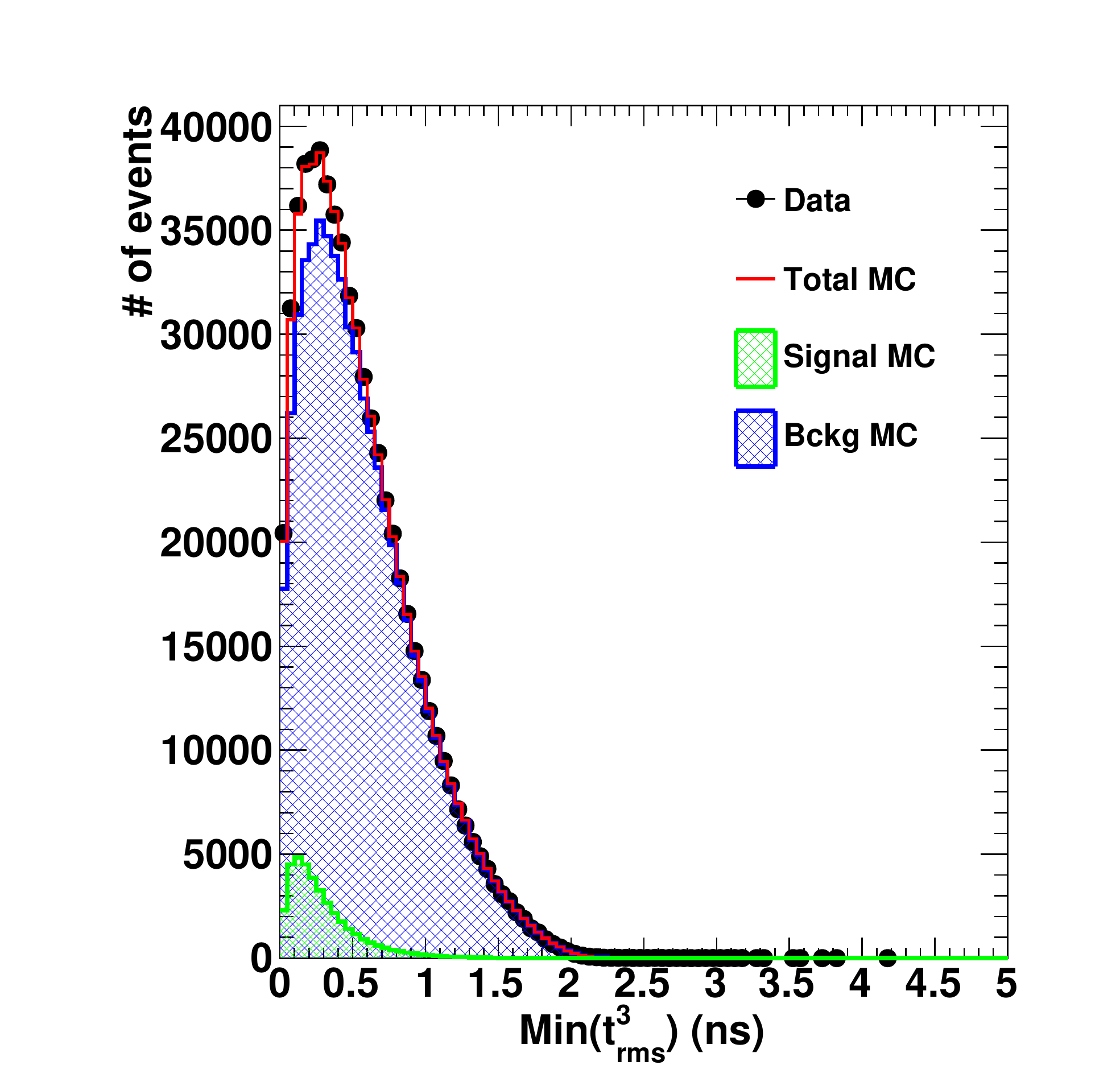}
  \end{center}
  \caption{Distributions of the root-mean-square of the ToF-subtracted 
    timing of candidate hits, $t_{\rm rms}$ (top), 
    the minimum root-mean-square of hit timing (min$(t_{\rm rms})$) 
    of clusters of 6 hits, $\min(t_\text{rms}^6)$ (bottom left),
    and 3 hits, min($t_{\rm rms}^3$) (bottom right).
    The hits from the signal are expected to be concentrated in time 
    and thus $t_{\rm rms}$, $\min(t^6_\text{rms})$,
    and min($t_{\rm rms}^3$) are smaller compared to the background.
  }
  \label{fig:neutron4x.trmsold_mintrms}
\end{figure}

\begin{figure}[!ht]
  \begin{center}
    \includegraphics[width=0.49\textwidth]{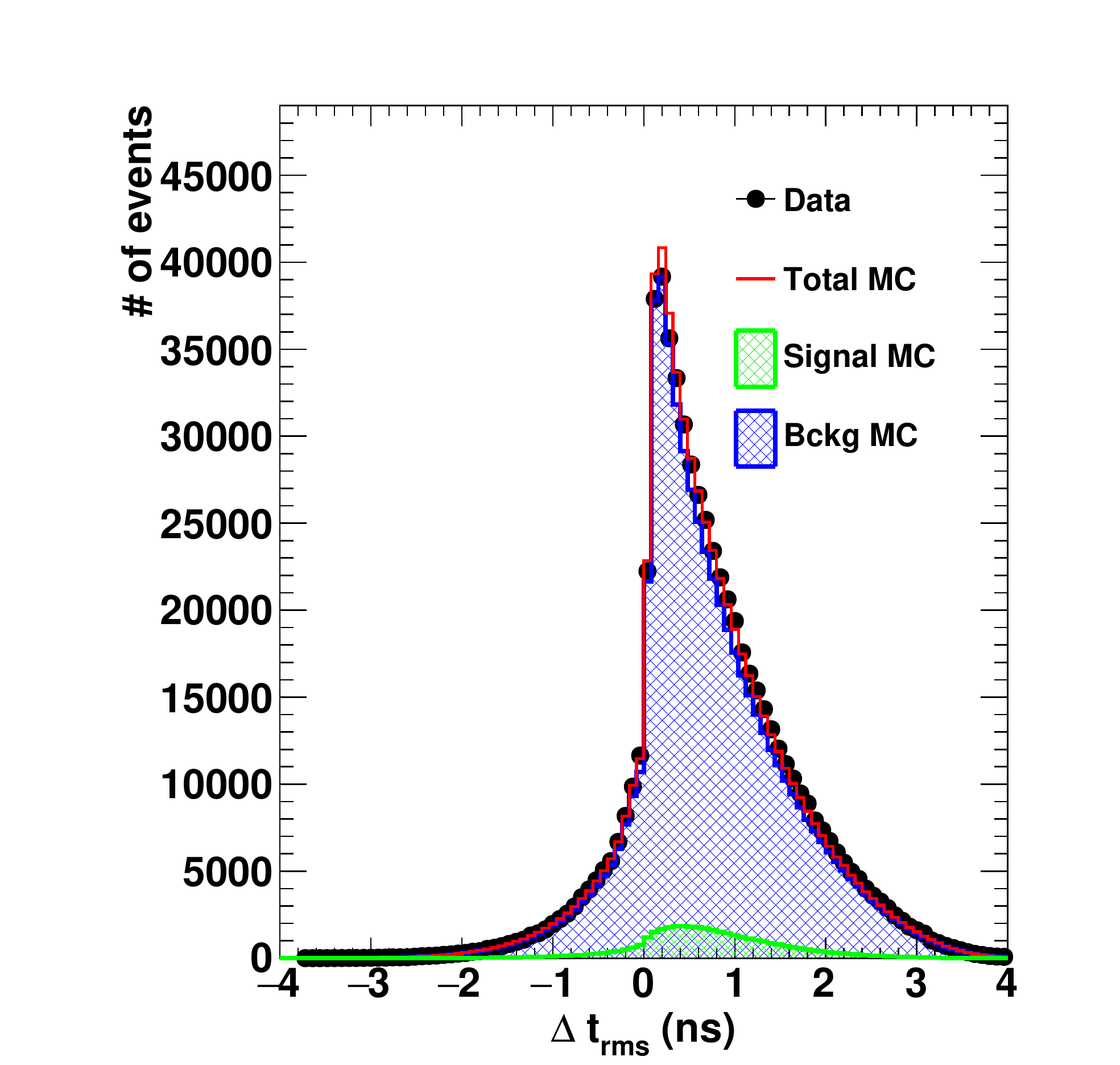}
    \includegraphics[width=0.49\textwidth]{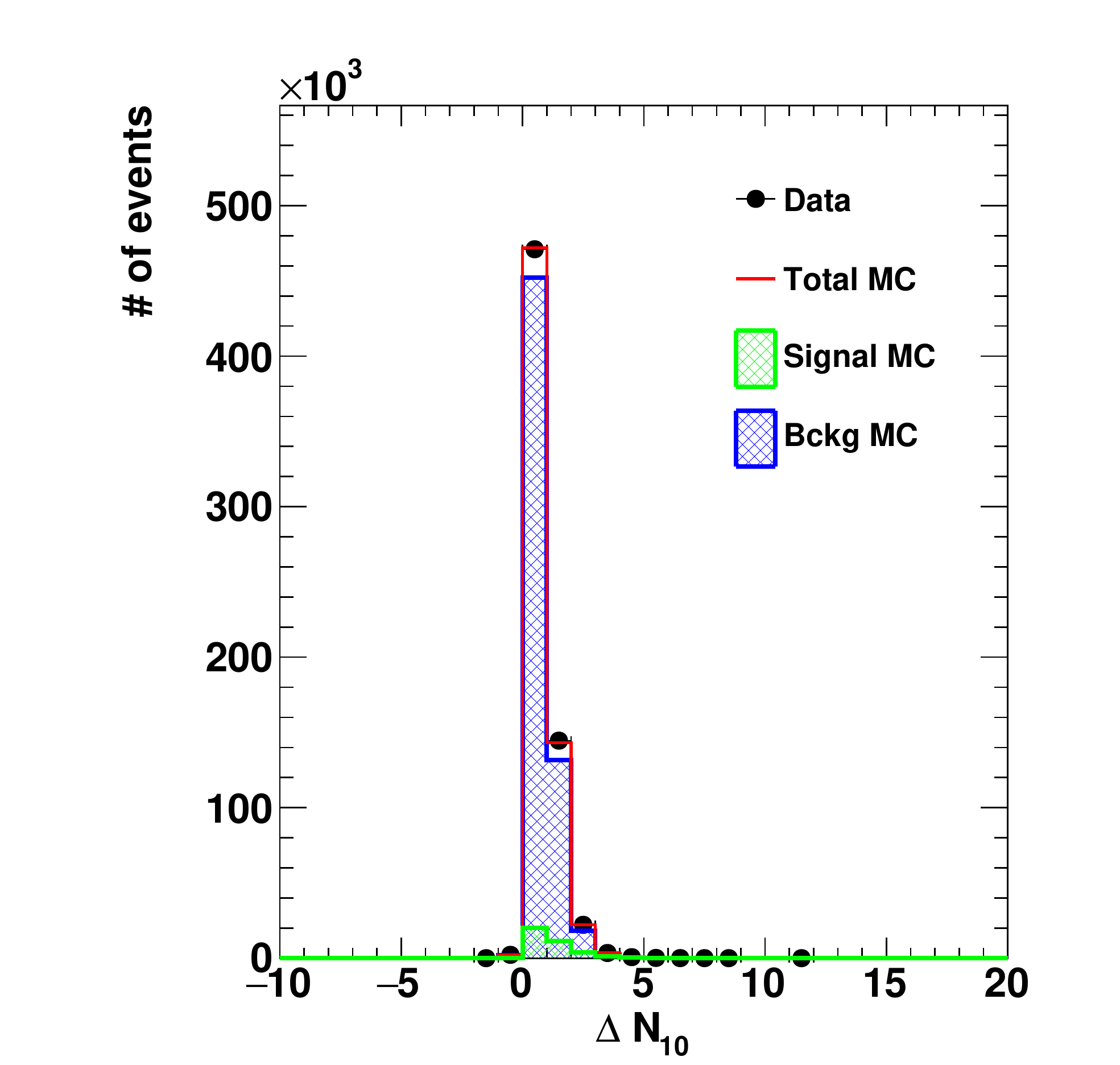}
  \end{center}
  \caption{Distribution of the Neut-Fit root-mean-square of hit timing
    difference variable, $\Delta t_{\rm rms}$ (left), and 
    the Neut-Fit number of hits in 10 ns difference variable, 
    $\Delta N_{10}$ (right).
    These two variables are expected to be close to 0 for the 
    signal because the vertexes between the two reconstructions 
    are expected to be close for signal but not necessarily same
    for the background. 
  }
  \label{fig:neutron4x.trmsdiff_n10d}
\end{figure}

\begin{figure}[htb]
  \begin{center}
    \includegraphics[width=0.49\textwidth]{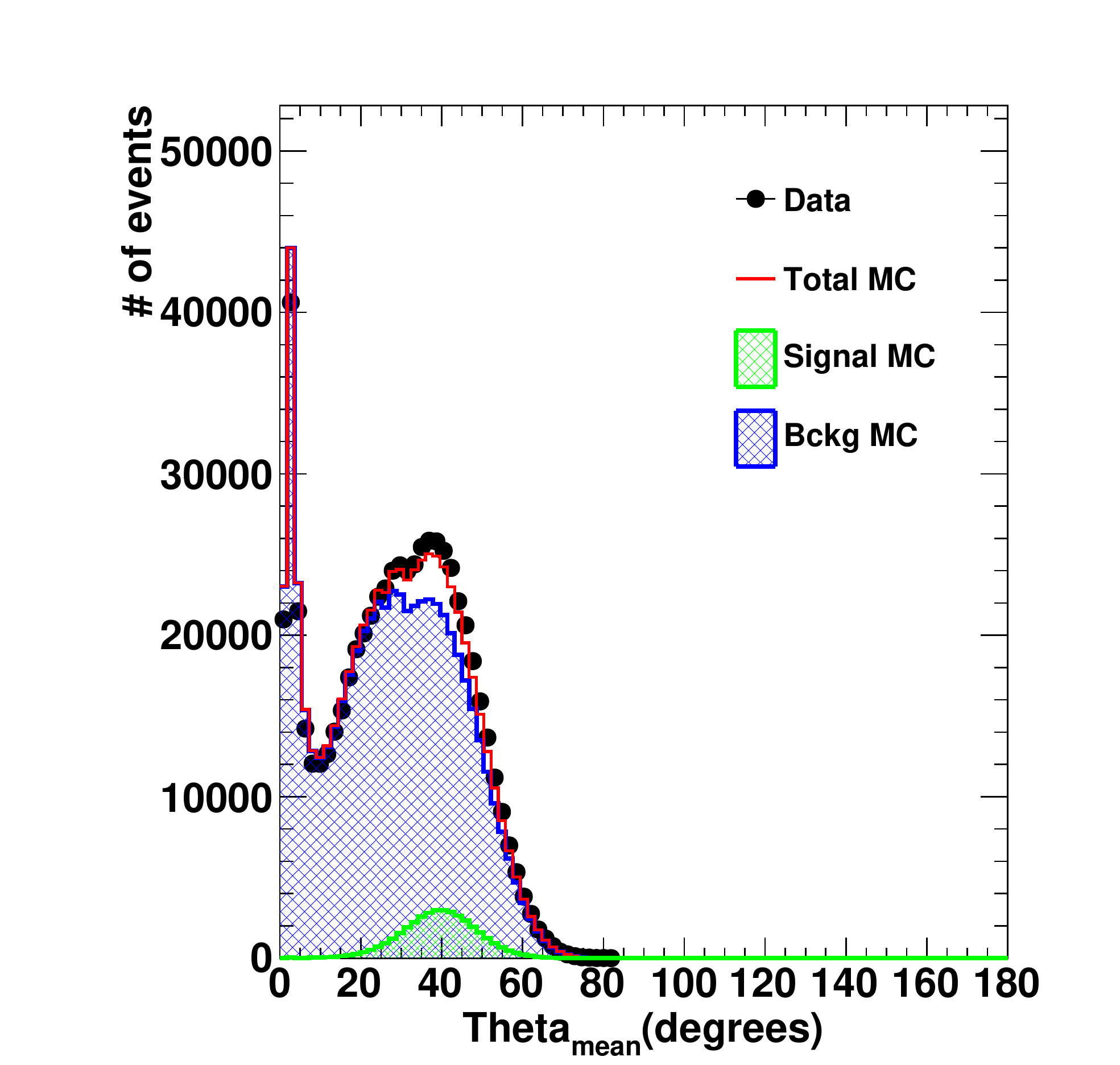}
    \includegraphics[width=0.49\textwidth]{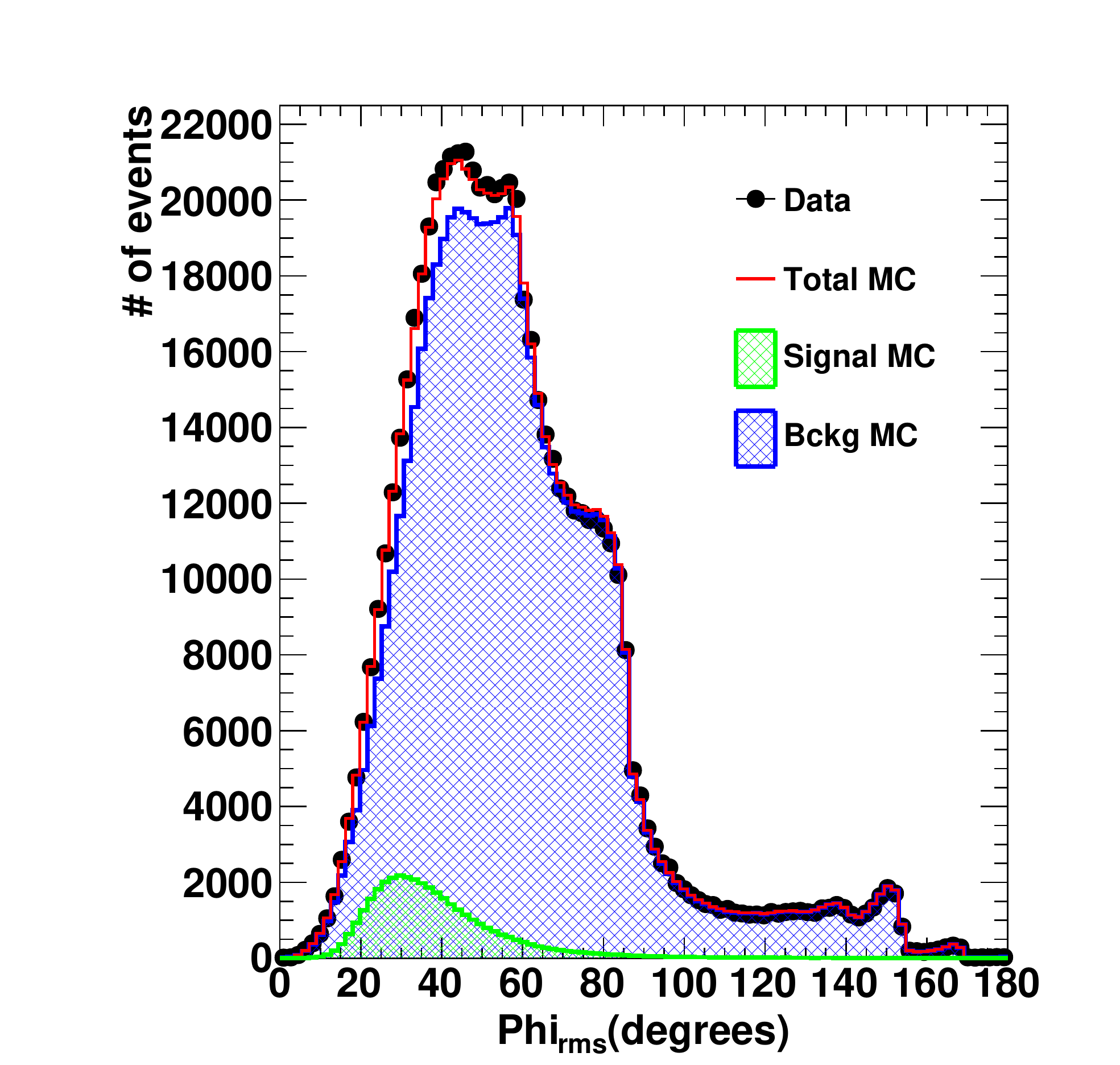}\\
    \includegraphics[width=0.49\textwidth]{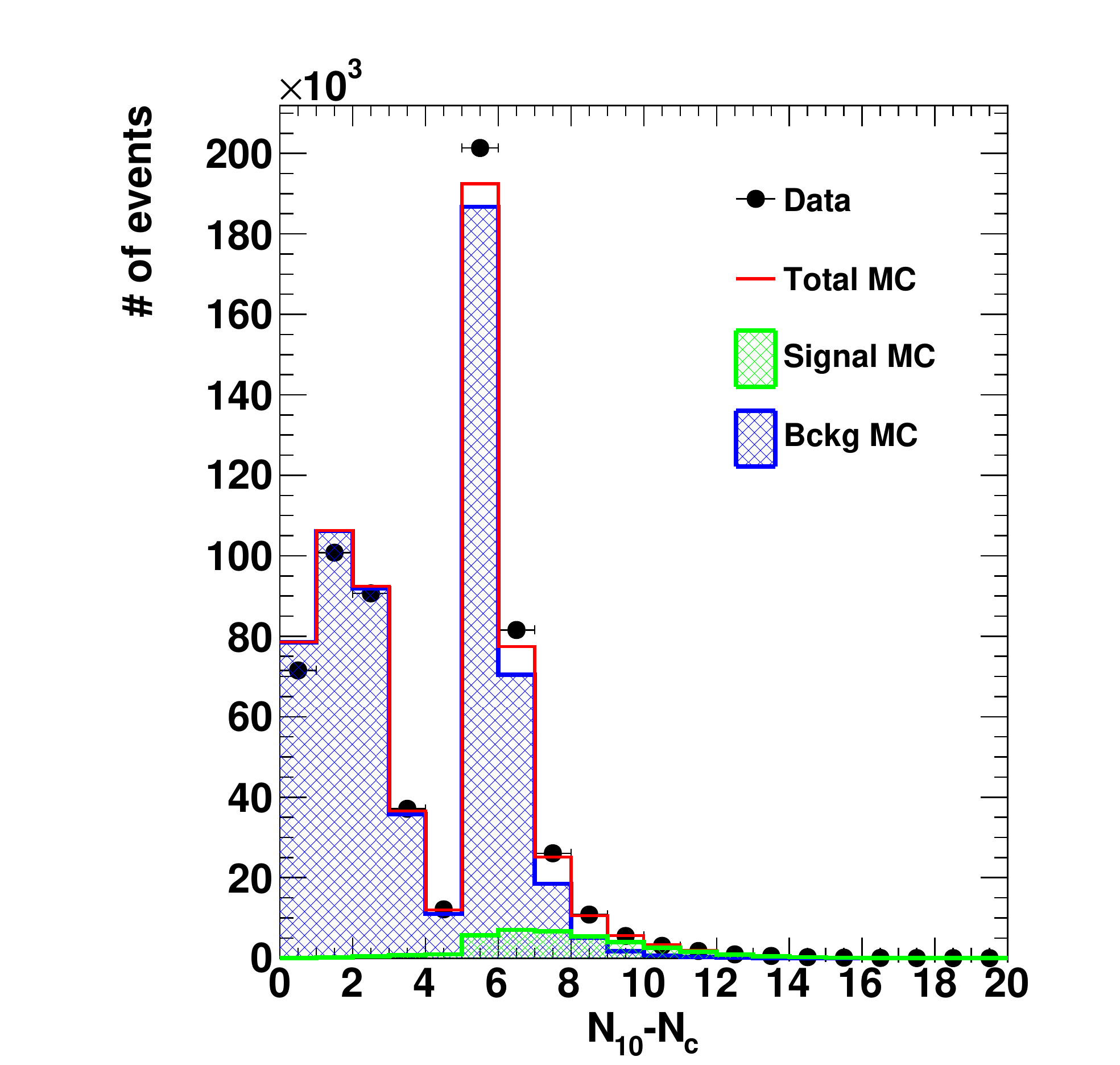}
  \end{center}
  \caption{Distributions of mean opening angle, 
    $\theta_{\rm mean}$ (top left),
    the hit vector root-mean-square of the azimuthal angle, 
    $\phi_{\rm rms}$ (top right),
    and 
    the clusters of hits variable, $N_{10} - N_{\rm c}$ (bottom).
    $\theta_{\rm mean}$ has a clear peak around 42 degrees for 
    signal as expected. $\phi_{\rm rms}$ is expected to be small for 
    signal than background. $N_{10} - N_{\rm c}$ 
  }\label{fig:neutron4x.theta_phi_nc}
\end{figure}

\begin{figure}
  \begin{center}
    \includegraphics[width=0.49\textwidth]{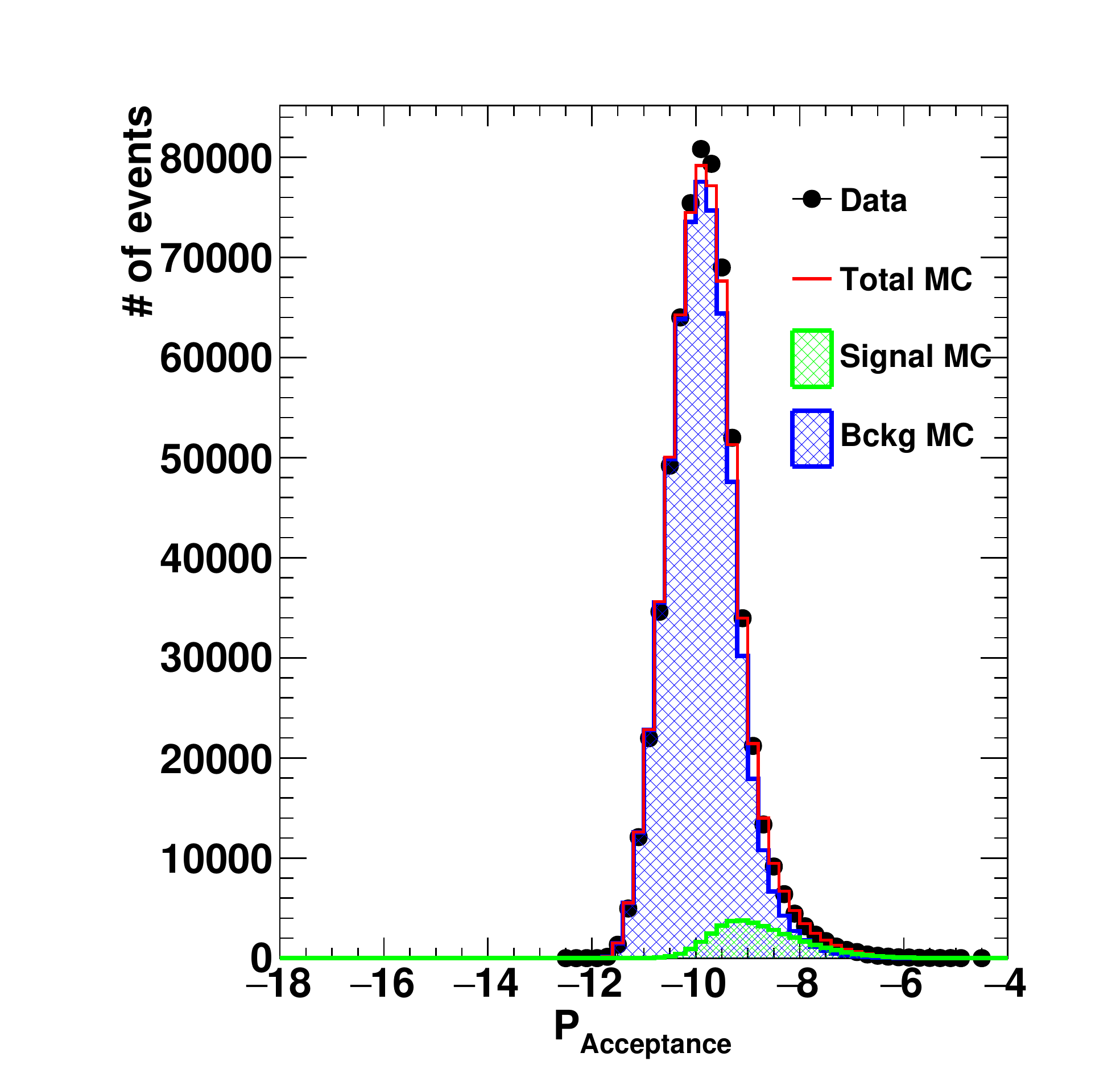}
    \includegraphics[width=0.49\textwidth]{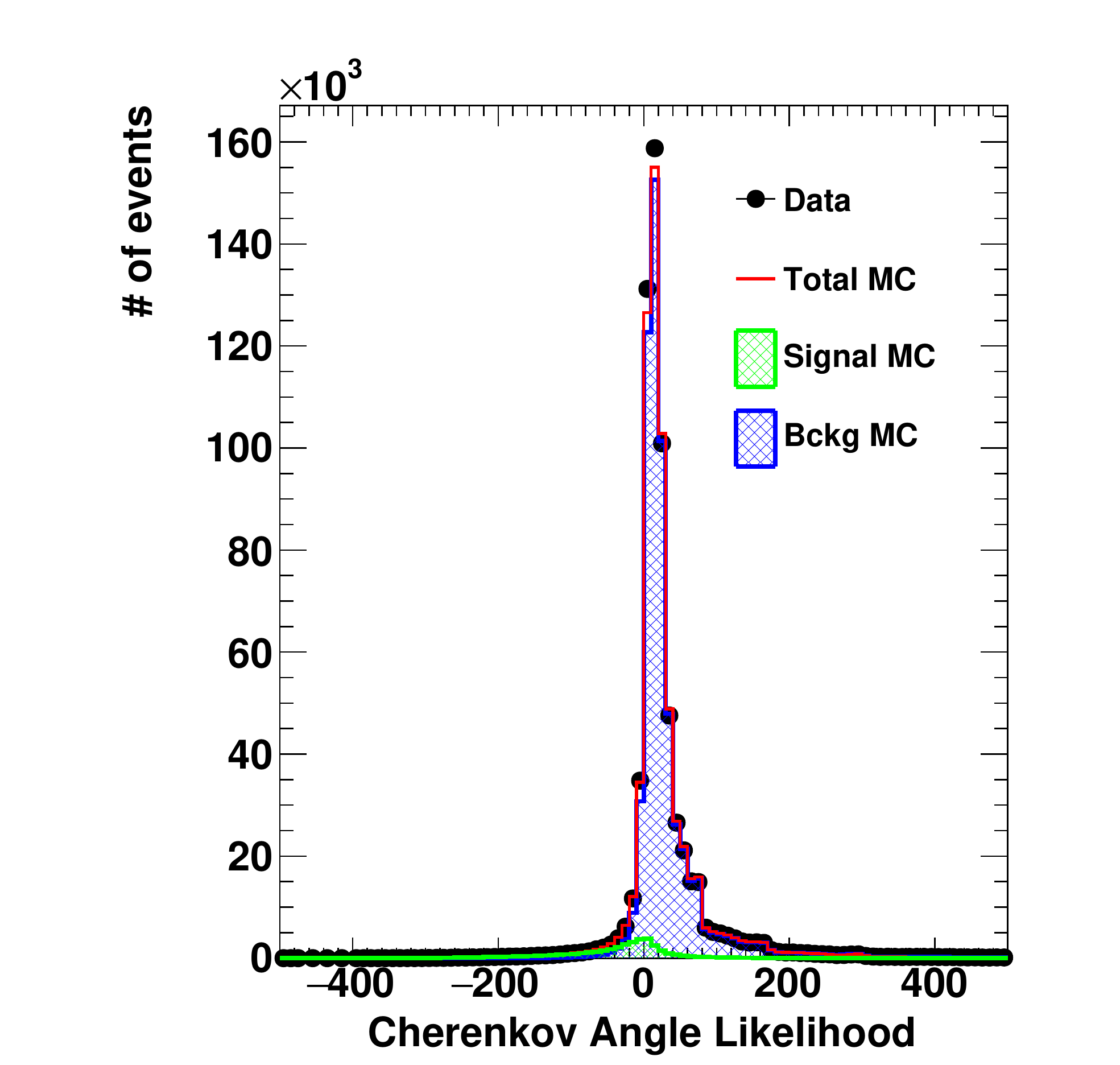}
  \end{center}  
  \caption{Distributions of the acceptance parameter,
    $P_{\rm{Acceptance}}$ (left) and 
    the Cherenkov angle likelihood parameter,
    $L_\text{Cherenkov}$ (right).
    $P_{\rm{Acceptance}}$ is larger for the signal events and 
    $L_\text{Cherenkov}$ is expected to be smaller for the signal,
    respectively.
  }\label{fig:neutron4x.accep_likelihood}
\end{figure}

\begin{figure}[!ht]
  \begin{center}
	\includegraphics[width=0.45\textwidth]{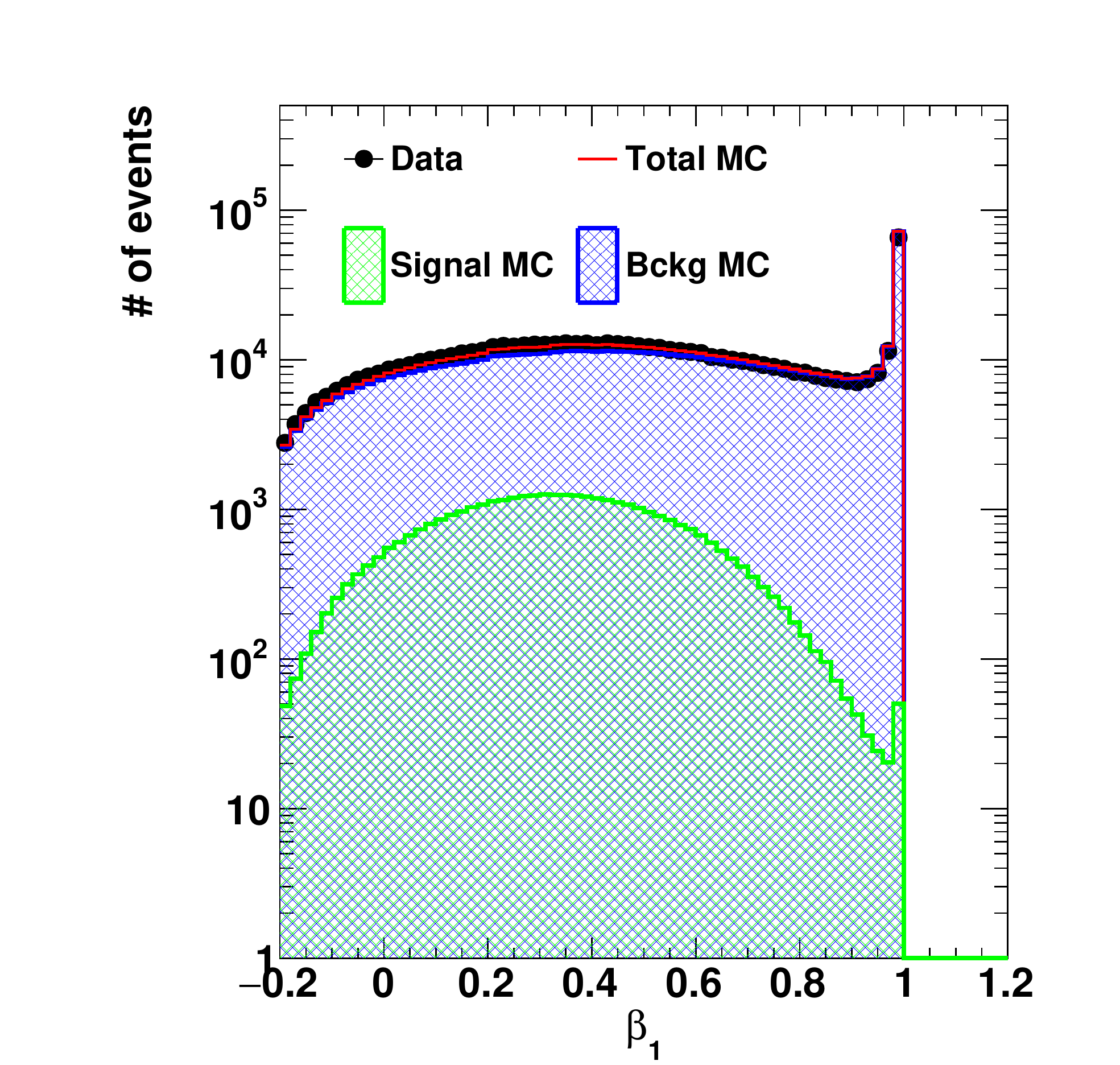}
	\includegraphics[width=0.45\textwidth]{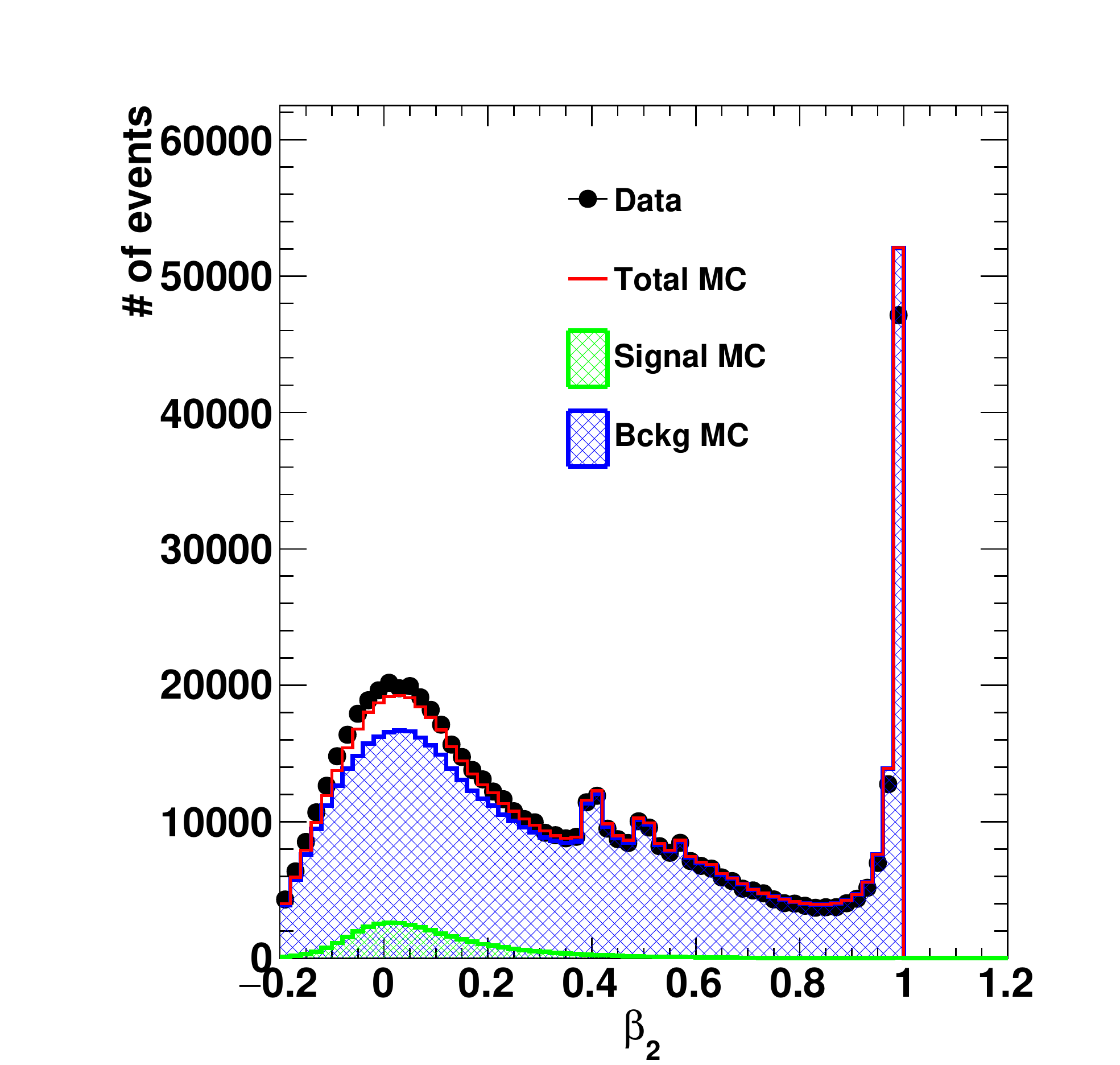}
	\includegraphics[width=0.45\textwidth]{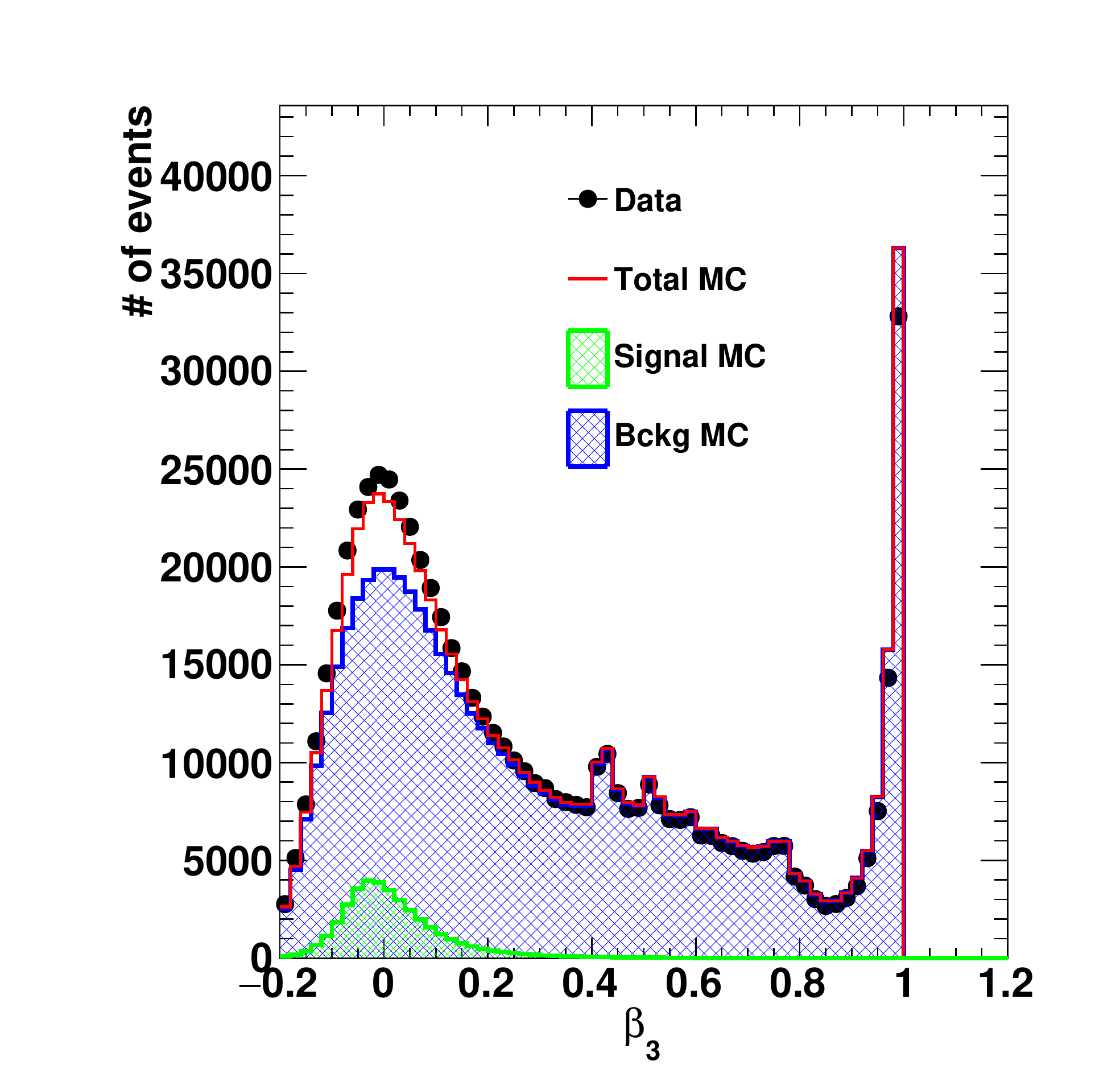}
	\includegraphics[width=0.45\textwidth]{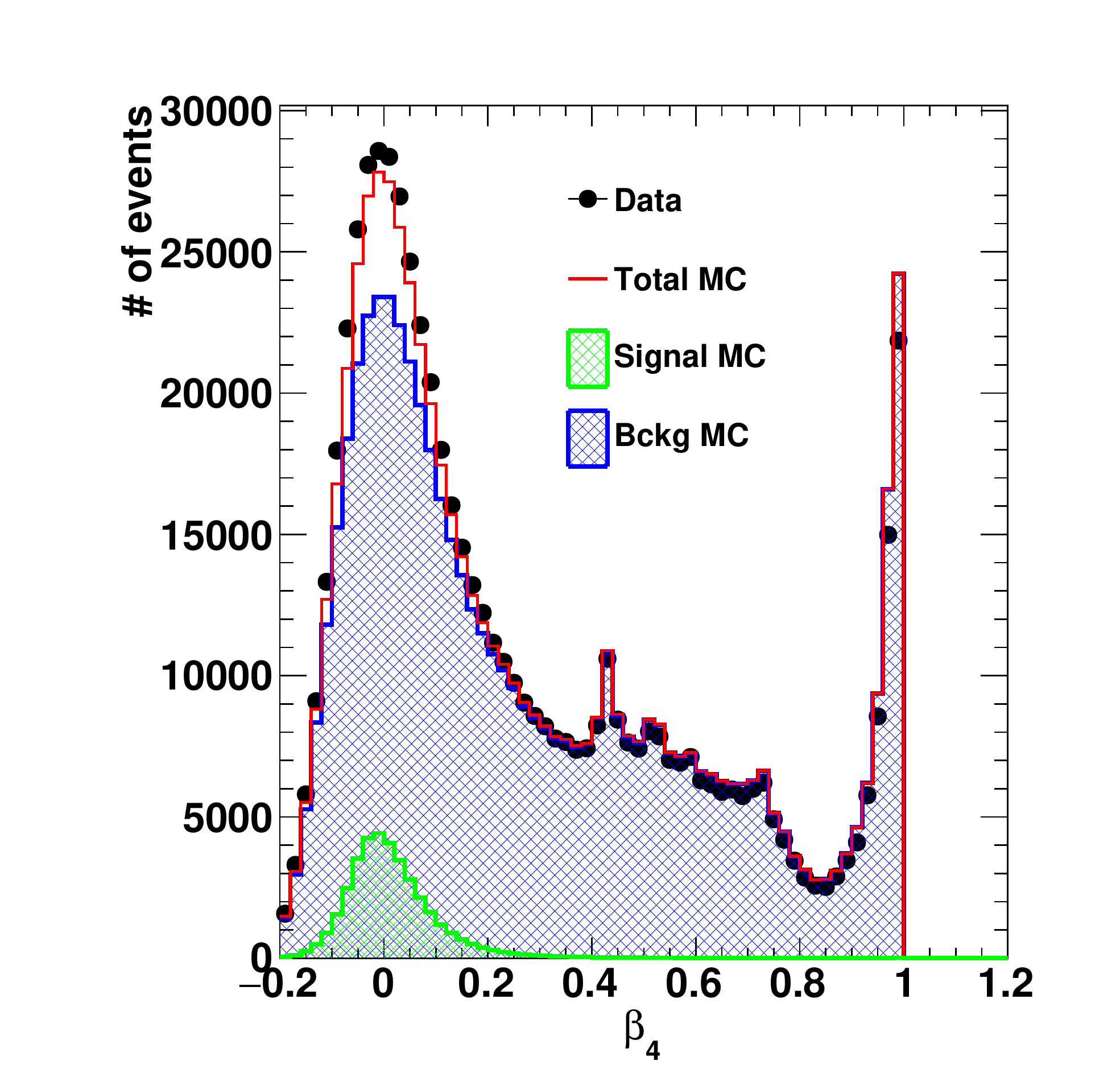}
	\includegraphics[width=0.45\textwidth]{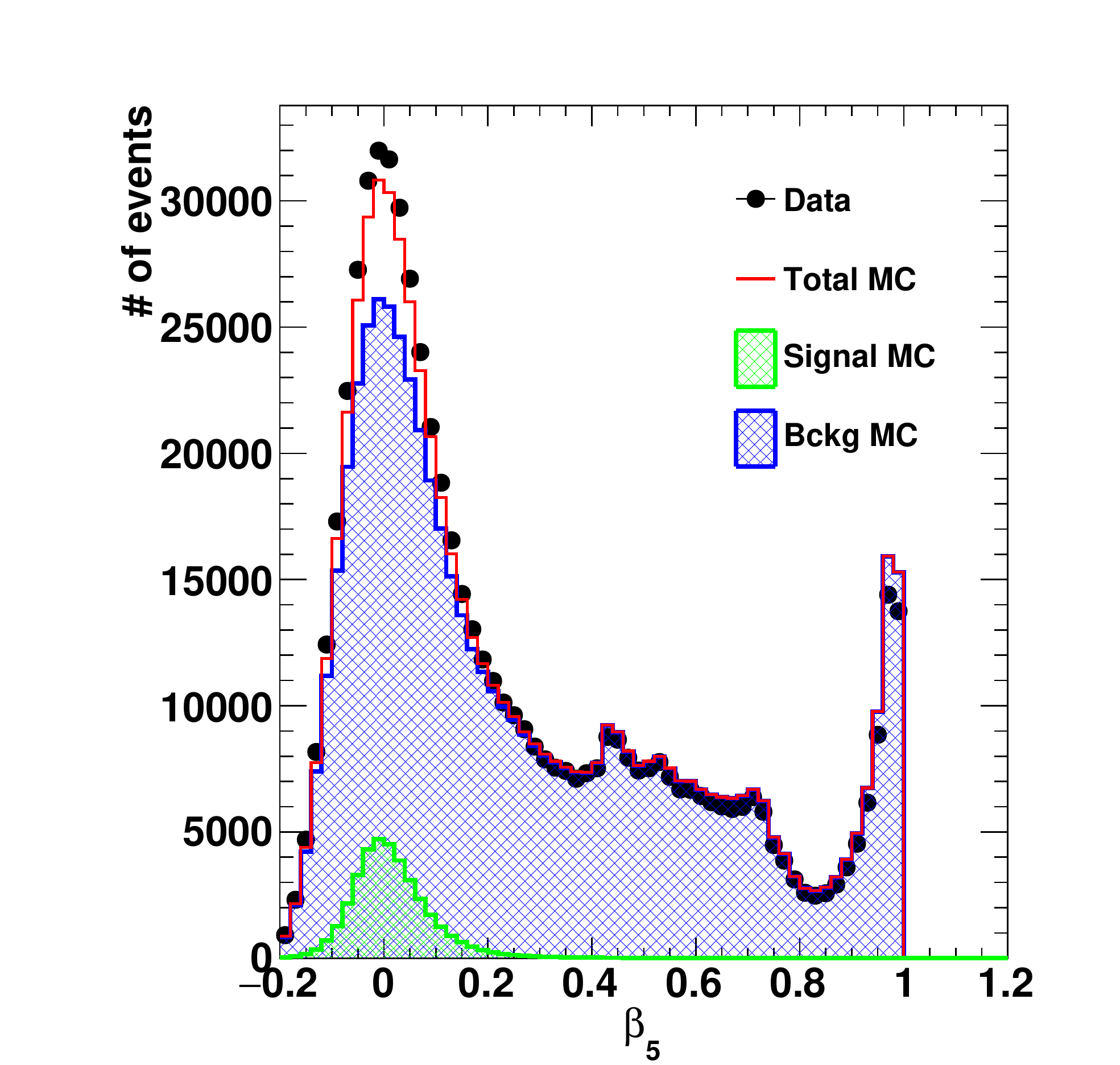}
  \end{center}
  \caption{Isotropy parameters $\beta_l$.
    The signal events concentrate around 0.35 for $\beta_1$ 
      and around 0 for other values of $l$, while backgrounds 
      have broader distributions and have peaks at around 1.}
  \label{fig:Isotropy}
\end{figure}

\begin{figure}[!ht]
  \begin{center}
    \includegraphics[width=0.49\textwidth]{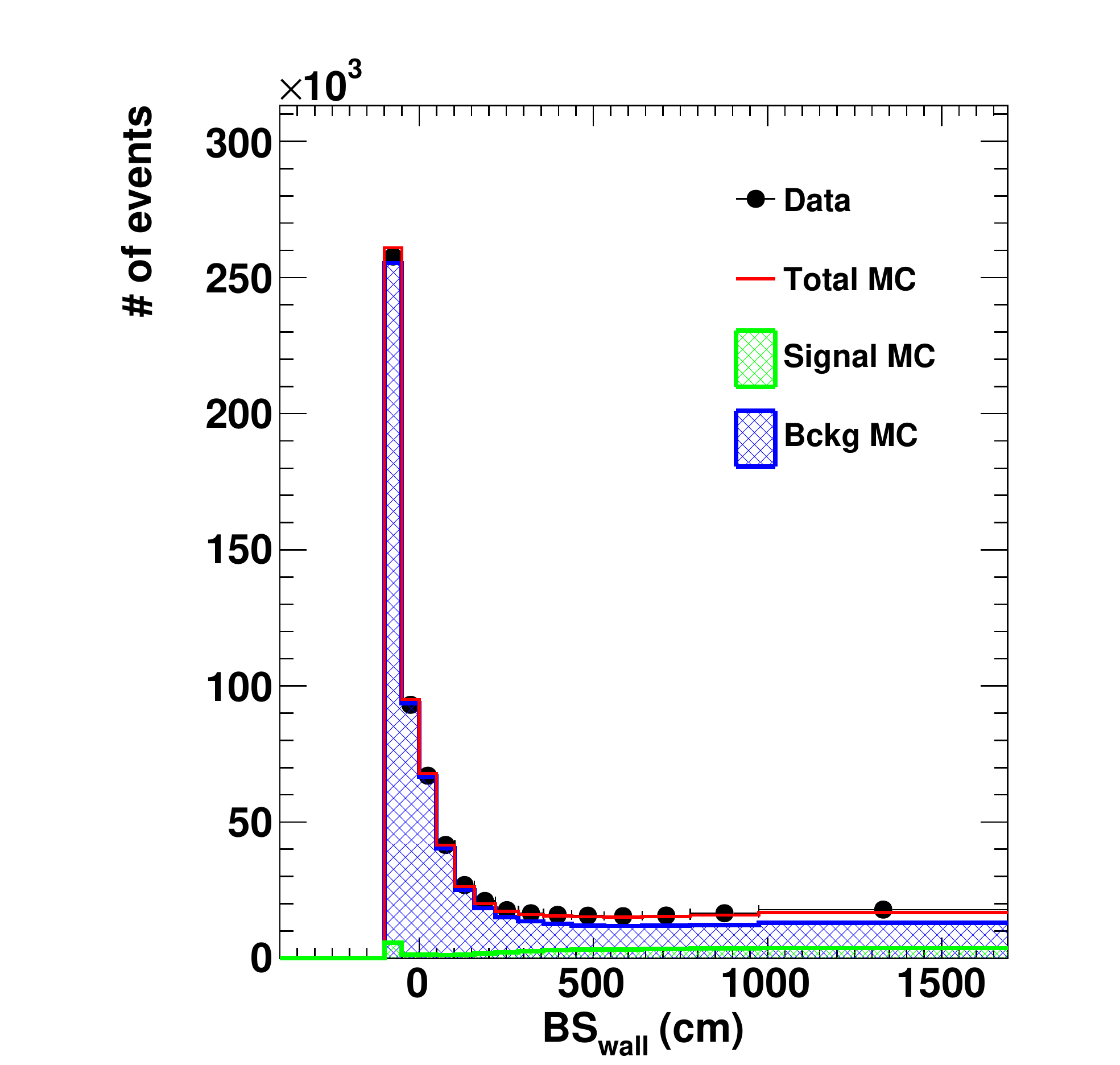}
    \includegraphics[width=0.49\textwidth]{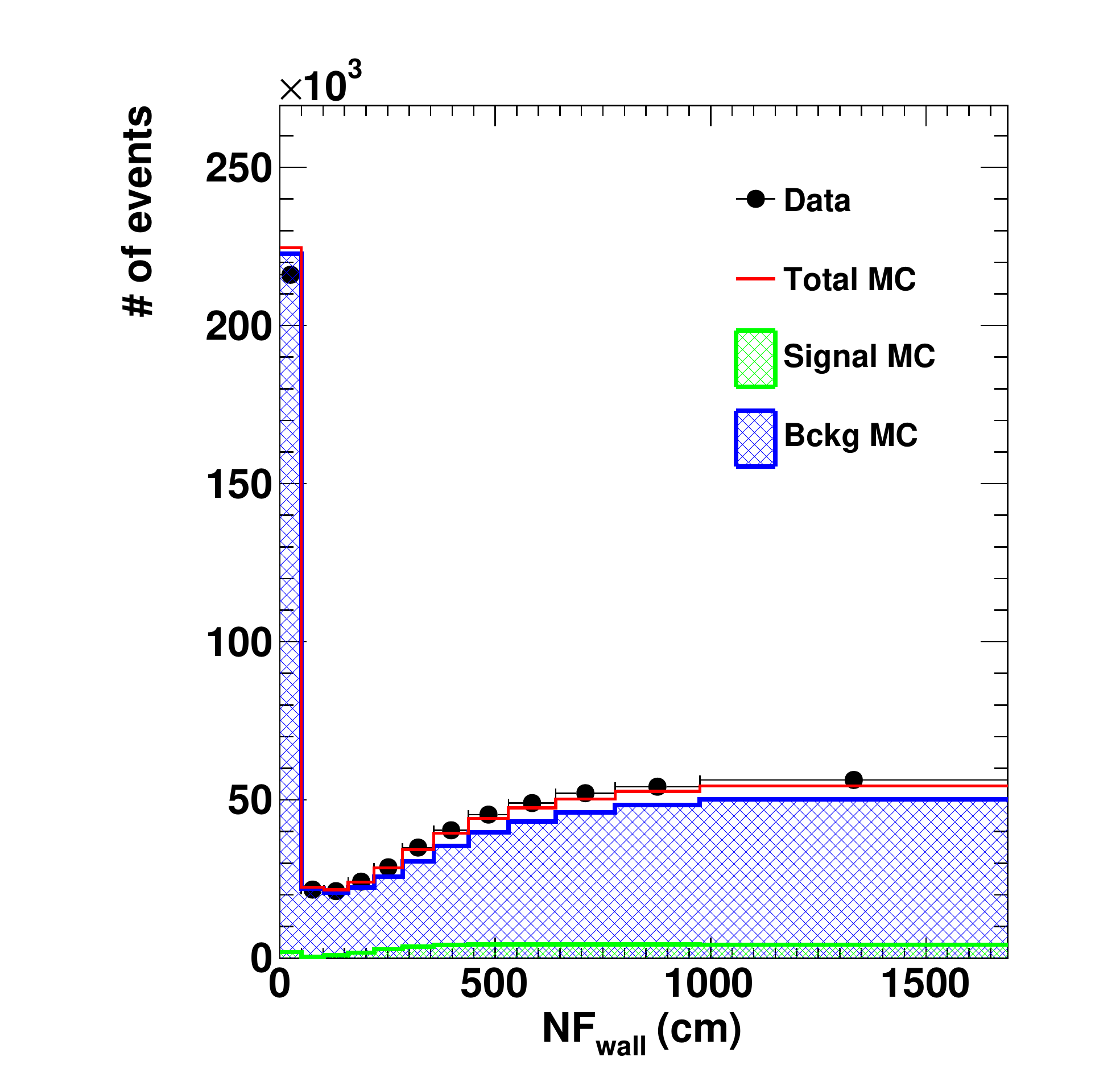}\\
    \includegraphics[width=0.49\textwidth]{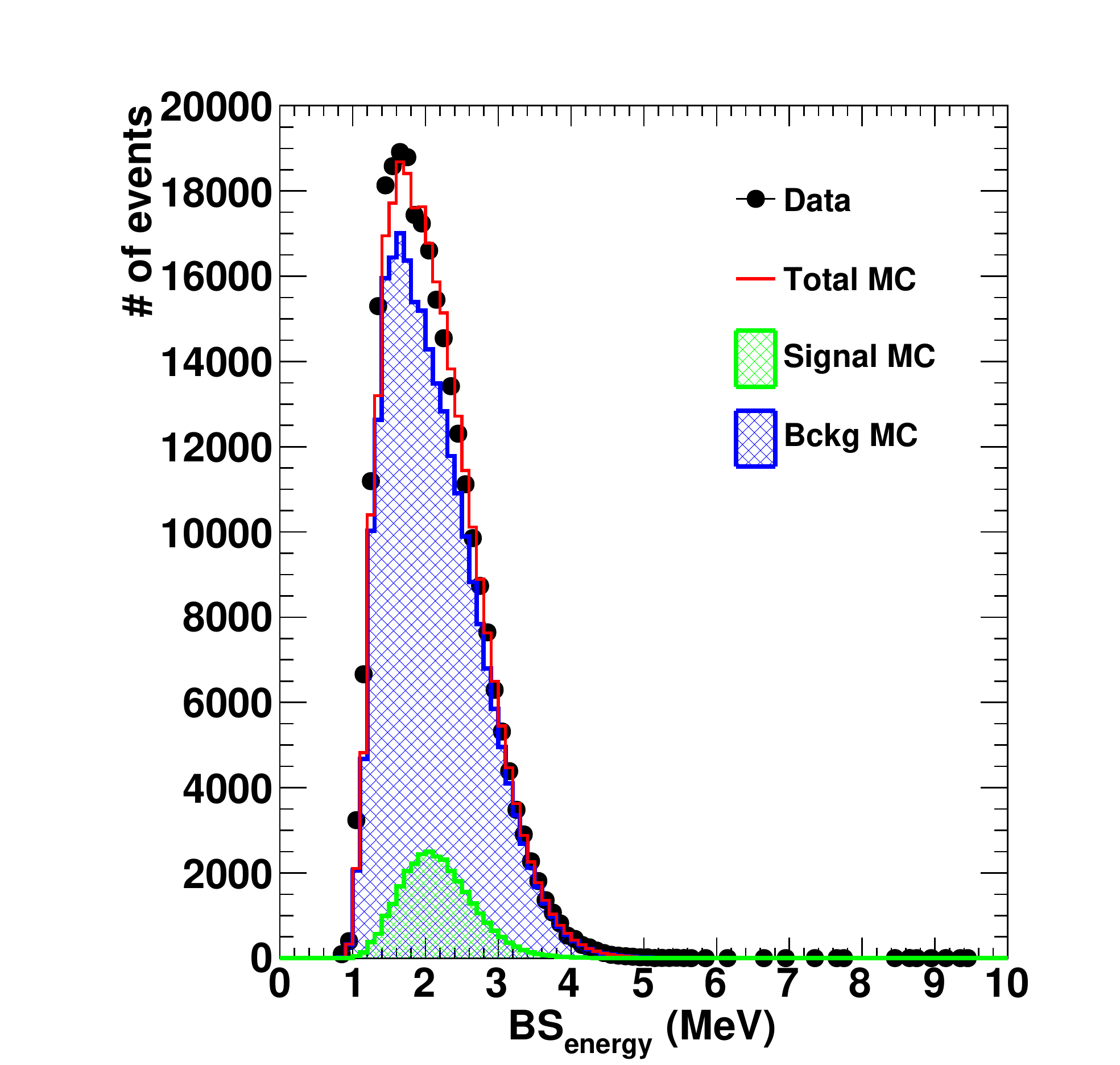}
  \end{center}
  \caption{Distributions of the distance between the neutron vertex 
    reconstructed by BONSAI and the nearest inner detector wall,
    $BS_{\rm wall}$ (top left), 
    the distance between the neutron vertex 
    reconstructed by Neut-Fit and the nearest inner detector wall,
    $NF_{\rm wall}$ (top left), 
    and the energy reconstructed by BONSAI,
    $BS_{\rm energy}$ (bottom).
    The normalization of $BS_{\rm energy}$ is 
    different for the signal and the background, as many background 
    events fail to be reconstructed at this stage.
    The background events concentrate close to the wall
    but the signal does not, as shown in both top plots, 
    $BS_{\rm wall}$ and $NF_{\rm wall}$, if signal events are 
    correctly reconstructed.
    The reconstructed energy, $BS_{\rm energy}$, is expected to have 
    peak at 2.2 MeV for signal but not for background.
  }\label{fig:neutron4x.bswall_fwall_bse}	 
\end{figure}

\begin{figure}[!ht]
  \begin{center}
  	\includegraphics[width=.49\textwidth]{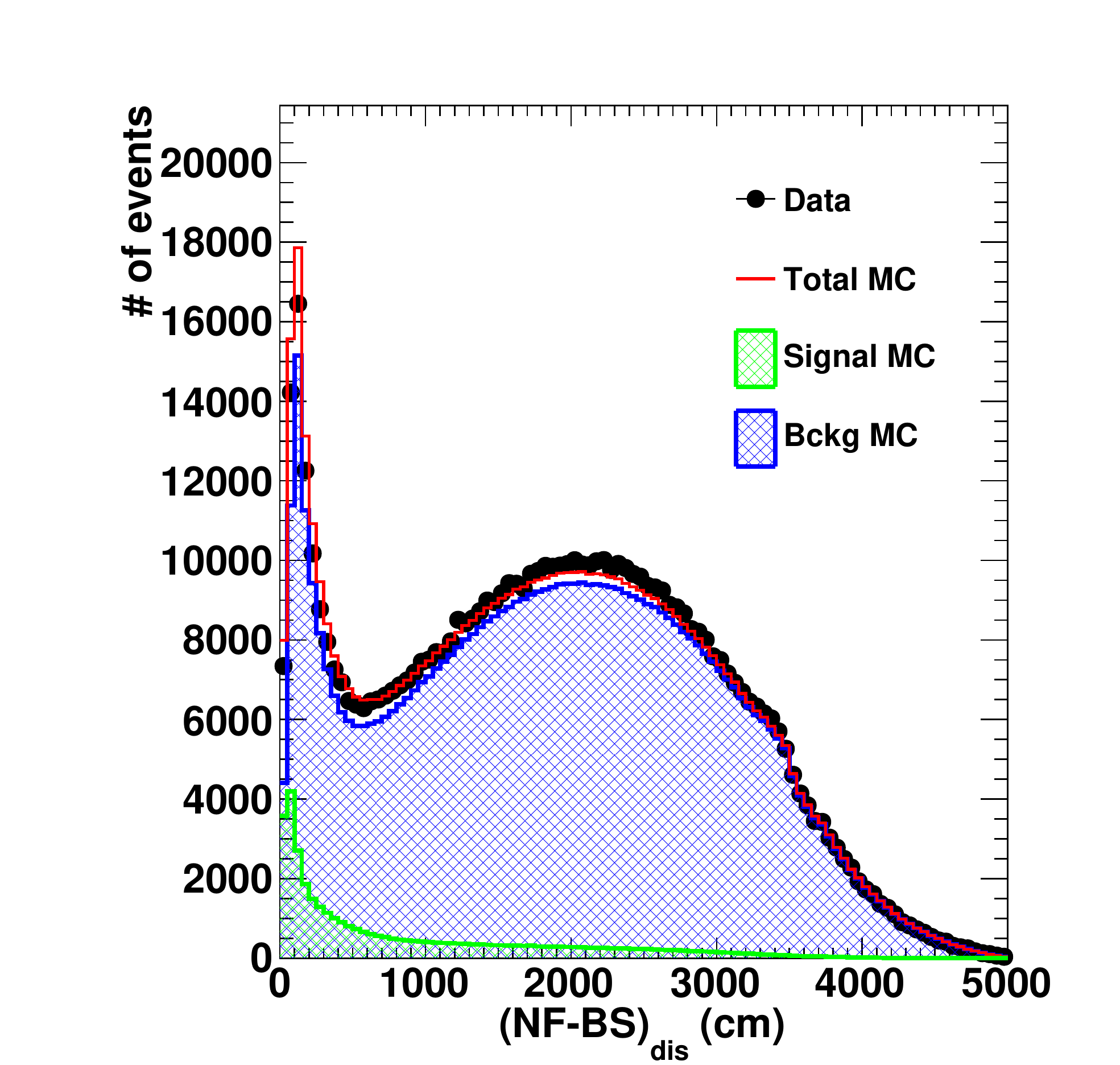}
  	\includegraphics[width=.49\textwidth]{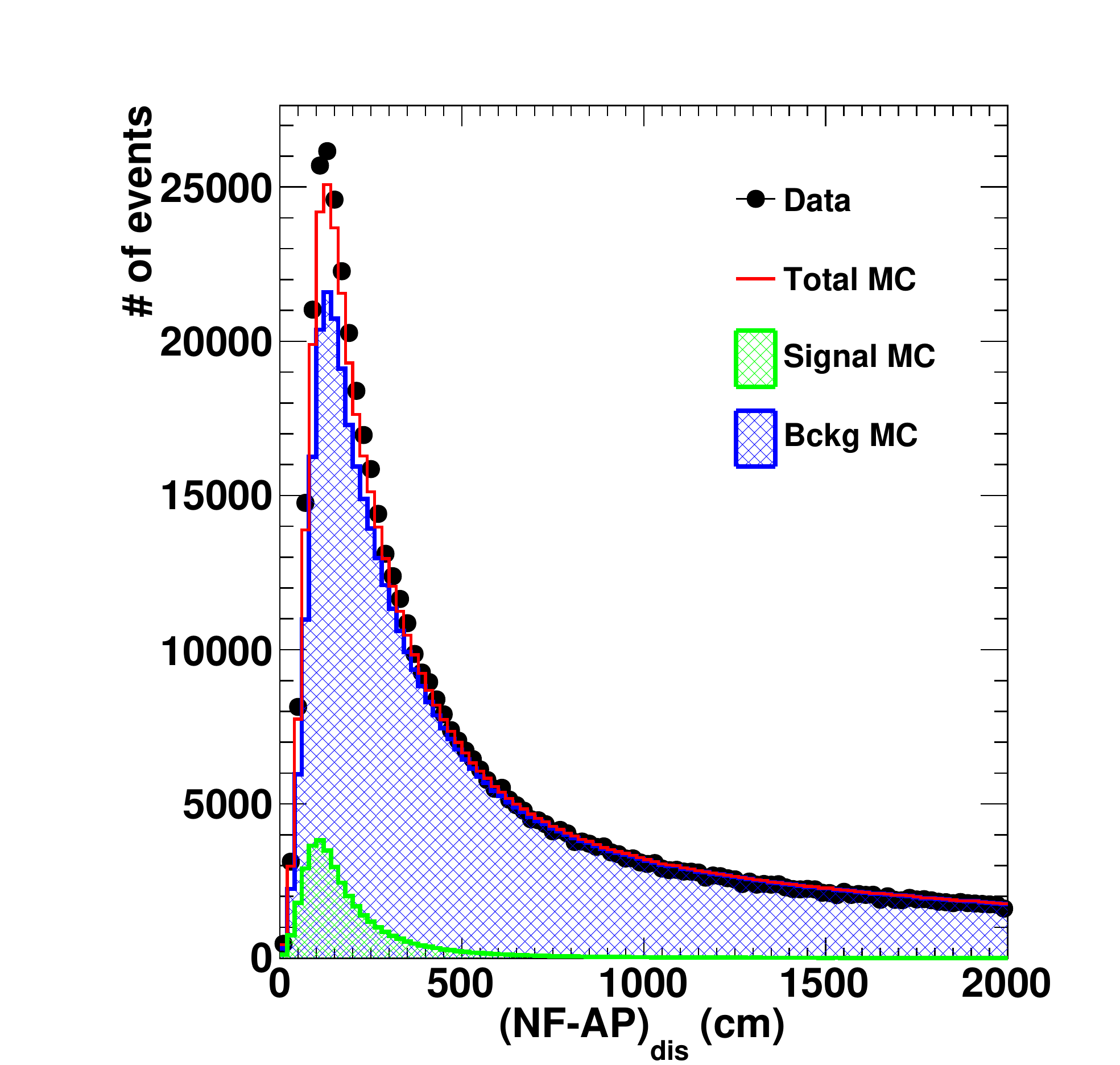}
  \end{center}
    \caption{Distributions of the distance between 
      the reconstructed verticies of Neut-Fit and BONSAI,
      $(NF-BS)_{\rm dis}$ (left), 
      and the distance between Neut-Fit the reconstructed 
      verticies of Neut-Fit and APFit,
      $(NF-AP)_{\rm dis}$, (right).
      These two values are expected to have a peak
      close to 0 because the vertexes from the two different
      reconstructions are expected to be similar for signal but 
      this is not necessarily true for the background. 
    }\label{fig:neutron4x.bfdis_fpdis}	
\end{figure}

\begin{figure}[!ht]
  \begin{center}
    \includegraphics[width=0.5\textwidth]{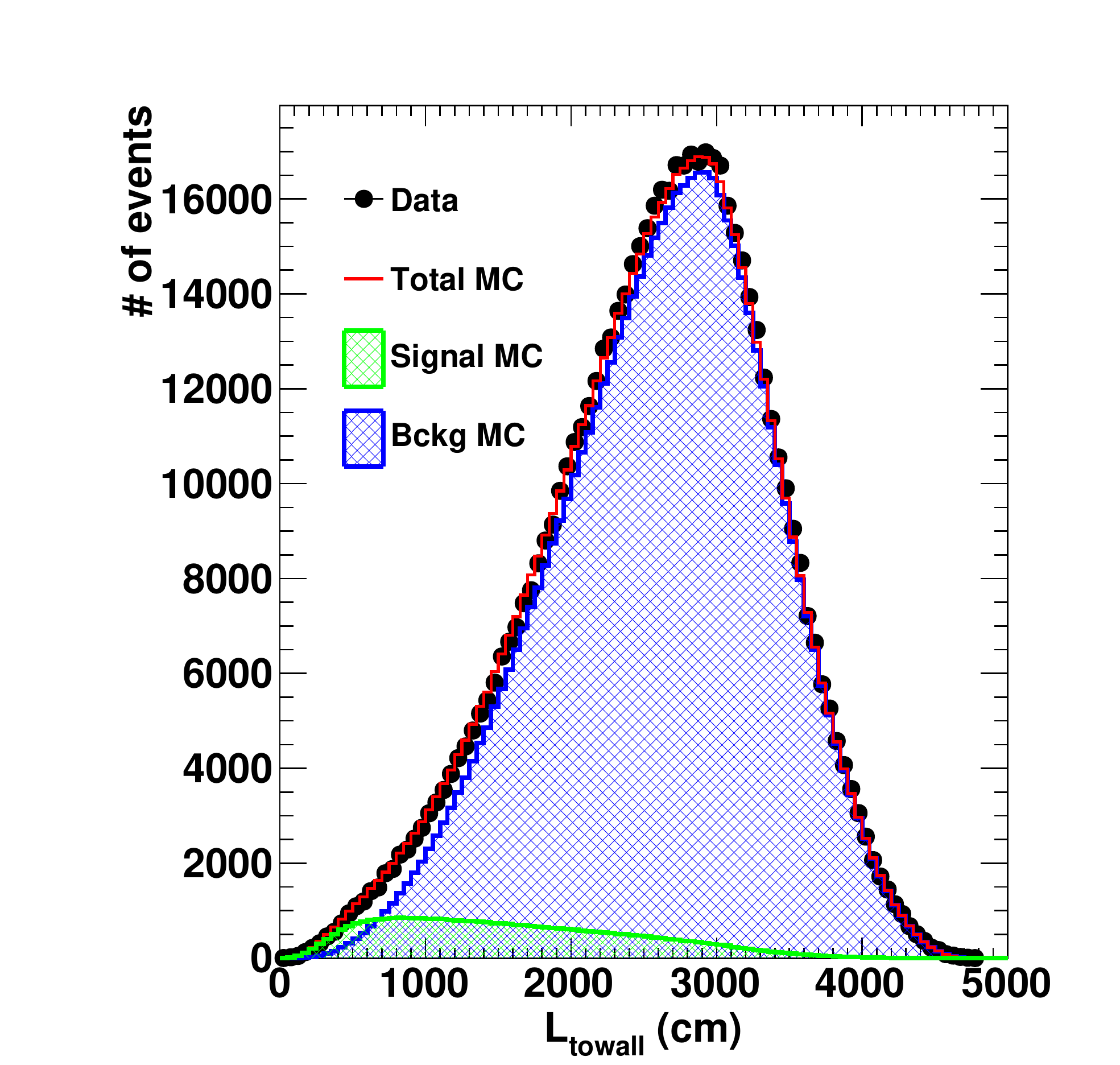}
  \end{center}
  \caption{Distribution of the distance from Neut-Fit's 
    reconstructed vertex to the ID wall in the direction 
    of the particle, $L_\text{towall}$.
    The background events tend to have larger value compared 
    to the signal.
  }\label{fig:neutron4x.towall}
\end{figure}

In this section, distributions of all the input variables to the 
neural net for neutron tagging are shown from \cref{fig:neutron4x.n10_n300} to \cref{fig:neutron4x.towall}. In each figure, both
data and the Monte-Carlo simulation outputs are shown.
In each plot, the green histogram corresponds to the neutron 
capture signal, the blue hatched area corresponds to the background,
the red histogram shows their sum and the overlaid black points 
show the data, respectively.
Also, the red histogram is normalized to the number of neutron 
capture events in the atmospheric neutrino event samples from 
the full SK-IV data set, which spanned 3,244.39 days between 
October 2008 and May 2018.


\clearpage

\section{Previous analysis method}\label{section:previous_method}
In previous analyses~\cite{Miura:2016krn,TheSuper-Kamiokande:2017tit,Kachulis:2017nci,Tanaka:2020emn} an older 
tagging algorithm was employed 
that did not make use of the time-clustered noise rejection and used 
$N^\text{RAW}_{10}$ instead of $N_{10}$ in its selections. 
The initial selection criteria required $N_{10}$ to be larger than 6.
In addition, the following parameters were not used in the neural network:
Acceptance, Cherenkov angle, isotropy and $L_\text{towall}$.
Instead, the number of hits on low-probability PMTs ($N_{\rm low}$, defined
below) was used.
The neutron tagging efficiency was estimated to be 20.7\%. 
With this algorithm, the neutron tagging efficiencies between Am-Be data
and the MC differed by at most 20\%.
This value was used as the systematic uncertainty for the algorithm.

\subsection{Number of Hits on Low-Probability PMTs: $N_{\rm low}$}

\begin{figure}[!ht]
  \begin{center}
	\includegraphics[width=0.6\textwidth]{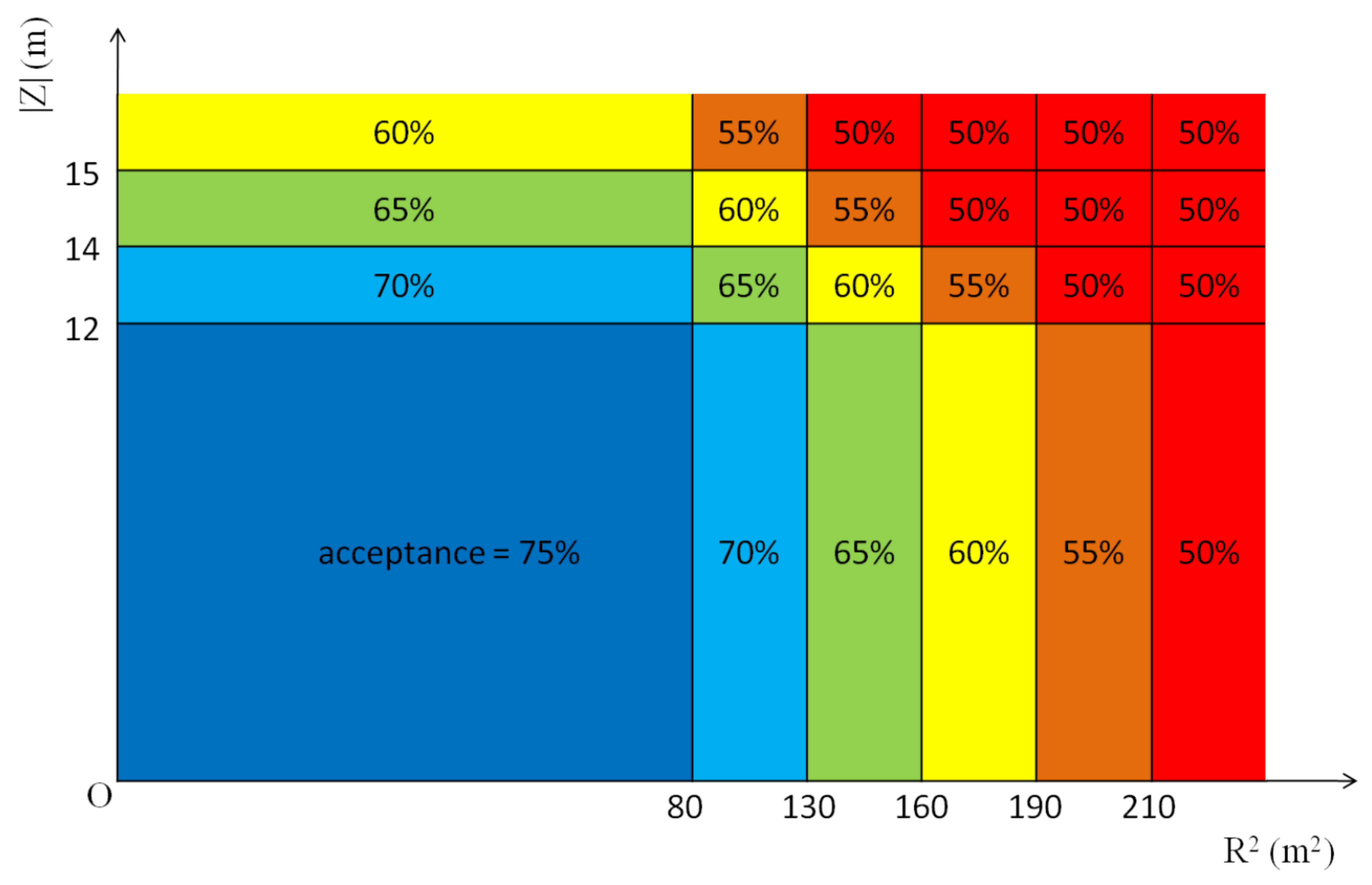}
  \end{center}
	\caption{Varying acceptance requirements for the $N_{\rm low}$
      cut as a function of tank coordinates.}
	\label{fig:nlowacceptance}
\end{figure}

Noise hits are expected to be distributed in a geometrically uniform manner while real photons from $\gamma$
induce a PMT position dependence because of their 
directionality. 
The PMT hit probability is defined in ~\cref{eqn:acceptance}.
In order to define a low-probability PMT, a threshold is defined
which depends on the vertex location in the detector, as shown in 
~\cref{fig:nlowacceptance}.

First, PMTs are sorted in order of their photon detection
probability.  
The probability value of each PMT is summed starting
from the highest value and the running sum is compared with a
threshold. 
When the sum exceeds this threshold the last PMT added and
any remaining PMTs are regarded as low-probability PMTs.  The
threshold table has a vertex position dependence.

\begin{figure}[!ht]
  \begin{center}
  \includegraphics[width=0.6\textwidth]{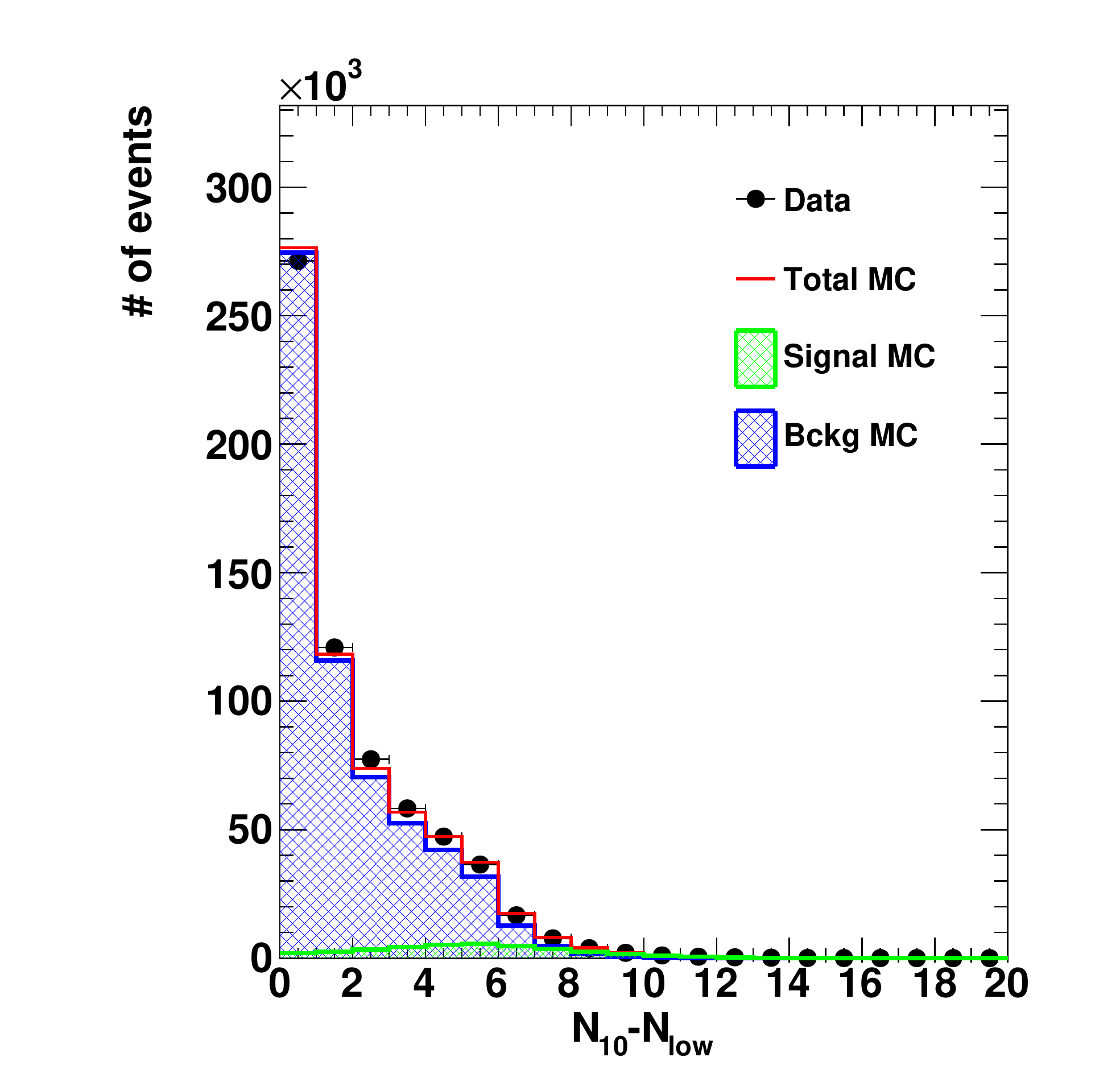}
  \end{center}
  \caption{Hits on low-probability PMTs, $N_{10} - N_{\rm
      low}$. The $N_{\rm low}$ parameter is expected to be small for signal
    events. The green histogram corresponds to the neutron capture
    signal and the hatched area shows the background.
    The red histogram shows the sum of the signal and the background, which has been 
    normalized to the number of capture events in data (black dots).}
  \label{fig:neutron4x.nlow}
\end{figure}

~\cref{fig:neutron4x.nlow} shows distributions of 
$N_{10} - N_{\rm low}$ which was used in as an input 
to the neural network.

\clearpage

\section{Glossary}
\begin{table}[htb]
\caption{Acronyms}
\label{ListOfSynonims}
\begin{center}
\begin{tabular}{|l|l|}
\hline
\multicolumn{2}{|l|}{General}\\
\hline
HV   & high voltage \\
MC   & Monte-Carlo \\
MLP  & mulit layer perception \\
PMT  & photo multiplier tube \\
ToF  & time of flight \\
\hline
\multicolumn{2}{|l|}{Super-Kamiokande detector}\\
\hline
SK   & Super-Kamiokande \\
ID   & inner detector \\
OD   & outer detector \\
\hline
\multicolumn{2}{|l|}{Event categories of SK}\\
\hline
FC   & fully-contained \\
PC   & partially-contained \\
UPMU & upward-going muon \\
\hline
\multicolumn{2}{|l|}{Names of the event triggers}\\
\hline
SLE  & super low enegry (trigger)\\
LE   & Low energy (trigger) \\
HE   & High energy (trigger) \\
SHE  & super high energy (trigger)\\
AFT  & after (trigger) \\
\hline
\multicolumn{2}{|l|}{Calibration sources}\\
\hline
Am-Be & americium-beryllium \\
BGO   & bismuth germanium oxide (inorganic scintillator) \\
\hline
\multicolumn{2}{|l|}{Event reconstruction tools}\\
\hline
APFit(AP) & An event reconstruciton software library developed for \\
          & the atmospheric neutrino and theproton decay analyses of SK \\
BONSAI(BS)& An event reconstruciton software library developed for \\
          & the solar and supernova neutrino analyses \\
Neut-Fit(NF) & Newly developed simple vertex reconstruction software \\
\hline
\multicolumn{2}{|l|}{Software libraries} \\
\hline
NEUT   & A neutrino-nucleus interaction simulation program library \\
GEANT3 & Software library of detector description and simulation \\
GEANT4 & Software library of detector description and simulation \\
FLUKA  & Fully integrated particle physics Monte-Carlo simulation package \\
GFLUKA & A versions of FLUKA designed to be used with GEANT3 \\
CALOR  & Hadronic interaction simulation package \\
MICAP  & Low energy neutron, ion and gamma-ray transport software library \\
SKDETSIM & A detctor simulation software libray developed for SK \\
ROOT   & A data analysis framework \\
TMLP   & TMulitLayerPerception (TMLP) library  \\
\hline
\end{tabular}
\end{center}
\end{table}
\clearpage
\acknowledgments
We gratefully acknowledge the cooperation of the Kamioka Mining and 
Smelting Company. The Super-Kamiokande experiment has been built and 
operated from funding by the Japanese Ministry of Education, Culture, 
Sports, Science and Technology, the U.S. Department of Energy, and 
the U.S. National Science Foundation. Some of us have been supported 
by funds from the National Research Foundation of Korea NRF-2009-0083526
(KNRC) funded by the Ministry of Science, ICT, and Future Planning 
and the Ministry of Education (2018R1D1A3B07050696, 2018R1D1A1B07049158), 
the Japan Society for the Promotion of Science, the National Natural 
Science Foundation of China under Grants No. 11235006, the Spanish 
Ministry of Science, Universities and Innovation 
(grant PGC2018-099388-B-I00), the Natural Sciences and Engineering
Research Council (NSERC) of Canada, the Scinet and Westgrid consortia 
of Compute Canada, the National Science Centre, Poland 
(2015/18/E/ST2/00758), the Science and Technology Facilities Council 
(STFC) and GridPP, UK, the European Union's Horizon 2020 Research 
and Innovation Programme under the Marie Sklodowska-Curie grant 
agreement no.754496, H2020-MSCA-RISE-2018 JENNIFER2 grant agreement 
no.822070, and H2020-MSCA-RISE-2019 SK2HK grant agreement no. 872549.

\bibliographystyle{JHEP}

\bibliography{ntag_for_JINST_rev1}

\end{document}